\documentclass[12pt]{article}
\pdfoutput=1
\usepackage{amssymb}
\usepackage{amsmath}
\usepackage{amsthm}
\usepackage{cancel}
\usepackage{slashed}
\usepackage{fullpage}
\usepackage{color}
\usepackage{braket}
\usepackage{graphicx}
\usepackage[small]{subfigure}
\usepackage{multirow}
\usepackage{verbatim}
\usepackage{cite}
\usepackage{bigints}
\usepackage{simplewick}
\usepackage{authblk}
\usepackage{hyperref}
\usepackage{empheq}
\usepackage{tikz}
\usetikzlibrary{calc}
\usepackage{bbm}
\hypersetup{
    linktocpage,
     colorlinks,
     citecolor=darkgreen,
     linkcolor= darkgreen,
     urlcolor=darkgreen
}

\usepackage{feynmp}
\newcommand{\mhpole}{ {m_h^{\text{pole}}}}
\newcommand{\mtpole}{ {m_t^{\text{pole}}}}

\newcommand{\LNP}{ \Lambda_{\text{NP}} }

\definecolor{darkred}{rgb}{0.5,0.0,0.0}
\definecolor{darkblue}{rgb}{0.0,0.0,0.9}
\definecolor{darkerblue}{rgb}{0.0,0.0,0.5}
\definecolor{purple}{rgb}{0.5,0.0,0.5}
\definecolor{darkgreen}{rgb}{0.0,0.5,0.0}
\definecolor{black}{rgb}{0.0,0.0,0.0}
\definecolor{brown}{rgb}{0.6,0.4,0.2}
\definecolor{newpurple}{rgb}{0.65, 0.38, 0.61}
\definecolor{newyellow}{rgb}{0.9718, 0.6093, 0.0759}
\definecolor{amber}{rgb}{1.0, 0.75, 0.0}
\definecolor{newblue}{rgb}{0.4, 0.52, 0.85}
\definecolor{newred}{rgb}{0.8524, 0.2595, 0.3294}
\definecolor{newgreen}{rgb}{0.2, 0.8, 0.2}
\definecolor{SMgreen}{rgb}{0.56, 0.69, 0.19}
\definecolor{neworange}{rgb}{0.94, 0.462, 0.162}
\newcommand{\red}{\color{darkred}}

\newcommand{\blue}{\color{darkblue}}
\newcommand{\green}{\color{darkgreen}}

\newcommand{\wh}{\widehat}
\newcommand{\wt}{\widetilde}
\newcommand{\abs}[1]{\left| #1 \right|}

\newcommand{\Tr}{\text{Tr}}

\def\bigtau{\mathcal{T}}

\def\be{\begin{equation}}
\def\ee{\end{equation}}
\def\cL{\mathcal{L}}
\def\cC{\mathcal{C}}
\def\cO{\mathcal{O}}
\def\cD{\mathcal{D}}
\def\cM{\mathcal{M}}
\def\cN{\mathcal{N}}
\def\cK{\mathcal{K}}
\def\cR{\mathcal{R}}

\def\msbar{\overline{\text{MS}}}

\newcommand{\kappag}{\kappa}

\newcommand{\RW}{\widetilde R}

\numberwithin{equation}{section}

\newcommand{\bps}{\beta_{0\star}^\prime}

\newcommand{\lol}{\lambda_{\text{1-loop}}}
\newcommand{\im}{\text{Im}}
\newcommand{\re}{\text{Re}}
\newcommand{\lstar}{\lambda_\star}
\newcommand{\mstar}{\mu_\star}

\newcommand{\FV}{\text{FV}}

\newcommand{\eps}{\varepsilon}

\newcommand{\Sfin}{S_\text{fin}}
\newcommand{\Ssub}{S_\text{sub}}

\newcommand{\GeV}{ {\text{GeV}}}

\newcommand{\GRind}{ \Gamma_\text{no-R}}
\newcommand{\GR}{ \Gamma_R}
\newcommand{\Sbs}{ S[\phib^\star] }

\newcommand{\erqef}[1]{\eqref{#1}}

\newcommand{\phib}{ \phi_b }

\title{{ 
\vspace{-1cm}Scale-invariant Instantons and \\[1mm] 
the Complete Lifetime of the Standard Model}\\[5mm]}

\author{Anders Andreassen\thanks{anders@physics.harvard.edu}  }
\author{William Frost\thanks{wfrost@physics.harvard.edu}  }
\author{Matthew D. Schwartz\thanks{schwartz@physics.harvard.edu} }

\affil{\emph{Department of Physics,
Harvard University, Cambridge, MA 02138, USA}}

\begin{document} 

\date{}
\maketitle
\thispagestyle{empty}

\begin{abstract}
In a classically scale-invariant quantum field theory, tunneling rates are infrared divergent due
to the existence of instantons of any size.
While one expects such divergences to be resolved by quantum effects, it has been unclear how 
higher-loop corrections can resolve a problem appearing already at one loop.
With a careful power counting, we uncover a series of loop contributions 
that dominate over the one-loop result and sum all the necessary terms. We also clarify 
previously incomplete treatments of related issues pertaining to global symmetries, gauge fixing and finite mass effects. 
In addition, we produce exact closed-form solutions for the functional determinants over scalars, fermions and vector bosons around the scale-invariant bounce, demonstrating manifest gauge invariance in the vector case.\\

With these problems solved, we produce the first complete calculation of the lifetime of our universe: $10^{161}$ years. With $95\%$ confidence, we
expect our universe to last more than $10^{65}$ years.
The uncertainty is part experimental uncertainty on the top quark mass and on $\alpha_s$
and part theory uncertainty from electroweak threshold corrections. Using our complete
result, we provide  phase diagrams in the $m_t/m_h$ and the $m_t/\alpha_s$ planes,  with uncertainty bands.
To rule out absolute stability to 3$\sigma$ confidence, the uncertainty on the top quark pole mass would have to be pushed below
250 MeV or the uncertainty on $\alpha_s(m_Z)$ pushed below $0.00025$.

\end{abstract}
\newpage

\tableofcontents
\newpage
\section{Introduction}
Tunneling through a barrier is a quintessentially quantum phenomenon. 
In quantum mechanics (QM), tunneling has been studied analytically, numerically and experimentally, leading to a consistent and comprehensive
picture of when and how fast tunneling occurs. In quantum field theory (QFT), much less is known. In QFT, one cannot solve
the Schr\"odinger equation, even numerically, due to the infinite dimensionality of the Hilbert space. The only approach to calculating
tunneling rates in QFT seems to be through the saddle point approximation of the path integral \cite{Langer:1967ax,Kobzarev:1974cp,Coleman:1977py,Callan:1977pt,Hammer:1978xu}. This approach involves analytic
continuation in an essential way. Because tunneling in QFT has important implications, such as for the stability of the Standard Model vacuum 
\cite{Frampton:1976kf,Frampton:1976pb,Sher:1988mj,Casas:1994qy,Espinosa:1995se,Isidori:2001bm,Espinosa:2007qp,
Ellis:2009tp,Degrassi:2012ry,Buttazzo:2013uya,Lalak:2014qua,Andreassen:2014gha,Bednyakov:2015sca,Branchina:2014rva,Iacobellis:2016eof}
and because QFT tunneling rates are nearly impossible to measure experimentally, it is critical to make sure the rather abstract formalism  is actually capable of calculating something physical. 

A number of the subtleties in going from QM to QFT were resolved long ago, some more recently, and some challenges still exist. 
For example, while tunneling rates are physical and therefore should be gauge invariant, it has been challenging to check directly that this is the case. Although exact non-perturbative proofs of gauge-invariance exist~\cite{Nielsen:1975fs,Fukuda:1975di} and there have been many investigations into gauge-dependence~\cite{Kang:1974yj,Dolan:1974gu,Frere:1974ia,Kunstatter:1985yt,DoNascimento:1987mn,Buchmuller:1994vy,Metaxas:1995ab,Boyanovsky:1996dc,Lin:1998up,Binosi:2005yk,Wainwright:2011qy,Patel:2011th,Garny:2012cg,Bednyakov:2015sca,Lalak:2016zlv} it has not been shown that gauge-invariance holds order-by-order in perturbation theory, as it does for $S$-matrix elements. For some context, recall that 
for the simpler question of whether a state is absolutely stable in the quantum theory, it was found that the corresponding bound {\it was} gauge-dependent with then-current perturbative methods \cite{Andreassen:2013hpa,Andreassen:2014eha,Andreassen:2014gha}. The problem was traced to an inconsistent power counting and improper use of the renormalization group equations. A consistent method was recently developed in~\cite{Andreassen:2014eha,Andreassen:2014gha}, with non-negligible implications for precision top and Higgs-boson mass bounds in the Standard Model.
  Recently, progress was made in understanding the gauge invariance of tunneling rates by Endo et al.~\cite{Endo:2017gal,Endo:2017tsz}; these authors showed explicitly that the rate is gauge-invariant to one loop for general massive scalar scalar field  theory backgrounds and we build upon their results. 
 
 In fact, gauge invariance is the least of our worries. 
In order to produce a precision calculation of the tunneling rate -- or even the leading order rate with the correct units -- one must understand a whole slew of subtleties not relevant for the absolute stability bound. First of all, there are suspicious elements in the common derivations~\cite{Kleinert,Muller-Kirsten:2012wla,Zinn-JustinPI,ZinnJustin:2002ru,Marino,Weinberg:2012pjx}
 of the Callan-Coleman decay rate formula \cite{Coleman:1977py,Callan:1977pt}. The leading-order confusion is that the rate is said to be determined, even in QM, by taking the imaginary part of $\langle a | e^{-H T} |a \rangle$, a manifestly real expression.  The resolution of this paradox involves not analytic continuation of the potential,
as is often cited, but the specification of complex paths to be integrated over in the path integral~\cite{Behtash:2015zha,Witten:2010cx}.  A more physical derivation of a decay rate in QFT was presented recently in~\cite{Andreassen:2016cff,Andreassen:2016cvx}. Some elements are reviewed in Section~\ref{sec:tunnel}.

Even if we ignore gauge-dependence and trust the decay rate formulas, we encounter a new roadblock in trying to evaluate tunneling rates in QFTs like the Standard Model,  due to classical scale invariance. 
The basic issue with scale invariance can be seen in the Gaussian approximation to the path integral around a reference field configuration $\phib$:
\be
\frac{\Gamma}{V} \sim \frac{1}{T V} \int \cD \phi e^{-S[\phib + \phi]} \approx \frac{1}{T V}  \int \cD \phi e^{-S[\phib] - \frac{1}{2} \phi S''[\phib] \phi} 
\label{saddle}
\ee
One typically evaluates the right-hand-side by expanding the fluctuations $\phi$ in a basis of eigenfunctions of the operator $S''[\phib]$.
If the action has a symmetry spontaneously broken by $\phib$, then there will be fluctuation directions $\phi_0$
with zero eigenvalue, that is, for which $S''[\phib] \phi_0 = 0$. Integrating over $d\xi_0$ in the field direction $\phi=\xi_0 \phi_0$ then leads to an infrared divergence in Eq.~\eqref{saddle}. Examples include the zero modes associated with translation invariance where $\phi_0 \propto \partial_\mu \phib$ or scale invariance
where  $\phi_0 \propto (1 + x^\mu \partial_\mu) \phib$. For translations, the infrared divergence is expected -- it generates a factor of $V T$ so that the rate is extrinsic, a decay rate per unit volume. For scale invariance, the infrared divergence has no natural volume cutoff and so the decay rate is apparently infinite.

Anyone with even minimal familiarity with QFT would immediately guess that the resolution to the 
scale-invariance divergence is related to dimensional transmutation~\cite{Coleman:1973jx}, that the classical scale invariance is broken by quantum effects. Unfortunately, connecting the $\beta$-functions to the decay rate calculation within a consistent perturbative framework has remained elusive. In fact, there are two related technical difficulties. 

First of all, to integrate over a zero-mode fluctuation, one must use a collective coordinate~\cite{Langer:1967ax,Gervais:1974dc,Gervais:1975pa,Callan:1975yy,Jevicki:1976kd} rather than an infinitesimal fluctuation. For example, with translations, one must integrate over $x_0^\mu$ parametrizing fields $\phib(x^\mu + x_0^\mu)$ in the path integral before the Gaussian approximation is applied (in the middle expression in Eq.~\eqref{saddle} not the rightmost one). The difference between $x_0^\mu$ and coefficients $\xi^\mu$ of fluctuations in the $\partial_\mu \phib$ direction is a Jacobian factor $J = \int d^4x (\partial_\mu \phi_b)^2$. For translations, this Jacobian
is finite. For scale transformations, one wants to move from linearized fluctuations $\phi = \xi_d \phi_d$ 
proportional to the dilatation mode $\phi_d = (1+ x^\mu \partial_\mu )\phib$ to a collective scale coordinate $R$. Unfortunately, in this case, the Jacobian factor $J =\int d^4x \phi_d^2$ is infinite. Related Jacobians for the spontaneously broken $SU(2)\times U(1)$ symmetry of the Standard Model are also infinite~\cite{Kusenko:1996bv,Branchina:2014rva,Isidori:2001bm}. 

The second problem is that even if one could regularize the Jacobian and go to collective coordinates, the resulting integral $\int dR$ over scales $R$ would still be infinite. While quantum corrections do break scale invariance at some order, they do not resolve the infinity in the one loop approximation. Indeed, the $R$ dependence of the integrand can be deduced from renormalization group invariance. As we review in Section~\ref{sec:scale} at one loop, the integral is still infinite. While there is $R$ dependence at higher-loop order, for the higher-loop effects to cancel the infinity form the one loop integrand would require a diversion from the usual loop power counting. This is certainly possible, as the unusual power counting of the Coleman-Weinberg model~\cite{Coleman:1973jx} is often required to extract physical predictions from the effective potential \cite{Kang:1974yj,Andreassen:2014eha,Andreassen:2014gha}, but a proper power counting for the deca rate does not seem to have been explored in the literature.

A number of unsatisfying approaches to resolve the two problems with the dilation mode have been used in the literature. One method is to impose a scale on the bounce by hand, by demanding a constraint be satisfied \cite{Affleck:1980mp,Glerfoss:2005si,Nielsen:1999vq,Frishman:1978xs}, such as $\langle \phi^3 \rangle  = \Lambda^3$ for some fixed $\Lambda$. Then one can try to split the path integral into integrations around the constrained instanton and integrations over $\Lambda$. This approach seems impractical, as explicit constrained instantons are hard to find~\cite{Nielsen:1999vq} and the Jacobian to go between $R$ and $\Lambda$ is no simpler than between $\xi_d$ and $R$. 
While constrained instantons are helpful in understanding how a scalar mass can be a small perturbation, as we discuss in Section~\ref{sec:mass}, they are irrelevant to resolving the integral over $R$.

In practice, for the decay rate in the Standard Model, people always just invoke dimensional analysis \cite{tHooft:1976snw,Isidori:2001bm,Branchina:2014rva,DiLuzio:2015iua}: cut off the divergence in the Jacobian by the Higgs mass and assume the integral over $R$ is
dominated by the bubble size $R_m$ with the maximal rate. This seems to us a bit cavalier. After all, the fate of the universe is on the line.

In this paper, we provide definitive resolutions to both challenges associated with the dilatation mode. To regularize the Jacobian, a powerful approach has been known for some time but has not been widely appreciated \cite{McKane:1978md,Drummond:1978pf,Shore:1978eq,Glerfoss:2005si}. 
The approach is a based on a powerful stereographic projection into five dimensions, where symmetries are manifest and the natural inner product is the conformal metric on the 4-sphere. 
By undoing this transformation, it becomes clear that the projection is not actually necessary. The key is simply that  a zero mode satisfying $S''[\phib] \phi_d = 0$ is also a zero mode of any rescaled operator $f(x) S''[\phib] \phi_d$. Since the rescaling factor $f(x)$ changes the norm on the eigenfunctions, one can choose $f(x)$ so that $\phi_d$ is normalizable and the Jacobian is finite. More explicitly, for $S''[\phib]  = - \Box - V''[\phib]$, a wise choice is $f(x)  = V''[\phib]^{-1}$ which allows the spectrum of eigenfunctions to be found in closed form, and the same basis to be used for fluctuations around $\phib$ and around $\phi=0$. We explain this procedure in Section~\ref{sec:jacprob}.

Once collective coordinates have been invoked, we address the issue of the proper power counting required to evaluate the integral over $R$. Indeed, at one loop, the integral over $R$ is infinite. At two loops, there is an $\exp( - \hbar \ln^2 R)$ term which makes the integral over $R$ finite. Despite the $\hbar$ suppression of this term, the integral scales as $\hbar^{-1/2}$ compared to the one loop integral, making it divergent as $\hbar \to 0$. In other words, the two loop result is parametrically more important than the one loop result, a scaling essential to regulating the divergence. At three loops, the integral scales like $\hbar^{-3/2}$, so that three loop is more important than two loops. Conveniently, for four loops and higher, the integral stills scales like $\hbar^{-3/2}$ compared to one loop. We show that the entire series of leading contributions can be summed in closed form.  

With these scale invariance problems solved, we proceed to compute the functional determinants around the bounce over real scalar, complex scalar, vector boson and fermionic fluctuations. We produce for the first time exact formulas for the path integrals in each case. In the gauge boson case, we work in a general 1-parameter family of Fermi gauges and show the result is gauge invariant and that it agrees with the result in $R_\xi$ gauges as well.

Applying our exact formulas to the Standard Model, we update the famous stability/metastability phase diagram. For the first time, we can give an exact NLO prediction for the instability/metastability phase boundary. We find that with current data, the dominant uncertainties are from the top quark mass and $\alpha_s$, and these are both comparable to the theory uncertainty from electroweak threshold corrections, currently known to NNLO. 

Although this paper is rather long, we have tried to compartmentalize it into more-or-less self-contained sections. Section 2 reviews how tunneling rates are computed. This section is very brief and the interested reader is referred to~\cite{Andreassen:2016cvx} for more details. Section~\ref{sec:scale} contains new results about resolving the problems associated with scale invariance of the classical action. Section~\ref{sec:gendet} introduces the methods we will use in later sections to generate exact expressions for functional determinants. The longest section is Section~\ref{sec:funcdet} which computes one-by-one the functional determinants for scalars, vectors and fermions. For a reader just interested in the final formulas, these are summarized in Section~\ref{sec:summ}. The application to the Standard Model is in Section~\ref{sec:SM}. We tie up one lose end about the finite Higgs boson mass in Section~\ref{sec:mass}. Our results are digested, including a summary of the SM bounds and limits in our conclusions, in Section~\ref{sec:conc}.

\section{Tunneling Formulas and Functional Determinants \label{sec:tunnel}}
In this section we review how to compute a decay rate in quantum field theory. We give some formal expressions for the rate
in Section~\ref{sec:precise} and show how to use the saddle point approximation and how to evaluate functional determinants 
around non-trivial backgrounds in Section~\ref{sec:saddle}. 

\subsection{Defining the Decay Rate \label{sec:precise}}
Suppose our QFT has a metastable extremum localized around the constant classical field configuration $\phi(x)={\green \phi_\FV}$. We would like to compute
the lifetime of ${\green \phi_\FV}$ or equivalently the rate to tunnel form ${\green \phi_\FV}$ through the energy barrier to any other field configuration. 
The basic tunneling formula was introduced into high-energy theory by Coleman and Callan in 1977 \cite{Callan:1977pt}, although
it has roots in earlier condensed matter treatments~(e.g.~\cite{Langer:1967ax}). The formula is explained at length in Coleman's famous Erice lectures~\cite{Coleman:1978ae}, as well as in numerous textbooks~\cite{ZinnJustin:2002ru,Marino,Weinberg:2012pjx,Weinberg:1996kr}.  
Most of these treatments start with the premise that the decay rate can be computed by evaluating
\be
\frac{\Gamma}{2} \sim \im  \lim_{T\to\infty}  \frac{1}{T}  \ln  \langle \phi_\FV | e^{-H T} | \phi_\FV \rangle
\sim   \im  \lim_{T\to\infty}  \frac{1}{T}  \ln \int_{\phi(-T/2)=\phi_\FV}^{\phi(T/2)=\phi_\FV} \cD \phi e^{- S[\phi]}
\label{Gnaive}
\ee
Unfortunately, this formula cannot be correct as written, since the matrix element, and path integral over real paths, are purely real. 

\begin{figure}[t]
\begin{center}
\includegraphics{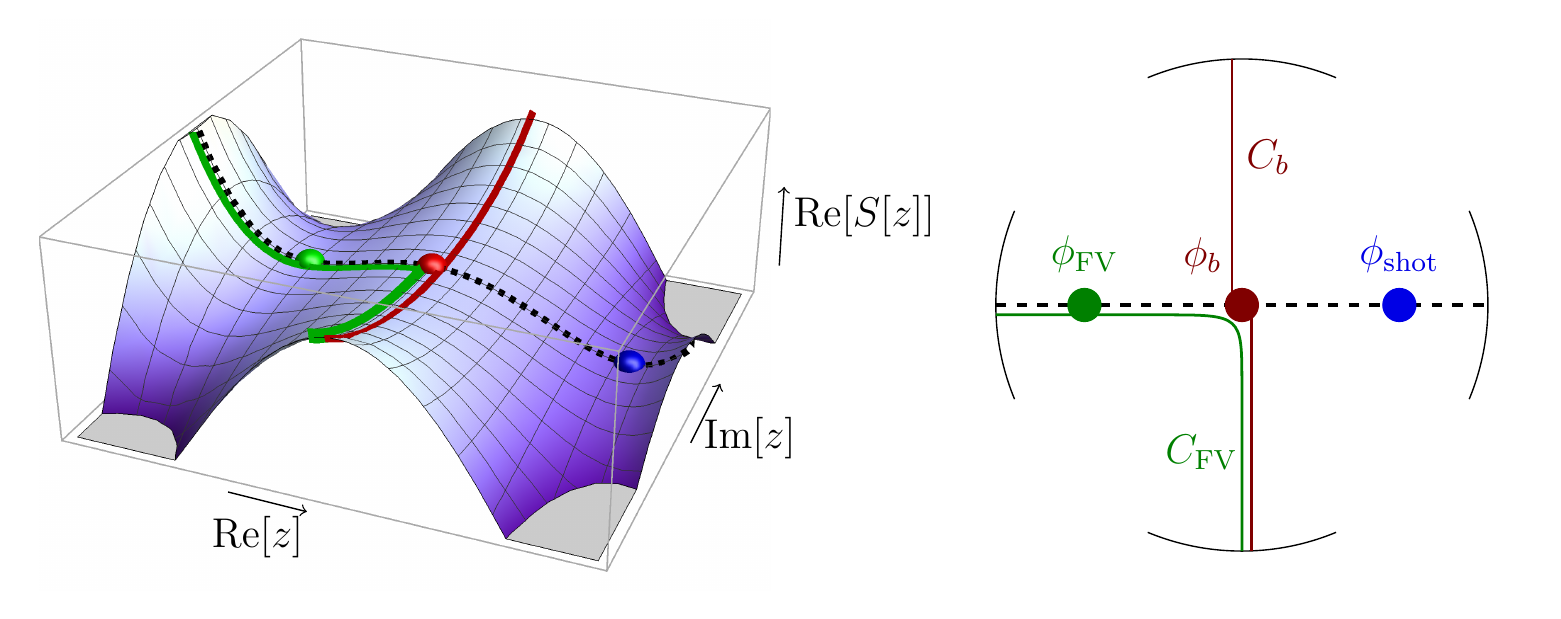}
\caption{
(Left) The real part of the action along  family of field configurations $\phi(z)$ parameterized by a complex parameter $z$. $z$ is chosen
so that real $z$ passes through ${\green \phi_\FV}$ (green dot),  ${\red \phi_b}$ (red dot), and ${\blue \phi_\text{shot}}$ (blue dot).
(Right) shows a top-down view. Integrating along real field configurations only (black dashed contour) makes the path integral real. The decay
rate must be calculated by integrating along steepest descent contours (red and green contours) which involve complex field configurations. 
\label{fig:saddleimaginaryparts}
}
\end{center}
\end{figure}

We would like the $T\to\infty$ limit to pick out the energy of an state localized near the false vacuum
whose imaginary part is to give the decay rate. This is certainly the intuition behind Eq.~\eqref{Gnaive}. Instead, Eq.~\eqref{Gnaive} picks out the true ground state energy $E_0$ which is real. To see this, we note that there are three relevant paths through field space satisfying the boundary conditions $\phi(-\frac{T}{2})=\phi(\frac{T}{2})={\green \phi_\FV}$: 1) The constant state $\phi={\green \phi_\FV}$; 2) the {\it bounce}, interpolating from ${\green \phi_\FV}$ at Euclidean time $\tau =\pm T/2$ to a bubble of shape ${\red \phib}(\vec{x})$ in a small time window (hence the name instanton) near $\tau=0$; and 3) the {\it shot}, ${\blue \phi_{\text{shot}}}$ matching ${\green \phi_\FV}$ at $\tau=\pm T/2$ but hovering near the true vacuum $\phi_0$ over most of Euclidean time~\cite{Andreassen:2016cff,Andreassen:2016cvx}. While the hope is for the path integral to pick out the {\it bounce} configuration at large $T$, instead it picks out the shot since the shot has smaller action, with the result that the path integral is real. In Fig.~\ref{fig:saddleimaginaryparts} we sketch of the real part of the action along paths $z$ passing through ${\green \phi_\FV}$, ${\red \phib}$ and ${\blue \phi_\text{shot}}$.

In order for the path integral and energies to be complex we must introduce a unitary-violating unphysical deformation of the theory.
This deformation should prevent flux from returning to the false vacuum so that the strict $T\to\infty$ limit can be taken. 
For example, we could impose Gamow's outgoing-wave-only boundary conditions to solve the Schr\"odinger equation~\cite{Gamow:1928}. A more formal deformation commonly used is analytic continuation of the classical potential.
 For example, with a potential $V(\phi) = m^2 \phi^2 + \lambda \phi^4$ if $\lambda$ is positive then  energies are real, but if $\lambda$ is negative energies are complex. 
The analog of this potential in non-relativistic quantum mechanics has been studied for
 decades~\cite{Banks:1973ps,Collins:1977dw,Bender:1984jc,Bender:1990pd,Jentschura:2010zza,Jentschura:2011zza}. Unfortunately, the analytic continuation method only works for (unphysical) situations in which the potential is unbounded from below.
When the potential is unbounded, no flux can return, but also the path integral over real field configurations is divergent so one has
no choice but to change the integration domain to over complex paths. If the path integral is finite, as in the Standard Model, then it is analytic in a domain around real paths and analytic continuation simply reproduces the original (real) result~\cite{Andreassen:2016cff,Andreassen:2016cvx}.

A useful way to define the rate for physical situations, where the action is bounded from below, is to change the integration domain
from real field configurations to field configurations associated with steepest descent contours (i.e. those for which $\im S[\phi] = 0$)~\cite{Behtash:2015zha,Witten:2010cx}.
To be precise,  the rate per unit volume for 
the formation of bubbles with the shape $\phib(\vec{x},0)$ is given by
\be
\frac{1}{2}\frac{\Gamma}{ V} =  \lim_{T\to\infty} \frac{1}{2}\frac{1}{ T V} \frac{ \im \int_{\red \cC_b} \cD \phi e^{-S[\phi]}}{ \re \int_{\green \cC_\FV} \cD \phi e^{-S[\phi]}}
\label{cc}
\ee
Here $S$ is the Euclidean action, $V$ the volume of space, and $T$ is a time for which the transition rate has exponential behavior (see discussion in \cite{Andreassen:2016cff,Andreassen:2016cvx}).
The contour ${\green \cC_\FV}$ (green contour in Fig.~\ref{fig:saddleimaginaryparts}) is the steepest descent trajectory through field space passing through ${\green \phi_\FV}$. 
The contour ${\red \cC_b}$ (red contour in Fig.~\ref{fig:saddleimaginaryparts}) is the steepest descent trajectory passing through the bounce.
Note that if we ignore these contour prescriptions and just integrate over real field configurations, 
along the black dashed line in Fig.~\ref{fig:saddleimaginaryparts}, passing through ${\green \phi_\FV}$, ${\red \phib}$ and ${\blue \phi_\text{shot}}$,
there is no imaginary part and the rate defined this way is zero, similar to Eq.~\eqref{Gnaive}.

For additional perspective, and insight into the factor of $\frac{1}{2}$, we can alternatively write the decay rate as
\be
\frac{1}{2}\frac{\Gamma}{ V} = \lim_{T\to\infty}  \frac{1}{ T V} \frac{ \im \int_{\green \cC_\FV} \cD \phi e^{-S[\phi]}}{\re \int_{\green \cC_\FV} \cD \phi e^{-S[\phi]}}
\label{ccfull}
\ee
The contour ${\green \cC_\FV}$  
 passes through real field configurations until the saddle point $\phi={\red \phib}$ is reached when it veers into complex field space (even for real $\phi$) traveling along ${\red \cC_b}$ as in Fig.~\ref{fig:saddleimaginaryparts}. In contrast, if we reverse the trajectory ${\red \cC_b}$, as it passes through ${\red \phib}$ it does not head towards ${\green \phi_\FV}$, since $\re(- S[\phi])$ increases in that direction, rather it continues into the conjugate complex field space. Thus integrating along ${\red \cC_b}$ gives twice the imaginary part of the integral along ${\green \cC_\FV}$. 
The doubling of the contour is the origin of the factor of $\frac{1}{2}$ in Eq.~\eqref{cc}

The explanation of why these arcane contour prescriptions produce the decay rate is given in~\cite{Andreassen:2016cff,Andreassen:2016cvx}.
Briefly, the idea is to start by relating the tunneling rate to the time derivative of the probability $\int_\cR d^3 x |\psi(x)|^2$ for a state to be found in a region $\cR$ on the other side of the energy barrier. This leads to the formula
\be
\Gamma =\lim_{\bigtau\to\infty}\abs{ \frac{2\im \int \cD \phi~e^{- S[\phi] }\delta(\tau_\Sigma[\phi])}{\int \cD \phi~e^{- S[\phi] }}}
\label{GammaRQFT}
\ee
Here, $\Sigma$ is codimension-1 surface bounding ${\mathcal R}$ comprising fields with the same energy density as the false vacuum $U[\phi]=U[\phi_\FV]$ and   $\tau_\Sigma[\phi]$ is the Euclidean time at which the field configuration $\phi(\vec{x},\tau)$ first passes through
$\Sigma$.  
Unlike Eq.~\eqref{ccfull}, this formula has the advantage that the left hand side can be shown to be the decay rate. In the
saddle point approximation, it reduces to Eq.~\eqref{cc}. 

\subsection{Functional Determinants and Zero  Modes \label{sec:saddle}}
Since the decay rate is defined by path integrals along steepest descent contours, we can compute these path integrals
in the saddle-point approximation. To quadratic order, Eq.~\eqref{cc} reduces to
\be
\frac{1}{2}\frac{\Gamma}{ V}  
=  \lim_{T\to\infty}  \frac{1}{2 T V} \frac
{ \im \int \cD \phi e^{- S[\phib] - \frac{1}{2} \phi S''[\phib] \phi }}
{\int \cD \phi e^{- S[\phi_\FV ]- \frac{1}{2} \phi S''[\phi_\FV] \phi}}
\label{ccg}
\ee

To evaluate the these path integrals, we must be precise about the integral measure. We do this by expanding the fields in some basis $\phi_j$:
\be
\phi = \phib(x) + \sum \xi_j \phi_j(x) \label{xip}\, .
\ee
The path integral measure can then be
defined as $\cD \phi = \prod_j d \xi_j$. 

An orthogonal basis is naturally provided by eigenfunctions of an operator. It is often convenient to take the operator to be $S''[\phib]$, so that
\be
S''[\phib] \phi_j = (-\Box + V''[\phib] )\phi_j = \lambda_j \phi_j
\ee
To find the inner product on these basis functions, we note that
 \be
 \lambda_ k \int d^4x \phi_j  \phi_k= \int d^4x \phi_j (-\Box + V''[\phib]) \phi_k = \lambda_ j \int d^4x \phi_j  \phi_k
 \label{inner1}
 \ee
 where integration-by-parts has been used in the last step. So functions with different eigenvalues are orthogonal according to the inner product
 $\langle \phi_i | \phi_j \rangle = \int d^4x \phi_i \phi_j$. 
 It is also convenient to normalize the fluctuations so that
 \be
 \langle \phi_i | \phi_j \rangle =2\pi \delta_{ij} \label{normconv}
 \ee
  Then we find
\be
  \int \cD \phi e^{- S[\phib] - \frac{1}{2} \phi S''[\phib] \phi} 
  =e^{- S[\phib]}\prod_j \int_{-\infty}^{\infty} d \xi_j e^{-\frac{1}{2}\lambda_j \xi_j^2 2\pi} 
  =    e^{- S[\phib]}  \prod_j \sqrt{ \frac{1}{\lambda_j }}  \label{lamprod}
\ee
  The point of the normalization convention in Eq.~\eqref{normconv} is to make removing a normalized fluctuation equivalent to removing its eigenvalue from the product in Eq.~\eqref{lamprod}.
 
 If one of the eigenvalues is negative, then this expression (after analytic continuation) will have an an imaginary part, as desired. There is 
at most a single negative eigenvalue\cite{Coleman:1987rm}. It corresponds to the bounce being a local maximum of the action on the direction from $\phi_{\FV}$ (see Fig.~\ref{fig:saddleimaginaryparts}) but a local minimum in all other directions. 
 
 If there are zero eigenvalues, then Eq.~\eqref{lamprod} is infinite. Examples are the translation modes, which are proportional to
 $ \partial_\mu \phib$. To check,
 using $S'[\phib]=0$ and that $S[\phi]$ has no explicit position dependence, we find
 \be
S''[\phib] \partial_\mu \phib
= \partial_\mu( S'[\phib])=  0
 \ee
 confirming that $\partial_\mu \phi$ are zero modes.
To set the normalization of these modes according to our convention, we note that
\be
\langle \partial_\mu \phib | \partial_\nu \phib \rangle 
=\frac{1}{4}\delta_{\mu\nu} \int d^4 x (  \partial_\lambda \phib )( \partial_\lambda \phib) = \delta_{\mu\nu}S[\phib]
\ee
Thus the rescaled  modes $\sqrt{\frac{2\pi}{S[\phi]}}\partial_\mu \phib$ are normalized according to Eq.~\eqref{normconv}. 

Separating out the translation modes Eq.~\eqref{xip} becomes
\be
\phi^{\xi} = \phib(x) +\xi^\mu\sqrt{\frac{2\pi}{S[\phi]}}\partial_\mu \phib(x) +  \sum \xi_j \phi_j(x) \label{xip2} 
\ee
To integrate over translations, we use collective coordinates~\cite{Langer:1967ax,Gervais:1974dc,Gervais:1975pa,Callan:1975yy,Jevicki:1976kd}, parametrizing fields with
 \be
 \phi^{x_0,\zeta} = \phib(x^\mu+x_0^\mu) + \sum \zeta_j \phi_j(x^\mu+x_0^\mu)
\label{xoz}
 \ee
By expanding Eq.~\eqref{xoz} for small $x_0^\mu$ and comparing to Eq.~\eqref{xip2} we see that the Jacobian to go from $\xi^\mu$ to $x_0^\mu$ is
 \be
J = \sqrt{\frac{S[\phib]}{2\pi}}
\label{Jnormal}
 \ee
 Then the path integral can be written as
 \begin{align}
  \int_{\cC_\FV} \cD \phi e^{-S[\phi]} 
  = \cN \left( \frac{S[\phib]}{2\pi}\right)^2 \int d^4 x_0   \sqrt{ \frac{1}{\det^\prime S''[\phib]}}
 \end{align}
where  $\det^\prime$ refers to the functional determinant with the zero eigenvalues taken out by hand
 and $\cN$ some (infinite) constant.  Noting that the integral over $d^4 x_0$ gives the volume of euclidean space time, we find
\be
\frac{\Gamma}{ V} =\left(\frac{S[\phib]}{2\pi}\right)^2e^{-S[\phib] + S[\phi_\FV]} \im \sqrt{\frac{\det[S''[\phi_\FV]]}{\det^\prime[S''[\phib]}}
\label{ccG}
\ee
 
 \section{Scale Invariance \label{sec:scale}}
Any classically scale invariant action will admit an infinite family of bounces related by scale transformations. To be explicit, we take the potential $V(\phi) = \frac{1}{4}\lambda \phi^4$ and assume throughout this paper that $\lambda <0$. Then there is a 5-parameter family of bounces given by
\be
\phib^{R,x_0^\mu}(x) = \sqrt{\frac{8}{-\lambda}} \frac{R}{R^2 + (x+x_0)^2}
\label{phibdef}
\ee
These Fubini-Lipatov instantons~\cite{Fubini:1976jm,Lipatov:1976ny} all satisfy $\Box\phib- \lambda \phib^3=0$ and
all have the same Euclidean action
\be
S[\phib] = \int d^4 x \left[\frac{1}{2} (\partial_\mu \phib)^2   + \frac{1}{4}\lambda \phib^4\right] 
=- \frac{8\pi^2}{3\lambda} >0
\ee
There are four normalizable fluctuations around the bounce corresponding to translations $\phi_\mu = \sqrt{\frac{2\pi}{S[\phib]}} \partial_\mu \phib$. These are handled using collective coordinates as discussed above. We therefore take $x_0=0$ without loss of generality.
 We also use $r =\sqrt{x_\mu x^\mu}$ as our radial coordinate so that the bounce is
  \be
 \phib(x) = \sqrt{\frac{8}{-\lambda}} \frac{R}{R^2 + r^2}
 \label{bounceform}
 \ee 
  
The (unnormalized) dilatation mode is
\be
\phi_d(x) =  \partial_R \phib = -\frac{1}{R}(1+ x^\mu \partial_\mu) \phib  
= \sqrt{\frac{8}{-\lambda}} \frac{r^2-R^2 }{(r^2+R^2)^2}
\label{phid}
\ee
Like the translation modes, it is an eigenfunction of the second variation of the action around the bounce with zero eigenvalue:
\be
S''[\phib] \phi_d = \Big[-\Box + V''(\phib)\Big] \phi_d = \left(-\Box + 3 \lambda \phib^2\right) \phi_d = 0
\ee
We would like to proceed, as with translations, by going from linearized fluctuations $\phi = \xi_d (N_d \phi_d)$, with $N_d$ a normalization factor, to the collective coordinate $R$ so that we can write
\be
\frac{\Gamma}{ V} = \left(\frac{S[\phib]}{2\pi}\right)^2e^{-S[\phib] + S[\phi_\FV]} \im 
\int dR J_d \sqrt{\frac{\det[S''[\phi_\FV]]}{\det^\prime[S''[\phib]}}
\label{ccD}
\ee
with $\det^\prime$ now having the $\phi_d$ dilatation mode removed and $J_d$ the Jacobian. 

There are two problems with this. First, the Jacobian is infinite:
\be
J_d^2 =  \frac{\langle \phi_d | \phi_d \rangle}{2\pi} = \frac{1}{2\pi}\int d^4x \phi_d^2 = \infty
\label{Jdinf}
\ee
Second, the integral over $R$ is divergent. Even including the one-loop $R$-dependence from dimensional transmutation, as required at this order, is not enough to remove this infrared divergence. 

In the literature, for the first problem $J_d$ is often assumed to be made finite by natural infrared cutoff of the scalar mass \cite{Isidori:2001bm,Branchina:2014rva}. Unfortunately, the scalar mass adds more problems than it solves -- adding a $m^2 \phi^2$ term to the Lagrangian removes all bounces from the solution to the equations of motion. Moreover, adding an infrared cutoff seems to miss the point. Why is the Jacobian infinite in the first place? Going from small linear fluctuations around a bounce to fluctuations corresponding to exact scale transformations seems perfectly reasonable and therefore should be non-singular. In fact, the mass term is irrelevant to the problem (see Section~\ref{sec:mass}). 

For the second problem, of the IR divergent integral over $R$, it is common to pick the scale $R_\star$ for which the leading order result $\Gamma = \exp(\frac{8\pi^2}{3\lambda(R_\star^{-1})})$ is maximal, with $\lambda(\mu)$ the running coupling, and evaluate the $R$ integral by dimensional
 analysis~\cite{tHooft:1976snw,Isidori:2001bm,Branchina:2014rva,DiLuzio:2015iua}. Although this ad hoc solution does produce an answer, we cannot assess its accuracy, since dimensional analysis has ignored rather than solved the problem. To get a dimensionally correct answer, one could try choosing $\mu=R_\star^{-1}$ or $\mu=R^{-1}$ before doing the $R$ integral, or doing the $R$-integral before choosing $R$ at all. None of these attempts are consistent with perturbation theory, and  in any case they all give a divergent answer. 

An alternative approach that is discussed in the literature invokes constrained instantons~\cite{Affleck:1980mp,Glerfoss:2005si,Nielsen:1999vq,Frishman:1978xs}. The idea of constrained instantons is to fix $R$ by demanding that some operator have a given expectation value, such as $\langle \phi^3 \rangle = \Lambda^3$ for some $\Lambda$. Fixing the scale in this way merely swaps the $R$ integral for an integral over $\Lambda^{-1}$, and the problem is still there.

One might hope these infinities from $J_d$ and the $R$ integral would cancel, but they do not (and should not). From a physical point of view, unless something makes the rate to produce different sizes bubbles different, the net decay rate should be infinite. It is quantum corrections which break the scale invariance, but the Jacobian is determined by the bounce from the classical theory. We will explain how to deal with the IR divergent integral over $R$ and what scale the couplings are evaluated at in Section \ref{sec:siprob} after we have solved the Jacobian problem in the next section. 

\subsection{Solving the Jacobian Problem \label{sec:jacprob}}
The Jacobian is infinite because the dilatation fluctuation $\phi_d$ is
not normalizable. Of course, the path integral is basis independent;
changing the normalization of a fluctuation $N_d \phi_d \to \phi_d$ in the expansion in Eq.~\eqref{xip2} can be compensated for by rescaling
$\xi_j \to \frac{1}{N_d}$ in the path integral measure. The problem with having a non-normalizable zero mode is that one cannot see clearly what happens when it is removed in computing $\det'$. In fact, the infinite Jacobian is secretly compensated by an infinity in $\det'$ (see Appendix \ref{app:norescale}). Here, we cleanly resolve the Jacobian problem by choosing a judicious basis in which the numerator and denominator path integrals can be computed exactly. 

 What basis would allow us to diagonalize fluctuations around the bounce and around the false vacuum simultaneously? Taking eigenfunctions of $S''[\phib]= -\Box +  V''[\phib]$ will not work, since these are not also eigenfunctions of $S''[\phi_\FV]= -\Box$. Instead, we use eigenfunctions of 
 \be
 \cO_\phi \equiv \frac{-1}{V''[\phib]}S''[\phib] = \frac{1}{3\lambda \phib^2}\Box - 1 
\label{Ob}
  \ee
for the scalar fluctuations around the bounce and eigenfunctions of
 \be
\wh \cO_\phi \equiv  \frac{-1}{V''[\phib]}S''[\phi_\FV] =   \frac{1}{3\lambda \phib^2}\Box 
  \ee
 for fluctuations around the false vacuum. Note that even for an arbitrary potential these operators will differ only by a constant and therefore  have the same eigenfunctions.
 
  An important feature of eigenfunctions of these operators is the inner product by which they are orthogonal. Similarly to Eq.~\eqref{inner1} we find that if $ \wh \cO_\phi \phi_j = \lambda_j \phi_j$ then 
  \be
 \lambda_ k \int  d^4x V''[\phib] \phi_j   \phi_k
 = \int d^4x \phi_j (\Box ) \phi_k = \lambda_ j \int d^4xV''[\phib]  \phi_j  \phi_k
 \label{inner2}
 \ee
 so that eigenfunctions are orthogonal according to 
 \be
\langle \phi_j | \phi_k \rangle_V  \equiv- \int d^4x V''[\phib] \phi_j \phi_k 
=  \langle \phi_j | \phi_j \rangle_V  \delta_{jk}
 \ee
Since we are using the same basis for both path integrals, we can normalize the eigenfunctions however we like. For example, we could choose $\langle \phi_j | \phi_j \rangle_V = 2\pi $, and indeed even the dilatation mode will be normalizable according to this metric. 
Furthermore, this basis still lets us evaluate the path integral, since
\be
\int d^4x  \phi_j S''[\phib] \phi_k = - \int d^4 x V''[\phib] \phi_j \wh \cO_\phi \phi_k = \lambda_j \langle \phi_j | \phi_j \rangle_V  \delta_{jk}
\ee
Thus the path integrals are still Gaussian in the fluctuations. The integral over a fluctuation normalized with $\langle \phi_j | \phi_j \rangle_V=2\pi$ then gives the usual factor of $\sqrt{\frac{1}{\lambda_j}}$. Note that these observations apply to any theory, not just a scale invariant one: one can always simultaneously diagonalize fluctuations around the bounce and fluctuations around the false vacuum. 

Now we restrict to the scale-invariant case, with $V(\phi) = \frac{1}{4}\lambda \phi^4$.  Explicitly, our eigenfunctions should satisfy 
\be
\cO_\phi \phi_n =  \lambda_n^\phi \phib, \qquad 
\cO_\phi  =\frac{1}{3\lambda \phib^2}\Box -1
\label{Obev}
\ee
and be orthogonal with respect to the inner product
\be
\langle \phi_j | \phi_k \rangle_V  
=- \int d^4x V''[\phib] \phi_j \phi_k 
= \int d^4x \frac{24 R^2}{(R^2 + r^2)^2} \phi_j(x) \phi_k (x)
\ee
Remarkably, we can find the solutions in closed form. For $x_0=0$ in Eq.~\eqref{phibdef}, they are
\be
\phi_{nslm}(r,\alpha,\theta,\phi) = \frac{1}{r} P_{n+1}^{-s-1}\!\!\left(\frac{R^2- r^2}{R^2 + r^2} \right) Y^{slm}(\alpha,\theta,\phi),
\label{efs}
\ee
with $P_l^m(x)$ the associated Legendre polynomial and $Y^{slm}(\alpha,\theta,\phi)$ are the 3D spherical harmonics:
\be
Y^{slm}(\alpha,\theta,\phi) = \frac{1}{\sqrt{\sin\alpha}} P^{-l-\frac{1}{2}}_{s+ \frac{1}{2}} ( \cos\alpha) P_l^m (\cos \theta) e^{- i m \phi}
\label{Yslm}
\ee
These spherical harmonics satisfy
\begin{align}
\vec{L}^2 Y^{slm} &=
-\frac{1}{\sin^2\alpha}\left[ \partial_\alpha (\sin^2\alpha \partial_\alpha \cdot) +\frac{1}{\sin\theta} \partial_\theta(\sin\theta \partial_\theta \cdot) 
 + \frac{1}{\sin^2\theta} 
 \partial_\phi^2 \right]Y^{slm}(\alpha,\theta,\phi) \\
 &= s(s+2)Y^{slm}(\alpha,\theta,\phi)
\end{align}
and are normalized as
\begin{align}
\langle Y^{slm} | Y^{s'l'm'} \rangle_\Omega &=
\int_0^\pi d\alpha \sin^2\alpha \int_0^\pi d \theta \sin \theta \int_0^{2\pi} d\phi Y^{slm} Y^{s'l'm'} \\
&=\frac{4\pi}{(2l+1)(s+1)} \frac{(s-l)!}{(s+l+1)!}\frac{(l+m)!}{(l-m)!} \delta_{ss'}\delta_{l l'}\delta_{m m'}
\label{YYnorm}
\end{align}

The full eigenfunctions in Eq.~\eqref{efs} satisfy  Eq.~\eqref{Obev}, i.e.
\be
\cO_\phi \phi_{nslm}=
\left[\frac{1}{3\lambda \phib^2}\left(\partial_r^2 + \frac{3}{r}\partial_r - \frac{\vec{L}^2}{r^2}\right)-1\right]\phi_{nslm} 
= \lambda_n^\phi \phi_{nslm} 
\ee
with $\vec{L}^2 \phi_{nslm} = s(s+2)\phi_{nslm}$. The  eigenvalues only depend on $n$:
\be
\lambda_n^\phi= \lambda_{nslm} = \frac{(n-1)(n+4)}{6} = -\frac{2}{3}, 0, 1, \frac{7}{3}, \cdots
\label{lvals}
\ee
The eigenvalues of $\wh \cO_\phi$ are
\be
\wh \lambda_n^\phi = \lambda^\phi_n+1 =\frac{(n+1)(n+2)}{6}= \frac{1}{3},1,2,\frac{10}{3}, \cdots
\label{lvalsfv}
\ee
The indices in $\phi_{nslm}$ are integers constrained by $0 \le |m| \le l \le s \le n =0,1,2,\cdots$. The degeneracy of each eigenvalue is therefore
\be
d_n = 
\frac{1}{6}(n+1)(n+2)(2n+3) = 1,5,14,30,\cdots
\label{dvals}
\ee
The eigenfunctions are normalized as
\be
  \langle \phi_{nslm} | \phi_{n's'l'm'} \rangle_V  = \frac{12}{2n+3}\frac{(n-s)!}{(n+s+2)!}   \delta_{nn'}\langle Y^{slm} | Y^{s'l'm'} \rangle_\Omega
\ee
with $\langle Y^{slm} | Y^{s'l'm'} \rangle_\Omega$ in Eq.~\eqref{YYnorm}.

The modes with $s=l=m=0$ are spherically symmetric, functions of only $r$.
The mode with $n=0$ that has $\lambda_0^\phi =  -\frac{2}{3}$ is
\be
\phi_{0000} = \sqrt{\frac{2}{\pi}}\frac{R}{R^2 + r^2}, \qquad \lambda_0^\phi = -\frac{2}{3},\qquad d_0= 1
\label{phi0000}
\ee
This mode is directly proportional to the bounce itself: $\phi_{0000} = \sqrt{\frac{-\lambda}{4\pi}} \phib$. The negative eigenvalue arises because the action has a local maximum at the bounce in the direction going from $\phi_\FV$ to $\phib$ (see Fig.~\ref{fig:saddleimaginaryparts}).

There are 5 modes with $n=1$, with $\lambda^\phi_1 =0$. The spherically symmetric one is
\be
\phi_{1000} =\sqrt{\frac{2}{\pi}} R\frac{R^2-r^2}{(R^2 + r^2)^2},\qquad \lambda_1^\phi = 0,\qquad d_1 = 5
\ee
This is proportional to the dilatation mode: $\phi_{1000}=-\sqrt{\frac{-\lambda}{4\pi}}R\phi_d $. The other $n=1$ modes, which also have
 $\lambda_1=0$, are the zero modes for translations. 

The modes with $n>1$ are not particularly interesting:
\be
\phi_{2000} =\sqrt{\frac{2}{\pi}} \frac{R \left(r^4+R^4-3 r^2 R^2\right)}{\left(r ^2+R^2\right)^3},\quad \lambda_2^\phi = 1,\qquad d_2 = 14
\ee
and so on.

Since the dilation and translation modes are both normalizable, computing the Jacobian is straightforward:
\be
J_d = \sqrt\frac{ \langle \phi_{d} | \phi_{d} \rangle_V}{2\pi}
= \frac{1}{ R} \sqrt{\frac{6 S[\phib]}{5\pi}}
\label{Jd}
\ee
The Jacobian for the translation modes with this metric differs from Eq.~\eqref{Jnormal}:
\be
J_T = \sqrt\frac{ \langle \partial_\mu \phib | \partial_\mu \phib \rangle_V}{2\pi}
=
\frac{1}{ R} \sqrt{\frac{6 S[\phib]}{5\pi}}
 \label{JT}
\ee
Note the important factors of $R$ in both Jacobians -- these are expected by dimensional analysis but obscure without the rescaling (cf. Appendix~\ref{app:norescale}). 
So we find
\be
\frac{\Gamma}{ V} =
e^{-S[\phib]}
\left(\frac{6 S[\phib]}{5\pi}\right)^{\frac{5}{2}}
  \im 
\int \frac{dR}{R^5}  \sqrt{\frac{\det \wh \cO_\phi}{\det^\prime \cO_\phi}}
\label{ccR}
\ee
Note that all the eigenvalues of $\wh\cO_\phi$ and $\cO_\phi$ are dimensionless, so this expression has the correct units. 

We have shown that by rescaling the operators for fluctuations around the bounce and the false vacuum, the natural basis for field fluctuations changes, and the Jacobian for going between this basis and the basis containing a collective coordinate for dilatations is finite. Since the final result 
in Eq.~\eqref{ccR} should be independent of this rescaling, there must be something that compensates for the infinite Jacobian if we do not rescale. In Appendix~\ref{app:norescale} we show that in fact without rescaling $\det'$ is infinite as well. 

\subsection{Solving the Scale Invariance Problem \label{sec:siprob}}
The next problem is that the integral over $R$ in Eq.~\eqref{ccR} is infinite. Even without evaluating the functional determinants, we can determine the $R$ dependence of the integrand in Eq.~\eqref{ccR} completely by 
exploiting renormalization group invariance of $\Gamma$.  To see this, and to resolve the infrared divergence issue, it is critical to be 
consistent in power counting the loop expansion, or equivalently, orders of $\hbar$. A similar consistency was essential to resolve the gauge invariance problem of the ground state energy density in~\cite{Andreassen:2014gha,Andreassen:2014eha}.
In the following, we insert appropriate factors of $\hbar$. Powers of $\hbar$ will always correspond to powers of couplings such as  $\lambda$ in this scalar field theory or $g^2$ in a gauge theory. 

To leading order (LO) in $\hbar$, the rate is determined entirely by the exponential factor in Eq.~\eqref{ccR}. 
Expanding this factor out explicitly we have
\be
\frac{\Gamma}{ V} =e^{\frac{1}{\hbar}\frac{8\pi^2}{ 3\lambda(\mu)}}
\int \frac{dR}{R^5} \cdots
\label{GRR}
\ee
where $\lambda(\mu)$ is the $\msbar$ coupling at the scale $\mu$ and $\cdots$ refer to the rest of Eq.~\eqref{ccR}. 
Everything after the exponential comes from a one loop  calculation and is subleading in $\hbar$. 
It is commonly said that the { leading order} prediction for the rate is $\Gamma/V= e^{\frac{8\pi^2}{\hbar 3\lambda(\mu)}}$. However,
 such a claim does not really make sense -- not only is this equation dimensionally inconsistent, there is no indication at what scale $\mu$ to choose -- so it is really no prediction at all. Indeed, the {\it leading} prediction must start at one loop. And, as we will see the leading prediction actually involves terms at two loops and higher. We will refer to the leading  finite prediction with correct units as the NLO rate. 

Now, $\Gamma$ is physical, so $\mu \frac{d}{d \mu} \Gamma=0$. This implies that the implicit $\mu$-dependence of $\lambda(\mu)$ must be compensated by explicit $\mu$-dependence in the NLO contribution.  In turn, the $\mu$ dependence of $\lambda(\mu)$ is fixed by its
RGE. Thus we know the exact $\mu$ dependence of the integrand to one-loop-higher order than we know the $\mu$-independent part. 
By dimensional analysis, the only scale around to compensate $\mu$ is $R$, and therefore we also know the full $R$ dependence 
of Eq.~\eqref{GRR} to one loop:
\be
\frac{\Gamma}{ V} =
\int_0^\infty \frac{dR}{R^5} e^{\frac{1}{\hbar}\frac{8\pi^2}{ 3\lambda(\mu)} - \frac{8\pi^2}{3} \frac{\beta(\mu)}{\lambda(\mu)^2} \ln (\mu R)}  
\left( \cdots \right) \label{IRprob}
\ee
At one loop, the terms in $(\cdots)$ have no explicit dependence on $\mu$, by RG invariance, and therefore no dependence on $R$ either, by dimensional analysis.
Here $\beta(\mu)$ is the $\beta$-function coefficient in the RGE for $\lambda$, 
$\mu\frac{d}{d \mu} \lambda =  \beta(\mu)$. 
Now we see clearly the IR divergence problem. All of the $R$ dependence in the one loop rate is explicit and the integral over $R$ is infinite.
 
 The only hope is for two-loop and higher-order contributions to come in and resolve the infinity. At first pass, this seems impossible, simply by counting factors of $\hbar$: terms in $(\cdots)$ at two loops and higher are necessarily $\hbar$ suppressed compared to the terms we have written. The resolution is that after the integral, superleading $\hbar$ dependence is generated, as we will now see.

First of all, let us assume the $\overline{\text{MS}}$ coupling $\lambda(\mu)$ has a minimum at some scale $\mu=\mu^\star$,
so $\beta(\mu^\star)=0$. If this is not true, then the running coupling $\lambda(\mu)$ is unbounded from below and rate is actually infinite. In fact, in this quartic scalar field theory, $\lambda(\mu)$ is monotonic, so we are going to have to assume there are other fields in the theory to continue. For a more general theory, we can perform the path integral over all other fields around the bounce, leading to a decay rate formula of exactly the same form as Eq.~\eqref{IRprob}, but with the $\beta$-function for $\lambda$ depending all all other couplings in the theory. In this case, $\beta(\mu)$ can vanish, as for example it does in the Standard Model for the Higgs quartic at the scale $\mu^\star \sim 10^{17} $ GeV.\footnote{
Note that the vanishing of the $\beta$ function can be achieved by balancing couplings $g^2 \sim \lambda$ at the same loop order. This is different from the requirement that the
effective potential have a minimum, which requires two-loop terms to cancel one-loop terms~\cite{Andreassen:2014eha}. The
scale $\mu^\star$ where the $\beta(\mu^\star)=0$ can be parametrically different from
the scale $\mu_X$ where $V_\text{eff}'(\mu_X)=0$. 
}

Since $\Gamma$ is independent of $\mu$ we are free to choose $\mu=\mu^\star$. Let us do so. Then the exponential in Eq.~\eqref{IRprob} has {\it no} $R$ dependence (since $\beta(\mu^\star)=0$) at all and the integral is surely infinite. With $\mu=\mu^\star$, the leading $R$ dependence in the exponential factor comes in at two loops, and has the form
\be
\frac{\Gamma}{ V} 
=
\int_0^\infty \frac{dR}{R^5} e^{-\frac{1}{\hbar}\Sbs  + \hbar  \Sbs \frac{\bps}{\lambda_\star}\ln^2 (\mu^\star R)}  
\left( \cdots \right)
 \label{IRprob2}
\ee 
where $\lambda_\star = \lambda(\mu^\star)$, $\Sbs = -\frac{8\pi^2}{3\lambda_\star}$,
$\beta' =\mu \frac{d}{d\mu} \beta(\mu)$ and $\bps = \beta'(\mstar)$ using the one loop $\beta$-function coefficients only
(cf. Eq.~\eqref{betaprimeSM} for its SM expression).
 At two loops, there is an additional single log term in the exponent scaling like $\hbar \ln(\mstar R)$. This will contribute subleading in $\hbar$ after the integral so we have  dropped it.

Now we observe that since $\bps >0$ and $\lambda^\star<0$ at the minimum, the integral over $R$ {\it is} finite:
\be
\Gamma_2 = \int_0^\infty \frac{dR}{R^5} e^{ \hbar \Sbs \frac{\bps}{\lambda_\star} \ln^2 R\mu^\star }
= \mstar^4 \sqrt{-\frac{\pi \lstar}{\hbar \Sbs \bps}}  e^{-\frac{4\lstar}{\hbar \Sbs \bps}}
\label{Gamma2def}
\ee
Note that this contribution is  parametrically more important as $\hbar\to 0$ than the one-loop correction
which scales like $\hbar^0$. Indeed, $\Gamma_2$ blows up as $\hbar\to0$, as it must to reproduce the divergence of the one loop integral. Thus, even though the two loop result is formally higher order in $\hbar$, we cannot justify expanding the exponential to provide only $\hbar$ corrections to $(\cdots)$ in Eq.~\eqref{IRprob}. 

Also note that the divergence returns if $\beta'_\star =0$. Thus the scale invariance is not
regulated by just dimensional transmutation (i.e. by $\beta \ne 0$)  but requires in addition that the $\beta$-function have a minimum.

A natural concern is that since the two-loop result parametrically dominates over the one-loop
result as $\hbar\to 0$, the three-loop result might dominate over two loops, and so on. 
To see if this happens, we examine possible
terms in the exponent, as allowed by RG invariance. At 
each logarithmic order, the coefficient of $\ln^n R\mu^\star$ is of order $\hbar^{n-1}$
plus terms suppressed by additional factors of $\hbar$. That is, we have
$ \ln R\mu^\star$, $\hbar \ln^2 R \mu^\star$, $\hbar^2 \ln^3 R \mu^\star$ and so on. 
Using the two-loop term to set up a Gaussian around which we perform a saddle-point approximation in $\hbar$, we find
a generic term becomes
\begin{align}
 \int \frac{dR}{R^5} e^{\hbar \Sbs \frac{\bps}{\lambda_\star} \ln^2 R\mu^\star +\hbar^{n-1} c_n \ln^n R\mu^\star }
 &=\int \frac{dR}{R^5} e^{  \hbar  \Sbs \frac{ \bps}{\lambda_\star} \ln^2 R\mu^\star }(1+ \hbar^{n-1} c_n \ln^n R\mu^\star + \cdots) 
 \nonumber \\
 &= \frac{1}{\hbar} \Gamma_2 c_n 
 \left( \frac{2 \lambda_\star}{\Sbs \bps} \right)^n + \cdots
 \label{gaussn3}
\end{align}
The $\cdots$ are all terms subleading in $\hbar$. 
So we see that in fact three-loop and higher-order contributions are more important than the one- and two-loop terms, but all terms at three loops and beyond are the same order in $\hbar$.

One might worry that since the $n=3$ term is more important than the $n=2$ term, the saddle point approximation cannot be justified. Note
however that expanding the exponential of the $n=3$ term to next order gives a term scaling like
$\hbar^4 \ln^6 R\mu^\star$. This term is  subleading by a factor of $\hbar$ to the terms we keep
from the expansion of the $n=6$ term. The same justification explains why we can  ignore $R$ dependence coming from the RG invariance of the non-logarithmic one-loop terms;  these are also subleading in $\hbar$.

Since an infinite number of terms are relevant, we have to  sum the series. Fortunately, this is possible since all of these terms depend on only the leading order $\beta$-function coefficients. In a pure scalar field theory, the one loop RGE is easy to solve in closed form. In a general theory, this will not be true. In any theory, we can write
\be
\hbar \lambda(\mu) =\hbar \lambda^\star +
 \sum_{n\ge2}\left( \hbar^{n+1} \kappa_{n}  + \cdots\right) \ln^n \frac{\mu}{\mu^\star}  
\label{beta01}
\ee
Note that we now sum over $n\ge2$ since $\lambda'(\mu^\star)=0$ by definition of $\mu^\star$ and,
matching our previous notation, $\kappa_2= \bps$. The $\kappa_n$ coefficients
are all determined by the $\beta$-functions in the theory evaluated at one loop. 
The terms denoted $\cdots$ all depend on $\beta$-function coefficients beyond one loop and are contribute to subleading order in $\hbar$ to the final answer, so we drop them. 

Eq.~\eqref{beta01} represents a perturbative solution of the coupled RGEs which is always possible to work out order-by-order in the couplings. It implies
\be
\frac{1}{\hbar \lambda(\mu)} = \frac{1}{\hbar \lambda_\star}
\sum_{m\ge 0} \left (-   \sum_{n\ge 2} \hbar^n \frac{\kappa_{n}}{\lambda_\star} \ln^n \frac{\mu}{\mu^\star} \right)^m + \cdots
\ee
Here the $\cdots$ have terms of the same logarithmic order but subdominant in $\hbar$ compared to the terms we keep. 
For the integrand to be RGE invariant, we know the rate can be written as
\be
\frac{\Gamma}{V} =\GRind \times  \GR
\ee
where $\GRind$ is $R$-independent. For example, from Eq.~\eqref{ccR}, 
\be
\GRind =\left[ e^{-S[\phib]} (R J_T)^4 (R J_d)
\im
\sqrt{\frac{\det \wh \cO_\phi}{\det^\prime \cO_\phi}}\right]_{R=\mu^{-1}=(\mu^\star)^{-1}}
\ee
and 
\be
\GR = \int \frac{d R}{R^5} \exp \left[ -\frac{ \Sbs }{\hbar}
\sum_{m\ge 1} \left (-   \sum_{n\ge 2} \hbar^n \frac{\kappa_{n}}{\lambda_\star} \ln^n \frac{1}{\mu^\star R} \right)^m
 \right] + \cdots
 \label{rgform}
 \ee
To repeat, there are corrections to this exponent that are subleading in $\hbar$ both from the 
higher-order terms in Eq.~\eqref{beta01} and from those fixed by the RG invariance of one-loop (or higher) non-logarithmic terms in the integrand. All of these terms necessarily make subleading contributions to the rate. 

Now, every term in the exponent in Eq.~\eqref{rgform} is proportional to $\hbar^{a-1} \ln^a (\mu^\star R)$ for some $a$.
There is only one term with $a=2$, corresponding to $n=2,m=1$. This term generates a factor of $\Gamma_2$ after integration,
as in Eq.~\erqef{Gamma2def}. All the other terms, including the cross terms, have $a>3$ and contribute to the same order in $\hbar$
after integration by Eq.~\eqref{gaussn3}. We can perform
the Gaussian integrals over all $n$ and $m$ by adding and subtracting the $n=2,m=1$ term. Then, as in Eq.~\eqref{gaussn3}, we get
\be
\Gamma_R = \Gamma_2 \left\{ 1 - \frac{1}{\hbar} \Sbs \sum_{m\ge 1} \left[ - \sum_{n \ge 2} \hbar^n \frac{\kappa_n}{\lambda_\star}
\left( \frac{2\lambda_\star}{\hbar \Sbs \bps} \right)^n \right]^m 
 - \frac{1}{\hbar} \Sbs  \frac{\kappa_2}{\lambda_\star}\left( \frac{2\lambda_\star}{ \Sbs \bps} \right)^2
\right\} + \cdots
\ee
The $1$ is subleading in $\hbar$, so we can drop it. The geometric series is easily resummed, giving
\be
\Gamma_R = \frac{1}{\hbar} \Sbs \Gamma_2 \left\{ \frac{\hbar \lambda_\star}{\hbar \lambda_\star +  \sum_{n \ge 2} \hbar^{n+1} \kappa_n
\left( \frac{2\lambda_\star}{ \hbar \Sbs \bps} \right)^n} -1  - \frac{4 \lambda_\star}{\Sbs^2 \bps}
\right\} + \cdots
\ee
Finally, we can express the answer in terms of the running coupling. The result is
\be
\Gamma_R = \frac{1}{\hbar} \Sbs \Gamma_2 \left[
\frac{\lambda_\star}{\lol(\hat\mu)} - 1 - \frac{4 \lambda_\star}{\Sbs^2 \bps}
\right] 
+ \cdots
\ee
In this expression $\lol(\hat\mu)$ means solve the coupled RGEs using the one loop $\beta$-function coefficients only and evaluate at the scale
\be
\hat{\mu} = \mu^\star\exp\left( \frac{2 \lambda_\star}{\hbar \Sbs \bps} \right)
\label{muhatdef}
\ee
For example, in a complex scalar theory, $\hat \mu = \mu_\star \exp(-\frac{\pi^2}{6\lambda})$. 
Note that for small coupling, the scales $\hat{\mu}$ and $\lambda_\star$ can be very far apart. 
 Keep in mind however that all the couplings in $\lol(\hat\mu)$ are evaluated at $\mu^\star$, so this resummation does not indicate sensitivity to high scales; it is merely shorthand for a series of terms all of the same order and the couplings.

Putting everything together and resetting $\hbar=1$, we find
\begin{multline}
\frac{\Gamma}{ V} =
\left[ e^{-S[\phib]} (R J_T)^4 (R J_d)
\im \sqrt{\frac{\det \wh \cO_\phi}{\det^\prime \cO_\phi}}
\prod_{{\text{fields}}~A}
\sqrt{\frac{\det \wh \cO_A}{\det^\prime \cO_A}} \right]_{R=\mu^{-1}=(\mu^\star)^{-1}}
\\
\times
\mstar^4\sqrt{-\frac{\pi \lstar \Sbs}{ \bps}}  e^{-\frac{4\lstar}{\hbar \Sbs \bps}}
 \left[
\frac{\lambda_\star}{\lol(\hat\mu)} - 1 - \frac{4 \lambda_\star}{\Sbs^2 \bps}
\right] 
\label{Rint}
\end{multline}
Here the extra determinants come from integrating over fluctuations of fields other that $\phi$ in the theory
around the bounce and false vacuum backgrounds. 
As long as $\mu^\star$ exists (meaning $\lambda(\mu)$ has a minimum), this is a finite expression derived with consistent power counting.
All of the singularities associated with scale invariance have been completely resolved. 

Finally, we point out that one does not have to choose $\mu= \mu^\star$. For $\mu \ne \mu^\star$ there are terms linear in $\ln(\mu R)$ in the exponent proportional to $\beta(\mu)$, which generate a slew of additional terms in Eq.~\eqref{Rint}. For example, Eq.~\eqref{Gamma2def} becomes
\be
\Gamma_2(\mu) =\mu^4  \sqrt{-\frac{\pi \lambda^2}{ S[\phib] ( \beta_0' \lambda - \beta_0^2)}} 
\exp \left[ -\frac{4}{S[\phib]} \frac{ \Big( \lambda + \frac{1}{4} S[\phib] \beta_0 \Big)^2}{\beta_0' \lambda - \beta_0^2} \right]
\label{Gamma2mu}
\ee
with $\beta_0$ the one loop $\beta$-function coefficient for $\lambda$.
The general expression including terms like this can be used to calculate the scale uncertainty on the final prediction. 

\section{Functional Determinants: General Results \label{sec:gendet}}
With the divergences associated with scale invariance understood, we now proceed to evaluating the functional determinants. 
We will be evaluating determinants for a number of cases: real scalar fluctuations, Goldstone boson fluctuations, gauge boson fluctuations (in general gauges), and Dirac fermion fluctuations. These are similar enough that is helpful to work out some results first that we can
then apply to the different examples of interest.

In this section we consider the general operator
\be
\cM(x) = -\Box - 3 x \lambda \phib^2
\label{Mx}
\ee
and will evaluate
\be
D(x) = \frac{\det \cM(x)}{\det \cM(0)}
\label{Dofx}
\ee
Comparing to Eq.~\eqref{Ob}, we see that the functional determinant for the bounce corresponds to $x=-1$. Later we will see
that Goldstone fluctuations have $x=-\frac{1}{3}$ and transverse gauge boson fluctuations in Fermi gauge have $x=-\frac{g^2}{3\lambda}$. 

\subsection{Regularized Sum}
The key to calculating $D(x)$ in Eq.~\eqref{Dofx} is that we know the spectrum exactly. Although we know it exactly in $d$ dimensions~\cite{Drummond:1978pf,McKane:1978md}, regularizing the eigenvalues does not necessarily correspond to a well-understood subtraction scheme. Instead, we will work in 4 dimensions and remove the divergences using Feynman diagrams.

Defining 
\be
\cO(x) = -\frac{1}{3\lambda \phib^2}\cM(x)  = \frac{1}{3\lambda \phib^2}\Box + x 
\ee
we know the spectrum of $\cO(x)$ exactly, as in Section~\ref{sec:jacprob};
\be
\cO(x) \phi_{nslm} = \lambda_n(x) \phi_{nslm}, 
\qquad \lambda_n(x) =\wh \lambda_n^\phi + x = \frac{(n+1)(n+2)}{6} + x 
\label{lambdax}
\ee
Then the determinant is
\be
\ln D(x) =
\ln \frac{ \prod_{n\ge0} [\lambda_n(x)]^{d_n}}{ \prod_{n \ge0} [\lambda_n(0)]^{d_n}} 
= \sum_{n\ge 0} d_n \ln \frac{\lambda_n(x)}{\lambda_n(0)}
\ee
where the degeneracies $d_n$ are in Eq.~\eqref{dvals}. 

This sum is UV divergent at large $n$. We regularize the sum by subtracting the terms of order $x$ and $x^2$ and then we will add those terms back in through dimensionally-regularized Feynman diagrams. Expanding at small $x$, we find
\be
S_\text{sub}^n(x) \equiv  \left[d_n \ln \frac{\lambda_n(x)}{\lambda_n(0)} \right]_{x,x^2}= (2n+3) x - \frac{9+6n}{n^2+3n + 2} x^2 
\label{Dsub}
\ee
Then we can perform the sum. That is, we compute
\be
\Sfin(x) =\sum_{n= 0}^\infty \left[ d_n \ln \frac{\lambda_n(x)}{\lambda_n(0)} - \Ssub^n(x) \right] 
\label{Dfindef}
\ee
finding
\begin{multline}
\label{Dfinis}
\Sfin(x)= (-3+6 \gamma_E)x^2 + \frac{11}{36} + \ln 2\pi + \frac{3}{4\pi^2} \zeta(3) - 4\zeta'(-1)\\
- x \kappa_x \left[  \psi ^{(-1)}\left(\frac{3+\kappa_x}{2}\right)-   \psi ^{(-1)}\left(\frac{3-\kappa_x}{2}\right)\right]
+\left(  6\, x-\frac{1}{6}  \right)
\left[\psi ^{(-2)}\left(\frac{3+\kappa_x}{2}\right)+\psi ^{(-2)}\left(\frac{3-\kappa_x}{2}\right)\right]
   \\
   +\kappa_x\left[\psi ^{(-3)}\left(\frac{3+\kappa_x}{2}\right)- \psi^{(-3)}\left(\frac{3-\kappa_x}{2}\right)\right]
   -2\left[ \psi^{(-4)}\left(\frac{3+\kappa_x}{2}\right)+\psi ^{(-4)}\left(\frac{3-\kappa_x}{2}\right)\right]
\end{multline}
where $\psi^{(n)}(x) = \frac{d^n}{dz^n} \psi(z)$ (defined by analytic continuation for complex $n$) with $\psi(z)$ the digamma function,
\be
\kappa_x \equiv \sqrt{1-24 x}
\label{kappaxdef}
\ee
and $\zeta'(-1) \approx -0.165$ is the derivative of the $\zeta$-function at $-1$.

\subsection{Divergent Parts}
To the subtracted part, we must add in dimensionally-regularized $\msbar$-subtracted divergent contributions. The subtractions were determined by removing the terms to second order in $x$. These terms can be reproduced by computing contributions to second order in $x$ to the effective action using Feynman diagrams.  The Euclidean action whose second variation gives $\cM(x)$  in Eq.~\eqref{Mx} is
\be
S =  \int d^4 x\left[ \frac{1}{2}( \partial_\mu \phi)^2-\frac{1}{2}( 3 x \lambda \phib^2) \phi^2 \right]
\ee
We want to treat the mass term, proportional to $x$, as an interaction to compute the divergent contribution to the effective action.

To compute Feynman diagrams in Euclidean space, we expand $e^{-S}$.  The $-$ sign in front of $S$ affects all the Feynman rules, and Feynman diagrams produce contributions to $-S_\text{eff}$; that is, in Euclidean space $-1$ serves the role that the $i$ prefactor does in Minkowski space. Thus, the interaction Feynman rule is
\be
\begin{gathered}
\includegraphics{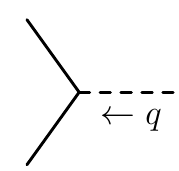}
\end{gathered}
\hspace{5mm}
=3x \lambda\widetilde{\phib^2}(q)
\ee
In our notation solid lines are propagating $\phi$ fields and dashed lines are background field insertions.
The injected momentum is distributed according to the Fourier transform of the bounce-squared~\cite{Isidori:2001bm}:
\be
\widetilde{\phib^2}(q) = \int d^4 x e^{i q^\mu x^\mu} \phib(x)^2 = \frac{16\pi^2 R^2}{-\lambda} \cK_0(\sqrt{q^2} R)
\label{FTbs}
\ee

At order $x$, there is only one graph, a tadpole
\be
-S_{x} = 
\begin{gathered}
\includegraphics{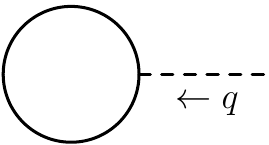}
\end{gathered}
\hspace{0mm}
=\frac{1}{2}\left(3x \lambda \right) \int \frac{d^d q}{(2\pi)^d} \widetilde{\phib^2}(q)  \int \frac{d^d k}{(2\pi)^d} \frac{1}{k^2} =0
\label{Sx1}
\ee
Here, the $\frac{1}{2}$ in front is a symmetry factor. 
This graph is scaleless and vanishes in dimensional regularization. Note that in 4D, the graph is quadratically divergent, in agreement
with the $\cO(n)$ term in Eq.~\eqref{Dsub} which is quadratically divergent when summed over $n$.  

There is one graph with two $x$ insertions:
\be
-S_{x^2} = 
\begin{gathered}
\includegraphics{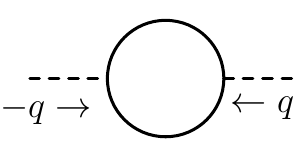}
\end{gathered}
\hspace{-1mm}
=\frac{1}{4}(3\lambda x)^2 \int \frac{d^d q}{(2\pi)^d} \widetilde{\phib^2}(q)   \widetilde{\phib^2}(-q)\int \frac{d^d k}{(2\pi)^d} \frac{1}{k^2}\frac{1}{(q+k)^2}
\ee
Here the $\frac{1}{4}$ is the symmetry factor. The $k$ integral can be done first giving
\be
B_0 =\mu^{d-4}\int \frac{d^d k}{(2\pi)^d} \frac{1}{k^2}\frac{1}{(k+q)^2}
=\frac{1}{16\pi^2}\left[\frac{1}{\eps} -\gamma_E + \ln (4\pi) + 2 - \ln \frac{q^2}{\mu^2} \right]
\label{B0}
\ee
Then we use
\begin{align}
\int \frac{d^4q}{(2\pi)^4}  \widetilde{\phib^2}(q)   \widetilde{\phib^2}(-q)  &= \frac{32\pi^2}{3\lambda^2} \label{usefulintegrals1}\\ 
\int \frac{d^4q}{(2\pi)^4}  \widetilde{\phib^2}(q)   \widetilde{\phib^2}(-q) \ln \frac{ q^2}{\mu^2}
  &= \frac{32\pi^2}{9\lambda^2} \left[1-6\gamma_E -6 \ln\frac{R\mu}{2}\right]
\label{usefulintegrals2}
\end{align}
to find that
\be
-S_{x^2} =
\left( \frac{3}{2\eps}  + \frac{5}{2} + \frac{3}{2} \gamma_E + \frac{3}{2}\ln \pi R^2\mu^2\right)x^2
\label{Sx2}
\ee
and therefore, rescaling $\mu^2 \to \frac{\mu^2}{4\pi e^{-\gamma_E}}$ to go to $\msbar$, we have
\be
-S_{\text{loops}}(x) = - S_{x} - S_{x^2} =  
\left( \frac{3}{2\eps}  + \frac{5}{2} + 3 \gamma_E + 3 \ln \frac{R\mu}{2}\right)x^2
\label{Ddiv}
\ee 
Note that the integrals in Eqs.~\eqref{usefulintegrals1} and \eqref{usefulintegrals2} are proportional to $S[\phib]$. This is not a coincidence, as the divergences must be canceled by renormalizing $\lambda$ in the tree-level action.\footnote{Technically, one should do the integrals in
Eqs.~\eqref{usefulintegrals1} and \eqref{usefulintegrals2} in $d$ dimensions, using a $d$-dimensional bounce, generating $\cO(\eps)$ terms  which contribute additional finite parts to Eq.~\eqref{Ddiv}. However, these finite parts must exactly cancel the finite contributions from the $d$-dimensional action renormalizing $\lambda$, since both multiply the same $\frac{1}{\eps}$ terms. Thus we can ignore both and use the 4D integrals as written.}

The full result in $\msbar$ for a bosonic functional determinant this combined with Eq.~\eqref{Dfindef}
\be
\sqrt{ \frac{\det \cM(0)}{\det \cM(x)}} = \exp\left[ -\frac{1}{2}\ln D(x) \right] = \exp\left[
-\frac{1}{2} S_{\text{fin}}(x) -S_{\text{loops}}(x) \right]
\ee
with $\Sfin(x)$ in Eq.~\eqref{Dfinis}.

\subsection{Angular Momentum Decomposition \label{sec:angular}}
We will also find it helpful to do the above sum in a different order, summing over $n$ first exactly and then regularizing the sum over angular momentum modes $s$. The operators whose determinants we are trying to calculate are spherically symmetric. Thus their eigenfunctions are separable and can be written as
\be
\phi(r,\alpha,\theta,\phi) = f_s(r) Y^{slm}(\alpha,\theta,\phi)
\ee
The 4D Laplacian then reduces to a 1D operator, $\Box\phi = \Delta_s \phi$, where
\be
\Delta_s \equiv \partial_r^2 + \frac{3}{r} \partial_r- \frac{s(s+2)}{r^2}
\label{Deltasdef}
\ee
and there is a $(1+s)^2$-fold degeneracy for each $s$. 

In terms of the angular momentum decomposition, the ratio of functional determinants in Eq.~\eqref{Dofx} becomes
\be
D(x) = \frac{\det \cM(x)}{\det \cM(0)}
=
\prod_s [R_s]^{(s+1)^2}
 \label{prodrs}
\ee
where
\be
R_s(x) 
= \frac{\det\left[ \frac{1}{3\lambda \phib^2} \Delta_s + x \right]}
{ \det \left[ \frac{1}{3\lambda \phib^2}\Delta_s\right]}
\label{Rsdef}
\ee

The exact radial eigenfunctions of these operators  are given in Eq.~\eqref{efs}:
\be
\phi_{ns}(r) =\frac{1}{r}P_{n+1}^{-s-1}\left(\frac{R^2-r^2}{R^2+r^2}\right)
\label{phins}
 \ee
These satisfy
\be
\left(\frac{1}{3\lambda \phib^2}\Delta_s +x\right)\phi_{ns} =
\lambda_n(x) \phi_{ns}
 \label{fns}
\ee
with $\lambda_n(x)$ in Eq.~\eqref{lambdax}. There are only eigenfunctions with $n\ge s$.  Thus,
\be
R_s(x)  = \frac{\det\left[ \frac{1}{3\lambda \phib^2} \Delta_s +x\right]}
{ \det \left[\frac{1}{3\lambda \phib^2}\Delta_s\right]}
=\prod_{n\geq s}\frac{\lambda_n(x)}{\lambda_n(0)}=\frac
{\Gamma(1+s) \Gamma(2+s)}
{
\Gamma\left(\frac{3}{2} + s - \frac{\kappa_x}{2}\right)
\Gamma\left(\frac{3}{2} + s +\frac{\kappa_x}{2}\right)
}
\label{RSx}
\ee
with $\kappa_x = \sqrt{1-24 x}$ as in Eq.~\eqref{kappaxdef}. 
In computing the product in Eq.~\eqref{prodrs}, the divergent contributions all appear at order $x$ and $x^2$ so we
compute subtraction terms
\be
 \Ssub^s(x) \equiv  \Big[(s+1)^2 \ln R_s(x) \Big]_{x,x^2}=
6(s+1) x + 18 \left[3 + 2s -2 (s+1)^2 \psi'(1+s) \right]x^2
\label{Dsubs}
\ee
The appearance of the digamma function $\psi(z)$ makes this subtraction more complicated than the subtraction in Eq.~\eqref{Dsub}. 
Note that $\psi'(s+1) \sim \frac{1}{s}$ at large $s$ so there is a logarithmic divergence encoded in this expression. 
Performing the sum, we find
\be
\sum_{s= 0}^\infty \Big[(s+1)^2 \ln R_s(x) - \Ssub^s(x) \Big]= \Sfin(x) 
\ee
in exact agreement with Eq.~\eqref{Dfinis}.

For some cases, the $s=0$ modes will be absent. For these cases, it is helpful to have a form for the finite contribution with the $s=0$ modes removed. To that end, we define
\be
\Sfin^+(x)=
\sum_{s= 1}^\infty \Big[(s+1)^2 \ln R_s(x) - \Ssub^s(x) \Big]
= \Sfin (x) - \ln\left[ \frac{\cos(\frac{\pi}{2} \kappa_x)}{6 \pi x}  \right] + 
6 x + 6 (9-\pi^2) x^2
\label{Splus}
\ee

\subsection{The Gelfand-Yaglom Method \label{sec:GY}}
There is a very powerful way of computing functional determinants, that does not require knowing the exact spectrum of the operators, called the Gelfand-Yaglom method \cite{Gelfand:1959nq}. Reviews and derivations of the method can be found in~\cite{Marino,Kirsten:2004qv,Kirsten:2003py,Endo:2017tsz,Dunne:2005rt}, here we just summarize its application to the scale-invariant potential of interest in this paper.

The Gelfand-Yaglom method says that functional determinants can be calculated by finding zero modes of the operators and evaluating their asymptotic behavior. 
For example, for  1-dimensional operators, like the ratio $R_s$ in Eq.~\eqref{RSx}, 
the method says that 
\be
R_s = \frac{\det\left[ \frac{1}{3\lambda \phib^2} \Delta_s+x\right]}
{ \det \left[ \frac{1}{3\lambda \phib^2}\Delta_s\right]}
=
\left[ \lim_{r \to 0} \frac{\phi_0^s(r)}{\phi_x^s(r)}\right]
\left[ \lim_{r \to \infty} \frac{\phi_x^s(r)}{\phi_0^s(r)}\right]
\equiv
 \frac{\phi_0^s(0)}{\phi_x^s(0)}
 \frac{\phi_x^s(\infty)}{\phi_0^s(\infty)}
\ee
where $\phi(\infty)$ and $\phi(0)$ are shorthand for the limits in this equation and the functions $\phi_x^s$ satisfy
\be
\label{GYdes}
\left[\frac{1}{3\lambda \phib^2}\Delta_s +x\right]\phi_x^s = 0
\ee
The boundary conditions are that the functions be regular at $r=0$.

 Note how powerful the method is: instead of finding an infinite number of eigenvalues and taking their product, we simply solve the differential equations in Eq.~\eqref{GYdes}, which can be done numerically, and evaluate the solutions as $r\to \infty$ and $r\to 0$.

To see how the Gelfand-Yaglom method works, consider the real scalar case $(x=-1)$ first.
First of all, we already know the answer from Eq.~\eqref{RSx}:
\be
R_s  = \frac{\det\left[ \frac{1}{3\lambda \phib^2} \Delta_s -1\right]}
{ \det \left[\frac{1}{3\lambda \phib^2}\Delta_s\right]}
=
 \prod_{n \ge s} \frac{\frac{1}{6}(n+1)(n+2)-1 }{\frac{1}{6}(n+1)(n+2)} = \frac{s(s-1)}{(s+2)(s+3)} \label{RSexact}
\ee
To use the  Gelfand-Yaglom method, we find the exact solution to Eq.~\eqref{GYdes} regular at $r=0$. It is
\be
\phi_0^s = r^s
\label{phifv}
\ee
The regular solution to Eq.~\eqref{GYdes} that reduces to $\phi_0^s$ at small $r$ is
\be
\phi_{-1}^s = \frac{r^s}{(R^2+r^2)^2} \left( R^4 + \frac{2 R^2 (s-1)}{s+2} r^2 + \frac{s(s-1)}{(s+2)(s+3)} r^4 \right)
\ee
Thus Gelfand-Yaglom predicts that
\be
R_s = \lim_{r \to \infty} \frac{\phi_{-1}^s(r)}{\phi_{0}^s(r)} =  \frac{s(s-1)}{(s+2)(s+3)}
\label{RsGY}
\ee
in exact agreement with Eq. \eqref{RSexact}. 

Note that for $n=1$, Eq.~\eqref{RSexact} gives $R_s=0$. This is because for $n=1$ there are zero modes. Indeed, the $n=1$, $s=0$ zero mode is the dilatation mode and the $n=1$, $s=1$ modes are the translations.  
For the $s=1$  case, the determinant ratio with eigenvalues removed is 
\be
R_1^\prime= \frac{\det^\prime\left[ \frac{1}{3\lambda \phi_{b}^2} \Delta_1-1\right]}
{ \det\left[\frac{1}{3\lambda \phib^2} \Delta_1\right]}
= 
\frac{ \prod_{n \ge 2}\frac{1}{6}(n+1)(n+2)-1 }{\prod_{n \ge 1}\frac{1}{6}(n+1)(n+2)}  = \frac{1}{10} \label{R11}
\ee
Similarly, $R_0^\prime =-\frac{1}{5}$ by multiplying the eigenvalues.

To use Gelfand-Yaglom to calculate the zero eigenvalues, we shift the operator by order $\epsilon$. That is, we replace Eq.~\eqref{GYdes} by
\be
\left[
\frac{1}{3\lambda \phib^2} \Delta_s-1 +\epsilon\right]\phi_{-1}^{1,\epsilon} = 0 \label{GYdes3}
\ee
Shifting the free-theory operator by $\epsilon$ is not necessary since all of its eigenvalues are nonzero.
Note that the zero mode $\phi_{11}$ is an eigenfunction of the shifted  operator with eigenvalue $\epsilon$. 
The function with eigenvalue $0$ is $\phi_{-1}^{1,\epsilon}(r)=\phi_{n_\epsilon,1}(r)$ with
$\phi_{ns}(r)$ in Eq.~\eqref{phins} and $n_\eps = -\frac{3}{2}+\frac{5}{2}\sqrt{1-\frac{24}{25} \epsilon}$.
Then, using $\phi_0^1(r) = r$, 
\be
R_1' = \lim_{\epsilon\to 0}  \frac{1}{\epsilon} \left[
\lim_{r\to \infty}
\frac{\phi_{{-1}}^{1,\epsilon}(r)}{r}\right]
\left[\lim_{r \to 0}
\frac{r}{\phi_{-1}^{1,\epsilon}(r)}
\right] = \frac{1}{10}
\label{R12}
\ee
in agreement with Eq.~\eqref{R11}. 

The $s=0$ functional determinant can be computed in exactly the same way without any additional complication, finding $R_0'=-\frac{1}{5}$, in agreement with the direct calculation.

\section{Functional Determinants \label{sec:funcdet}}
In this section we compute the functional determinants for the fluctuation of scalars, Goldstone bosons, vector bosons and Dirac fermions around the scale-invariant bounce configuration. We produce analytic formulas for all the cases. In the vector boson case, we check explicitly that the result is gauge invariant by using a generic value of $\xi$ in Fermi gauges, and also show agreement between Fermi and $R_\xi$ gauges.

\subsection{Real Scalars\label{sec:bouncedet}}
The case of a single  scalar field was introduced in Section~\ref{sec:scale}. The Euclidean Lagrangian is
\be
\cL = \frac{1}{2}(\partial_\mu \phi)^2 + \frac{\lambda}{4} \phi^4
\ee
The Euclidean equations of motion $-\Box\phi + \lambda \phi^3 =0$ are solved by $\phi=\phib$. The operator for quadratic fluctuations around $\phib$ is
\be
\cM_\phi = - \Box + 3 \lambda \phib^2
\ee
Thus real scalar fluctuations correspond to the case studied in Section~\ref{sec:gendet} with $x=-1$. 

For $x=-1$,  the finite contribution from the sum over $n \ge 0$ is singular, ($\Sfin(x)$ in Eq.~\eqref{Dfinis} is singular as $x\to -1$).
This is due to the zero modes at $n=1$ corresponding dilatations and translations around the bounce.  
To compute the determinant with zero modes removed, we must first rescale the operator. We therefore define
\be
\cO_\phi =- \frac{1}{3\lambda \phib^2} \cM_\phi =  \frac{1}{3\lambda \phib^2} \Box - 1
\ee
Recall from Section~\ref{sec:jacprob} that this rescaling allows the change to collective coordinates to have a finite Jacobian. 
The Jacobians for dilatation and translations are given in Eqs.~\eqref{Jd} and~\eqref{JT}. 

To compute $\det^\prime \cO_\phi$ we must remove these modes from the sum in Eq.~\eqref{Dfindef} and add in only the $n=1$ contributions to the false vacuum fluctuations. We note the only zero mode for $n=1$ in Eq.~\eqref{Dfindef} is in
\be
d_1 \ln \lambda_1(x)= 5 \ln (x+1)
\ee
which is singular at $x=-1$. Removing this term from the sum, we find a smooth limit as $x\to -1$:
\be
\Sfin^\phi =
\lim_{x\to -1} \left[\Sfin(x) - 5 \ln(x+1)\right] = \frac{15}{2} + 6 \gamma_E  - i \pi -12 \zeta'(-1) + \ln \frac{7776}{3125} 
\label{Dfinphi}
\ee
Note that we should leave the $x^2$ terms in the subtraction at $n=1$ to avoid overcounting, since these are included in the loops. 
Also note that we only remove the $n=1$ term from $\det^\prime \cO_\phi$, while the $n=1$ term from $\det \wh \cO_\phi$ is finite and should not be removed.

Combining with the divergent part from Eq.~\eqref{Ddiv}, we then have
\be
\im \sqrt{\frac{\det \wh \cO_\phi}{\det^\prime \cO_\phi} }  =
 \frac{25}{36} \sqrt{\frac{5}{6}} 
\exp \left[ \frac{3}{2\eps}- \frac{5}{4}  + 6 \zeta'(-1) +3  \ln \frac{R\mu}{2} \right]
\label{imform}
\ee
Here, $\wh \cO_\phi$ means the operator with $\phib=0$, corresponding to fluctuations around the false vacuum. 
Note how the factors of $\gamma_E$ have dropped out. 

The remaining task is to renormalize. In $\msbar$ the $Z$-factor for $\lambda$ at one loop is
\be
Z_\lambda = 1+\frac{9\lambda_R}{16\pi^2}\frac{1}{\eps}
\ee
The renormalized action on the bounce then becomes
\be
S[\phib ] =-\frac{8\pi^2}{3 \lambda_0} =-\frac{8\pi^2}{3Z_\lambda \lambda_R} = -\frac{8\pi^2}{3 \lambda_R} + \frac{3}{2\eps} + \cdots
\label{Sbdb}
\ee
Combining with Eq. ~\eqref{imform}, we get
\be
e^{-S[\phib]}
\im \sqrt{\frac{\det \wh \cO_\phi}{\det^\prime \cO_\phi} }
 =
 e^{\frac{8\pi^2}{3\lambda_R} }
 \frac{25}{36} \sqrt{\frac{5}{6}} 
\exp\left[
- \frac{5}{4}  + 6 \zeta'(-1) +3  \ln \frac{R\mu}{2}
\right]
\label{fdetrat}
\ee


\subsection{Complex Scalars and Global Symmetries \label{sec:global}}
Next, we discuss the case when the false vacuum admits a continuous global symmetry that is spontaneously broken by the bounce.
For concreteness, we take the simplest example, a field theory of a complex scalar field $\Phi$ with a global $U(1)$ symmetry. The Euclidean
Lagrangian density is
\be
\cL =| \partial_\mu \Phi|^2  + V(\Phi)
\ee
where $V(\phi) =  \lambda |\Phi|^4$. We expand the field as
\be
\Phi = \frac{1}{\sqrt{2}} \left(\phib + \phi + i G\right)
\label{phiexp}
\ee
With this normalization for a complex field, the bounce is the same as Eq.~\eqref{bounceform} and still
satisfies $-\Box \phib  + \lambda \phib^3= 0$.

Expanding around the bounce background to quadratic order, the scalar and Goldstone modes satisfy
\begin{align}
(-\Box + 3\lambda \phib^2)\phi &= 0\\
(-\Box + \lambda \phib^2)G &= 0
\end{align}
Both of these equations are special cases of Eq.~\eqref{Mx}, with $x=-1$ for $\phi$ and $x=-\frac{1}{3}$ for $G$. 
The scalar fluctuations $\phi$ we have already discussed:  there are 5 zero modes with $n=1$ (since $\lambda_1(-1)=0$ with $\lambda_n(x)$ in Eq.~\eqref{lambdax}), corresponding to translations and dilatations. The Jacobians for removing these zero modes in conformal coordinates
are given in Eq.~\eqref{Jd} and~\eqref{JT} and the functional determinant with zero modes removed is in Eq.~\eqref{fdetrat}

For $G$ with $x=-\frac{1}{3}$ there is a single zero mode ($\lambda_0(-\frac{1}{3})=0$) corresponding to phase rotations $\Phi \to e^{i \alpha} \Phi$.
 The $n=0$ mode has no degeneracy and the eigenfunction is
\be
G_0 = \phi_{0000} \propto \phib
\ee
as in Eq.~\eqref{phi0000}. Infinitesimally along this direction, $\Phi = \frac{1+ i \alpha }{\sqrt{2}}\phib$ which has the same action as $\Phi$ up to order $\alpha^2$.

As with $\phi$, we have to remove the fluctuations along the zero-mode direction exactly using collective coordinates. With the measure determined by the operator $\cO_G = \frac{1}{3 \lambda \phib^2} (\Box- \lambda \phib^2)$ we have
\be
J_G = \sqrt\frac{ \langle \phi_{b} | \phi_{b} \rangle_V}{2\pi}
= \sqrt{-\frac{1}{2\pi}\int d^4x V''[\phib] \phib^2} = \sqrt{\frac{16 \pi}{-\lambda}} = \sqrt{\frac{6 S[\phib]}{\pi}}
\label{JG}
\ee

To calculate the determinant with zero mode removed we follow the procedure in Section~\ref{sec:bouncedet}. The $n=0$ mode  in Eq.~\eqref{Dfindef} contributes
\be
d_0 \ln \lambda_0(x)=  \ln \left(x+\frac{1}{3}\right)
\ee
Note the singularity as $x\to -\frac{1}{3}$. Removing this from the sum, we find a smooth limit as $x\to - \frac{1}{3}$:
\be
\Sfin^G
= \lim_{x\to - \frac{1}{3}} \Big[\Sfin(x) - \ln \left(x+\frac{1}{3}\right) \Big] =
\frac{13}{18} + \frac{2}{3} \gamma_E - 4 \zeta'(-1)+ \ln 2 
\label{DfinG}
\ee
Adding in the divergent piece, in Eq.~\eqref{Ddiv} with $x=-\frac{1}{3}$, we get
\be
\sqrt{\frac{\det \wh \cO_G}{\det^\prime \cO_G}}
=
\sqrt{\frac{1}{2}} 
\exp \left[ \frac{1}{6\eps}  -\frac{1}{12} + 2  \zeta'(-1) + \frac{1}{3}  \ln \frac{R\mu}{2} \right]
\label{detGfinal}
\ee

The full contribution from the Goldstone fluctuations is therefore
\be
\int d\theta J_G \sqrt{\frac{\det \wh \cO_G}{\det^\prime \cO_G}}  = 2\pi
\sqrt{\frac{6 S[\phib]}{\pi}}
\sqrt{\frac{1}{2}} 
\exp \left[ \frac{1}{6\eps}  -\frac{1}{12} + 2  \zeta'(-1) +\frac{1}{3}  \ln \frac{R\mu}{2} \right]
\label{GJsum}
\ee
where $\int d\theta = 2\pi$ gives the volume of $U(1)$. 

For other gauge groups, the procedure is identical up to the group volume factor. Indeed, at quadratic order, all of the Goldstone boson directions decouple and the path integral over each direction gives a factor of Eq.~\eqref{detGfinal} and a Jacobian. All that needs to be changed is the group volume factor. For $SU(2)$, this is $2\pi^2$.

Putting the results for the Goldstone fluctuations together with the scalar fluctuations, we get for the complex scalar theory
\begin{align}
\frac{\Gamma}{V} &=\frac{1}{VT}e^{-S[\phib]}
\int d^4x \int dR  \int d\theta 
J_d J_T^4 J_G
\im \sqrt{\frac{\det \wh\cO_\phi}{\det^\prime \cO_\phi} }
\sqrt{\frac{\det \wh \cO_G}{\det^\prime \cO_G}}
\\
&= e^{-S[\phib]}\frac{2\sqrt{3}}{\pi^2} S[\phib]^3 \int \frac{dR}{R^5} \exp\left[\frac{5}{3\eps} -\frac{4}{3} + 8 \zeta'(-1)  + \frac{10}{3} \ln \frac{R \mu}{2} \right]
\end{align}
The $R$ integral is would be cutoff by higher order effects if $\lambda(\mu)$ were bounded from below (which it is not in this theory).
 The UV divergence is canceled by the rernormalization of $\lambda$, as in the real scalar theory. In this case, 
the action on the bounce in renormalized perturbation theory is (cf. Eq.~\eqref{Sbdb})
\begin{align}
S[\phib]  
&=- \frac{8\pi^2}{3\lambda_R} +\frac{5}{3 \eps} 
 + \cO(\lambda_R)
\end{align}
So that,
\be
\frac{\Gamma}{V} =e^{-S[\phib]}\frac{2\sqrt{3}}{\pi^2}S[\phib]^3 \int \frac{dR}{R^5} \exp\left[ -\frac{4}{3} + 8 \zeta'(-1)  + \frac{10}{3} \ln \frac{R \mu}{2} \right]
\ee
which is UV finite. 

We note for future reference that at each $s$ the contribution to the functional determinant for the Goldstone modes follows
from Eq.~\eqref{RSx}
\be
R_s^G = R_s\left(-\frac{1}{3}\right) =  \frac{s}{s+2}
\label{RsG}
\ee
We also note that for $s=0$ there is a zero mode. Removing the zero mode we find $R_0^G{}'=1$.
We also note that since $n=0$ implies $s=0$ if we remove the $s=0$ modes, we automatically remove the zero modes. Thus,
if we sum over only $s>0$, the Goldstone contribution is simply
\be
\Sfin^{G+}=
\lim_{x\to - \frac{1}{3}} \Sfin^+(x) = \frac{85}{18} -\frac{2\pi^2}{3}+ \frac{2}{3} \gamma_E - 4 \zeta'(-1)+ \ln 2 
\label{SfinGp1}
\ee

\subsection{Vector Fields and Local  Symmetries}
Next, we discuss the contribution of gauge bosons to the decay rate. We continue the $U(1)$ case, but now with Euclidean Lagrangian
\be
\cL =\frac{1}{4}F_{\mu\nu}^2+(\partial_\mu \Phi^\star + i g A_\mu \Phi^\star)(\partial_\mu \Phi- i g A_\mu \Phi)  + V(\Phi) + \cL_{\text{GF}} + \cL_{\text{ghost}}
\ee
where $V=\lambda |\Phi|^4$ as before and we expand $\Phi=\frac{1}{\sqrt{2}}(\phib + \phi  + i G)$ as in Eq.~\eqref{phiexp}.
 For the gauge fixing term, we can consider the $R_\xi$ gauges, as in~\cite{Isidori:2001bm,Branchina:2014rva},
where  
\be
\cL^{R_\xi}_{\text{GF}} = \frac{1}{2\xi}(\partial_\mu A_\mu - g \phib G)^2
\ee
So that at quadratic order, 
\begin{multline}
\cL_{R_\xi} = \frac{1}{2}A_\mu \left[(-\Box +g^2 \phib^2)\delta_{\mu\nu}  + \frac{\xi-1}{\xi} \partial_\mu \partial_\nu \right]A_\nu
+\frac{1}{2}G \left[-\Box +\left(\frac{g^2}{\xi}+\lambda\right) \phib^2\right] G  \\
+\left(\frac{1}{\xi}+1\right) g A_\mu  (\partial_\mu \phib) G+\left(\frac{1}{\xi}-1\right) g A_\mu \phib (\partial_\mu G ) 
 + \bar{c} \left[ -\Box + g^2 \phib^2\right] c
 \label{LRxi}
\end{multline}
While these gauges have some convenient features, particular for $\xi=1$, they have a very serious drawback: they break the global $U(1)$ symmetry. As a consequence there is no zero mode for the Goldstone fluctuations. 
Because of the missing zero mode, we are unable to reproduce the results of~\cite{Isidori:2001bm,Branchina:2014rva}.
Although we are able to get sensible results in $R_\xi$ gauges using a numerical implementation of the  the Gelfand-Yaglom method, 
we find Fermi gauges~\cite{Endo:2017tsz} more convenient for deriving exact analytic results.

In Fermi gauges, the gauge fixing term is
\be
\cL_{\text{GF}} = \frac{1}{2\xi}(\partial_\mu A_\mu)^2
\ee
So that at quadratic order
\begin{multline}
\cL_{\text{Fermi}} =\frac{1}{2}A_\mu \left[(-\Box +g^2 \phib^2)\delta_{\mu\nu}  + \frac{\xi-1}{\xi} \partial_\mu \partial_\nu \right]A_\nu
+\frac{1}{2}G \left[-\Box +\lambda \phib^2\right]G  \\
+ g A_\mu  (\partial_\mu \phib) G - g \phib A_\mu \partial_\mu G  -\bar{c} \, \Box \,  c
\label{LFermi}
\end{multline}
Fermi gauges leave the global $U(1)$ symmetry of the Lagrangian intact (the action is invariant under $\Phi \to e^{i\alpha} \Phi$, $A_\mu \to A_\mu$). 
Note that since the ghost Lagrangian is independent of the bounce, the functional determinant over ghosts normalized to the false vacuum is just 1.

In  Fermi gauges, the equations of motion for $A_\mu$ and $G$ are coupled. At quadratic order
\begin{align}
&(-\Box +g^2 \phib^2) A_\mu + \left(1-\frac{1}{\xi}\right) \partial_\mu \partial_\nu A_\nu + g (\partial_\mu \phib) G - g \phib \partial_\mu G = 0
\\
&(-\Box +\lambda \phib^2) G + 2 g (\partial_\mu \phib) A_\mu + g \phib \partial_\mu A_\mu = 0
\end{align}
Following~\cite{Isidori:2001bm, Branchina:2014rva, Endo:2017tsz}, we then exploit the spherical symmetry, expanding $A_\mu$ as
\be
A_\mu = \!\!\!\!
\sum_{s=0,1,2,\cdots} \left[ a_S(r) \frac{x_\mu}{r}  + a_L(r) \frac{r}{\sqrt{s(s+2)}}\partial_\mu +
\Big(a_{T1}(r) V^{(1)}_\nu +a_{T2}(r) V^{(2)}_\nu\Big)
 \epsilon_{\mu\nu \rho\sigma} x_\rho \partial_\sigma \right] Y_{slm}(\alpha,\theta, \phi)
\label{eq:ExpandASpH}
\ee
with $V^{(1)}_\mu$ and $V^{(2)}_\mu$ two independent generic vectors and $Y_{slm}(\alpha,\theta, \phi)$ the 3D spherical harmonics in
Eq.~\eqref{Yslm}. In this basis, and writing $G(x) = G(r) Y_{slm}(\alpha,\theta,\phi)$ the fluctuation operators decouple for each $s,l,m$ and the resulting operators depend only on $s$. After some algebra (see~\cite{Isidori:2001bm} for some details), we find for $\xi=1$ that
the $a_S$ and $a_L$ modes couple to $G$, through the operator
\be
\label{Mslg}
\cM_s^{SLG} = \begin{pmatrix}
-\Delta_s + \frac{3}{r^2} + g^2 \phib^2 & -\frac{ 2\sqrt{s(s+2)}}{r^2} & g \phib' - g \phib \partial_r \\
- \frac{2\sqrt{s(s+2)}}{r^2} &  -\Delta_s - \frac{1}{r^2} + g^2 \phib^2 &- \frac{ \sqrt{s(s+2)}}{r} g \phib \\ 
 2 g \phib' + g \phib \partial_r + \frac{3}{r} g \phib & - \frac{ \sqrt{s(s+2)}}{r} g \phib  & -\Delta_s  + \lambda \phib^2
 \end{pmatrix}
\\
+\cM^\xi_s
\ee
where the gauge-dependent piece is
\be
\cM^\xi_s = \left(1-\frac{1}{\xi}\right)
\begin{pmatrix} 
\partial_r^2 + \frac{3}{r}\partial_r - \frac{3}{r^2}
&
- \frac{\sqrt{s(s+2)}}{r}\left(\partial_r - \frac{1}{r}\right) & 0 \\
\frac{\sqrt{s(s+2)}}{r}\left(\partial_r +\frac{3}{r}\right) & - \frac{s(s+2)}{r^2}& 0 \\
 0 & 0 &0 
 \end{pmatrix}
 \ee
and $\Delta_s$ is in Eq.~\eqref{Deltasdef}.
The corresponding false-vacuum operator is $\cM_s^{SLG}$ with $\phib=0$:
\be
\widehat{\cM}_s^{SLG} = \begin{pmatrix}
-\Delta_s + \frac{3}{r^2} & -\frac{ 2\sqrt{s(s+2)}}{r^2} &0 \\
- \frac{ 2\sqrt{s(s+2)}}{r^2} &  -\Delta_s - \frac{1}{r^2} &0\\ 
 0 & 0  & -\Delta_s  
\end{pmatrix} +\cM^\xi_s
\label{Mhatslg}
\ee
Note that in Fermi gauges the gauge-dependent part $\cM^\xi_s$ does not depend on $\phib$ so contributes in the same way to 
$\cM_s^{SLG}$ and $\wh \cM_s^{SLG}$.  This is very useful for establishing gauge invariance of the result, as we will see. 
 
In Fermi gauges, the transverse modes fluctuate independently, through
\be
\cM_s^T = -\Delta_s + g^2 \phib^2
\label{MsTdef}
\ee
or more simply,  they satisfy the Lorentz-invariant equation with operator
\be
\cM^T = -\Box + g^2 \phib^2
\label{MTdef}
\ee

\subsubsection*{Transverse Fluctuations}
For the transverse fluctuations in Fermi gauge, we can calculate the determinant exactly. In fact we already have, in Section~\ref{sec:gendet}.
The transverse fluctuation operator $\cM^T$ is the same as in Eq.~\eqref{Mx} with $x=-\frac{g^2}{3\lambda}$.
The determinant at each $s$ value is in Eq.~\eqref{RSx}:
\be
R_s^T  =R_s\left(-\frac{g^2}{3\lambda}\right) 
=\frac
{\Gamma(1+s) \Gamma(2+s)}
{
\Gamma\left(\frac{3}{2} + s - \frac{\kappa}{2}\right)
\Gamma\left(\frac{3}{2} + s +\frac{\kappa}{2}\right)
}
\label{RsT}
\ee
where
\be
\kappag = \sqrt{1+ \frac{8g^2}{\lambda}}
\label{kappagdef}
\ee

For the full  contribution of the transverse modes we must take the product over all $s>0$. Note that there is no $s=0$ contribution since
$Y_{000}$ is constant and therefore the transverse contribution at $s=0$ is absent in Eq.~\eqref{eq:ExpandASpH}. 
Subtracting the $s=0$ contributions from our previous results for the sum over all $s$, we then get for the transverse polarizations
\be
\ln\frac{\det \cM^T}{\det \wh \cM^T} =\Sfin^+\left(-\frac{g^2}{3\lambda}\right)
+2 S_{\text{loops}}\left(-\frac{g^2}{3\lambda}\right)
\label{detMT}
\ee
where $S_\text{loops}(x)$ is in Eq.~\eqref{Ddiv} and  $\Sfin^+(x)$ in Eq.~\eqref{Splus}. Explicitly:
\be
\Sfin^+\left(-\frac{g^2}{3\lambda}\right)
=\Sfin\left(-\frac{g^2}{3\lambda}\right)
-\ln\left[-\frac{\lambda}{2\pi g^2}\cos\left(\frac{\pi}{2}\sqrt{1+\frac{8g^2}{\lambda}}\right)\right]-2\frac{g^2}{\lambda}-\frac{2(\pi^2-9)}{3}\frac{g^4}{\lambda^2}
\label{eq:sfinT}
\ee
Note that there may be $s=0$ contributions included in $S_\text{loops}(x)$. We will account
for this by including the full gauge-invariant loop corrections for all modes at once when we discuss renormalization below.

\subsubsection*{Fluctuations with $s=0$}
For $s=0$, $Y_{000}$ is constant and so the transverse and longitudinal modes decouple. In this case, only the scalar vector boson polarization and the Goldstone mode remain. The fluctuation operator is
\be
\cM_0^{SG} = 
\begin{pmatrix}
\frac{1}{\xi} \left(-\Delta_0 + \frac{3}{r^2}\right) +  g^2 \phib^2  & g \phib' - g \phib \partial_r \\
2 g \phib' + g \phib \partial_r + \frac{3}{r}g \phib & - \Delta_0 + \lambda \phib^2 
\end{pmatrix}
\ee
The corresponding operator with $\phib=0$ is
\be
\widehat{\cM}_0^{SG} = 
\begin{pmatrix}
\frac{1}{\xi} \left(-\Delta_0 + \frac{3}{r^2}\right)   &  0 \\
0 & - \Delta_0
\end{pmatrix}
\ee
Note that $\cM_0^{SG}$ has two zero modes regular at $r=0$:
\be
\Psi_1 =
\begin{pmatrix}
0 \\
\phib
\end{pmatrix}
\qquad
\text{and}
\qquad
\Psi_2 =
\begin{pmatrix}
r\\
\frac{g}{2} r^2 \phib
\end{pmatrix}
\ee
The first zero mode  corresponds to the global $U(1)$ invariance we saw already in the $g=0$ case.  
The two zero modes for $\widehat \cM_0^{SG}$ regular at $r=0$ are
\be
\widehat{\Psi}_1 =
\begin{pmatrix}
0\\
1
\end{pmatrix}
\qquad
\text{and}
\qquad
\widehat{\Psi}_2 =
\begin{pmatrix}
r\\
0
\end{pmatrix}
\ee

We need to remove the zero modes by going to collective coordinates, just as in Section~\ref{sec:global}. 
Since we do not know the eigenfunctinos of $\cM_0^{SG}$ exactly we will use the Gelfand-Yaglom method. After rescaling our operator we add to it a shift of order $\epsilon$. Then we need to compute
\be
R_0^{ SG}{}'
=
 \frac
{\det' \left(\frac{-1}{3\lambda \phib^2}{\cM}_0^{SG} \right)}
 {\det \left(\frac{-1}{3\lambda \phib^2}\wh\cM_0^{SG} \right)}
= \lim_{\eps \to 0} \frac{1}{\eps}
 \frac
{\det \left( \frac{-1}{3\lambda \phib^2}{\cM}_0^{SG} + \eps\cdot {\mathbbm 1}\right)}
 {\det\left(\frac{-1}{3\lambda \phib^2}\wh\cM_0^{SG}\right)}
\ee
We can compute the zero modes of $\frac{-1}{3\lambda \phib^2}\cM_0^{SG} + \eps\cdot\mathbbm{1}$ perturbatively. Since only $\Psi_1$ goes to zero at $r\to \infty$, we only need the corrections to it. So we expand, following~\cite{Endo:2017tsz},
\be
\Psi_1^{(\eps)}  = \Psi_1 + \eps \check \Psi + \cO(\eps^2)
\ee
If the function $\check\Psi$ satisfies
\be
\frac{-1}{3\lambda \phib^2}\cM_0^{SG} \check\Psi  = -\Psi_1
\ee
then we will have $(\frac{-1}{3\lambda \phib^2}\cM_0^{SG} + \eps\cdot \mathbbm{1})\Psi_1^{(\eps)} = \cO(\eps^2)$ as desired. 

We can find the solution to this differential equation exactly by integration, following~\cite{Endo:2017tsz}:
\be
\check\Psi 
= 
\begin{pmatrix}
-\xi \sqrt{\frac{2g^2}{-\lambda}}\frac{r^3}{R^3} \phib \\
\left[\frac{r^2}{R^2}\left(1+ \frac{2g^2}{\lambda} \xi \right)+ \left(2-2 \frac{g^2}{\lambda} \xi\right) \ln(1+\frac{r^2}{R^2})
\right]\phib
\end{pmatrix}
\ee
The result is 
\be
R_0^{ SG}{}'
=\left[
\lim_{r\to 0} \frac
{\det(\widehat\Psi_1 \widehat \Psi_2)}
{ \det(\check\Psi \Psi_2)}
\right]
\left[
\lim_{r\to \infty} \frac
{ \det(\Psi_1 \Psi_2)}
{\det(\widehat\Psi_1 \widehat \Psi_2)}
\right]
= 1\
\label{eq:rsgzero}
\ee
Note that the result is gauge invariant, and its (trivial) $g\to0$ limit agrees with $R_0^{G}{}'=1$ computed at the end of Section~\ref{sec:global}.

\subsubsection*{Fluctuations with $s>0$ \label{sec:gts}}
Now let's consider the $s>0$ fluctuations.  We need to find three independent solutions to $\cM_s^{SLG} \Psi=0$ 
and $\widehat{\cM}_s^{SLG}\widehat\Psi=0$ with  $\cM_s^{SLG}$ and $\widehat{\cM}_s^{SLG}$ given in Eqs.~\eqref{Mslg} and~\eqref{Mhatslg}. The solutions need to be regular at the origin, but can have arbitrary normalization. The determinant is then
\be
R_s^{SLG} = \frac{\det \cM_s^{SLG}}{\det {\wh\cM}_s^{SLG}} = 
\frac{\det\wh \Psi(0)}{\det \Psi(0)} 
\frac{\det  \Psi(\infty)}{\det \wh\Psi( \infty)} 
\ee
where $\det \Psi = \det (\Psi_i^j)$ where $\Psi_i$ are the 3 solutions and $\Psi_i^j$ are the components of those solutions. 
Here and in the following, 
when we write $\Psi(0)$ or $\Psi(\infty)$ we mean the leading behavior as $r\to0$ or $r\to \infty$ respectively. 

The functions of $\wh\Psi$ are easy to find. They are
\be
\wh\Psi_1 = 
\begin{pmatrix}
s\, r^{s-1} \\
\sqrt{s(s+2)} \, r^{s-1}\\
0
\end{pmatrix},
\quad
\wh\Psi_2 = 
\begin{pmatrix}
\sqrt{s(s+2)}(s-s\xi - 2 \xi)\, r^{s+1} \\
(s^2 + 4s - 2 s \xi - s^2 \xi)\, r^{s+1}\\
0
\end{pmatrix},
\quad
\wh\Psi_3 = 
\begin{pmatrix}
0 \\
0 \\
r^s
\end{pmatrix},
\ee
So
\be
\det \widehat\Psi(r)= 2s (s + s\xi+2\xi) r^{3s} 
\ee
Note that the $\xi$ dependence of $\det \widehat\Psi$
 is only in the normalization, so it will drop out in the ratio of the determinant at $r=0$ and $r=\infty$. 

To find the solutions $\Psi$, as discussed in~\cite{Endo:2017tsz}, an immensely useful observation is that they can be expressed in terms of 3 auxiliary functions $\eta, \chi$ and $\zeta$ as
\be
\Psi = 
\begin{pmatrix}
\partial_r \chi + \frac{1}{r g^2 \phib^2} \eta - 2 \frac{\phib'}{g^2 \phib^3} \zeta \\
\frac{\sqrt{s(s+2)}}{r} \chi + \frac{1}{\sqrt{s(s+2)}r^2g^2 \phib^2} \partial_r(r^2 \eta)\\
g \phib \chi + \frac{1}{g \phib} \zeta
\end{pmatrix}
\label{Psiofeta}
\ee
where the auxiliary functions satisfy
\begin{align}
\label{chieq}
&\Delta_s \chi - \frac{2 \phib'}{r g^2 \phib^3}\eta - \frac{2}{r^3} \partial_r\left(\frac{r^3 \phib'}{g^2 \phib^3} \zeta  \right) + \xi   \zeta = 0\\
\label{etaeq}
&(\Delta_s - g^2 \phib^2)\eta - \frac{2\phib'}{r^2 \phib} \partial_r(r^2 \eta) + \frac{2 s(s+2)\phib'}{r \phib}\zeta =0\\
&\Delta_s \zeta =0
\label{zetaeq}
\end{align}
We define $\Psi_1$ as the solution with $\zeta=\eta=0$, $\Psi_2$ is the solution with $\zeta=0$ and $\eta \ne 0$ and $\Psi_3$ as the solution with $\zeta \ne 0$. Note that only $\Psi_3$ can be gauge-dependent. 

The exact form of $\Psi_1$ is easy to find. With $\zeta=\eta=0$ we find $\chi = r^s$ and so
\be
\Psi_1(r) = 
\begin{pmatrix}
s \,r^{s-1} \\
\sqrt{s(s+2)} \,r^{s-1} \\
g \phib \, r^s
\end{pmatrix},
\ee

For $\Psi_2$ which has $\zeta=0$ but $\eta \ne 0$, we can solve Eq.~\eqref{etaeq} exactly. We find that the non-zero solution regular at $r=0$ is
\be
\eta_2(r) = r^s \left(r^2+R^2\right)^\frac{\kappag-1}{2}
 {}_2F_1\left(\frac{1+\kappag}{2},s+ \frac{3+\kappag}{2},2+s,-\frac{r^2}{R^2}\right)
\ee
with $\kappag$ in Eq.~\eqref{kappagdef}.  At small and large $r$
\be
\eta_2(0) \sim r^s R^{\kappag-1}\left(1 - \frac{r^2}{R^2} \frac{7+4s +\kappag^2}{4(s+2)}\right),
\qquad
\qquad
\eta_2(\infty) \sim
C_\eta r^{s-2} R^{\kappag+1}
\ee
where
\be
C_\eta = \frac
{\Gamma(1+s) \Gamma(2+s)}
{\Gamma(s+ \frac{3}{2}  - \frac{\kappag}{2})\Gamma(s+ \frac{3}{2}+\frac{\kappag}{2})}
\ee
Note the reappearance of the ratio in Eq.~\eqref{RsT}.

Now given $\eta_2$, we can solve for $\chi$ using Eq.~\eqref{chieq}. 
 Conveniently, we do not need the full solution for all $r$, only its small $r$ and large $r$ behavior. Eq.~\eqref{chieq} simplifies in these limits and we find
 \be
\chi_2(0) \sim  \frac{\lambda}{8 g^2}\frac{R^{\kappag-1}}{s+2} r^{s+2},\qquad
\chi_2(\infty) \sim  C_\eta
\frac{\lambda}{8 g^2}
\frac{R^{\kappag-1}}{s+2} r^{s+2}
\ee
To these we could add a homogeneous solution of the form $\chi = r^s$. However, this is exactly the $\Psi_1$ solution which is orthogonal, so adding a $\Psi_1$ component to $\Psi_2$ will not affect the functional determinant. Dropping the homogeneous solutions is extremely important -- it is the essential simplification allowed by using these auxiliary functions.
Using the limiting forms of $\eta_2$ and  $\chi_2$, following the procedure outlined in~\cite{Endo:2017tsz}, we find
\be
\Psi_2(0) \sim
- \frac{\lambda}{8g^2} R^{\kappag+1} r^{s-1}
\begin{pmatrix}
1- \frac{r^2}{R^2} \frac{g^2}{\lambda}\frac{2}{s+2}
\\
\sqrt{\frac{s+2}{s}}\left[1- \frac{r^2}{R^2} \frac{g^2}{\lambda}\frac{2(s+4)}{(s+2)^2}\right]
\\
0
\end{pmatrix},
\quad
\Psi_2( \infty) \sim 
C_\eta
R^{\kappag+1} r^{s-1} \begin{pmatrix}
\frac{1}{4(s+2)}
\\
\sqrt{\frac{s+2}{s}}\frac{s+4}{4(s+2)^2}
\\
-\sqrt{\frac{-\lambda}{g^2}} \frac{r}{R}\frac{1}{s+2}
\end{pmatrix}
\ee
Here we have written only the terms that contribute at leading non-vanishing order to the determinant.

For $\Psi_3$, defined to have $\zeta \ne 0$, we can solve Eq.~\eqref{zetaeq} exactly for $\zeta= r^s$. 
Proceeding as for $\Psi_2$, we find
\be
\Psi_3(0) \sim
-\frac{\lambda}{8g^2} r^{s-1} R^2
\begin{pmatrix}
\frac{s(s+2)\frac{\lambda}{g^2}}{
(s+2)\frac{\lambda}{g^2} + 2} \left[1 - 2\frac{r^2}{R^2} \frac{g^2}{\lambda}
\frac{s-(s+2)\xi - 2\xi \frac{g^2}{\lambda}}{s(s+2)}
\right]
\\
\sqrt{\frac{s}{s+2}} \frac{(s+2)^2}{(s+2)+2\frac{g^2}{\lambda}}\left[
1 - 2\frac{r^2}{R^2} \frac{g^2}{\lambda}\frac{s+4-(s+2)\xi - 2 \frac{g^2}{\lambda}\xi}{(s+2)^2}
\right]
\\
\sqrt{\frac{8g^2}{-\lambda}} \frac{r}{R}
\end{pmatrix}
\ee
and
\be
\Psi_3(\infty) \sim
r^{s+1}
 \begin{pmatrix}
 \frac{s-(s+2)\xi }{4(s+2)}\\
\sqrt{s(s+2)} \frac{4+s-(s+2)\xi}{4(s+2)^2}
\\
0
\end{pmatrix}
\ee

Putting these solutions together we find
\be
\det \Psi (0) =
\sqrt{\frac{-\lambda}{8g^2}}R^\kappag r^{3s}
\frac{s+(s+2)\xi}{2\sqrt{s}(s+2)^{3/2}},
\quad
\det \Psi (\infty) =
C_\eta \sqrt{\frac{-\lambda}{8g^2}}R^\kappag r^{3s}
\frac{\sqrt{s}[s+(s+2)\xi]}{2(s+2)^{5/2}},
\ee
and so
\be
R_s^{SLG} = \frac{\det \cM_s^{SLG}}{\det {\wh\cM}_s^{SLG}} = 
\frac{\det \wh\Psi(0)}{\det \Psi( 0)} 
\frac{\det  \Psi( \infty)}{\det \wh\Psi(\infty)} 
 ={C_\eta} \frac{s}{s+2}
 \ee
 which is manifestly gauge invariant. 
 
 Comparing to Eqs.~\eqref{RsG} and \eqref{RsT} we see that
 \be
R_s^{SLG} = R_s^T R_s^G 
  \ee
Thus the scalar and longitudinal vector modes together contribute the same as a transverse mode to the determinant.  As a check, at $g= 0$, the vector bosons become free $R_s^T =1$, and $R_s^{SLG}=R_s^G$ as expected. 
Combining with the transverse modes the full determinant for $s>0$ is
   \be
 \begin{boxed} {
  R_s^{AG} = (R_s^T)^3 R_s^G,\qquad R_0^{AG'} = 1 
}  \end{boxed}
 \label{Rrel}
  \ee
  where $R_s^{AG}$ means all the gauge boson ($A^\mu$), ghost, and Goldstone boson contributions are included.
We have shown this result to be manifestly gauge invariant ($\xi$-independent in Fermi gauges). We also checked through a numerical implementation of the Gelfand-Yaglom method that the same formula emergences in $R_\xi$ gauges for each $s$.

The full functional determinant requires summing over $s$. We note that at large $s$,
\be
(s+1)^2 \ln R_s^{AG} \sim -2s\left(1+\frac{3g^2}{\lambda}\right) -2\left(1+\frac{3g^2}{\lambda}\right) - \frac{2}{3s} \left(1+ \frac{3g^4}{\lambda^2}\right)
\label{Rsubfull}
\ee
thus there are quadratic, linear and logarithmic divergences in the sum.

\subsubsection*{Renormalization}
To regulate the sum over $s$, we will subtract the divergent terms and add in dimensionally regulated Feynman diagrams, as explained in Section~\ref{sec:gendet}. 
An important cross check on the result is that the UV divergences should cancel those from $-S[\phib]$ using the renormalized coupling.
In scalar QED, the one loop $Z$-factor for $\lambda$ is
\be
Z_\lambda
=1+\frac{1}{16\pi^2}\frac{1}{\eps}\left(10 \lambda_R-6 g_R^2 +3 \frac{g_R^4}{\lambda_R}\right) 
\ee
The action on the bounce then becomes 
\be
S[\phib]  
=-\frac{8\pi^2}{3\lambda_R} +\frac{1}{\eps}\left(\frac{5}{3} -\frac{g_R^2}{\lambda_R}+ \frac{g_R^4}{2\lambda_R^2} \right)
 +\cdots
 \label{Sbd}
\ee
Thus we need the UV divergences in Eq.~\eqref{Sbd} to be matched by the functional determinant over scalar, gauge and Goldstone modes.

To proceed, we want to 
 compute the divergent contributions with Feynman diagrams in $d$ dimensions and subtract the corresponding contribution from the 4D result to sum over $s$. 
 Unfortunately, performing the subtractions in Fermi gauge is difficult. In Fermi gauge, due to the $g \phib A_\mu \partial_\mu G$ term in Eq.~\eqref{LFermi} there is a Feynman rule picking up the momentum of virtual Goldstones. This extra loop momentum generates new UV divergences and makes the diagrams difficult. This is explained in more detail in Appendix~\ref{app:Fermi} where we compute all the divergent parts (but not the finite parts). These divergences exactly correpond to those in Eq.~\eqref{Sbd} as expected.
 
Fortunately, we can compute the regularized contribution in any gauge. Indeed we have checked through a numerical implementation of the Gelfand-Yaglom method
that our result for the Fermi-gauge functional determinant 
$R_s^{AG}$ in Eq.~\eqref{Rrel} is reproduced in $R_\xi$ gauges for any $s$. 
  In $R_\xi$ gauges, for $s=0$, the transverse and
longitudinal model decouple, and we are left with the scalar, Goldstone and ghost fluctuations. The ghost contribution in $R_\xi$ gauges
corresponds to $x=-\frac{g^2}{\lambda}$ (like the transverse modes in Fermi gauges).
At $s=0$ we have checked numerically that this ghost contribution exactly cancels the contribution of the scalar and Goldstone contributions to $R_0^{AG'}$, giving
$R_0^{AG'}=1$ in $R_\xi$ gauges in agreement with our analytic results in Fermi gauges.

In $R_\xi$ gauge with $\xi=1$ with Lagrangian in Eq.~\eqref{LRxi}, the Feynman rules are 
\be
\begin{gathered}
\includegraphics{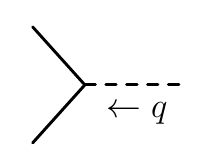}
\end{gathered}
=-(g^2+\lambda)\widetilde{\phib^2}(q),
~~
\begin{gathered}
\includegraphics{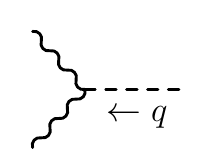}
\end{gathered}
=-g^2 \widetilde{\phib^2}(q) \delta_{\mu\nu},
~~
\begin{gathered}
\includegraphics{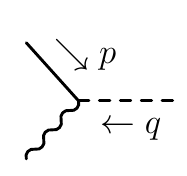}
\end{gathered}
=2i gq^\mu \widetilde{\phib}(q)
\ee
In addition, in $R_\xi$ gauge the ghosts do not decouple and have an interaction
\be
\begin{gathered}
\includegraphics{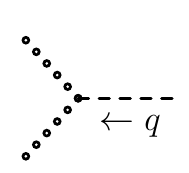}
\end{gathered}
=- g^2\widetilde{\phib^2}(q),
\ee
Then we find
\begin{align}
-S_{GG}^{R_\xi}&= 
\begin{gathered}
\includegraphics{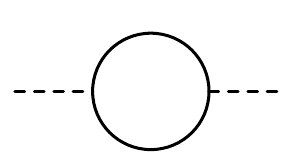}
\end{gathered}
= \left(\frac{g^2+\lambda}{\lambda}\right)^2\left[ \frac{1}{6\eps}  + \frac{5}{18} + \frac{1}{6} \gamma_E + \frac{1}{6}\ln \pi R^2\mu^2\right]
\\
-S_{AA}^{R_\xi} &= 
\begin{gathered}
\includegraphics{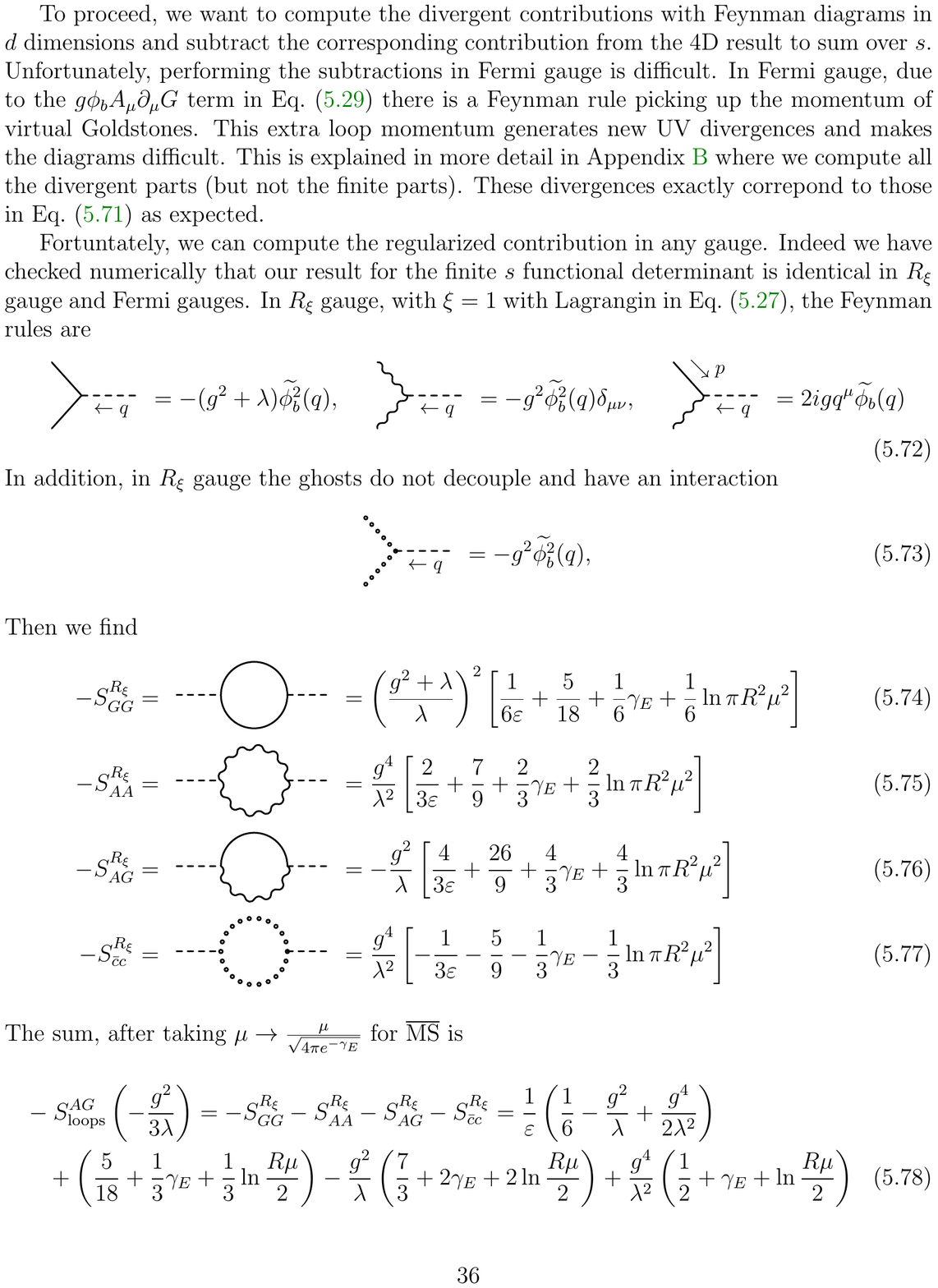}
\end{gathered}
= \frac{g^4}{\lambda^2} \left[ \frac{2}{3\eps}  + \frac{7}{9} + \frac{2}{3} \gamma_E + \frac{2}{3}\ln \pi R^2\mu^2 \right]
\\
-S_{AG}^{R_\xi} &= 
\begin{gathered}
\includegraphics{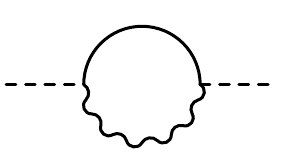}
\end{gathered}= 
-\frac{g^2}{\lambda} \left[ \frac{4}{3\eps}  + \frac{26}{9} + \frac{4}{3} \gamma_E + \frac{4}{3}\ln \pi R^2\mu^2 \right]
\\
-S_{\bar{c}c}^{R_\xi} &= 
\begin{gathered}
\includegraphics{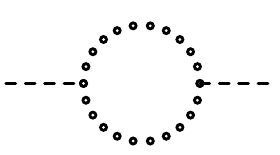}
\end{gathered}= \frac{g^4}{\lambda^2} \left[ -\frac{1}{3\eps}  - \frac{5}{9} - \frac{1}{3} \gamma_E - \frac{1}{3}\ln \pi R^2\mu^2 \right]
 \end{align} 
 The sum, after taking $\mu \to \frac{\mu}{\sqrt{4 \pi e^{-\gamma_E}}}$ for $\msbar$ is
 \begin{multline}
 -S_\text{loops}^{AG}\left(-\frac{g^2}{3\lambda}\right) =
- S_{GG}^{R_\xi} - S_{AA}^{R_\xi} - S_{AG}^{R_\xi} - S_{\bar{c}c}^{R_\xi} 
 = \frac{1}{\eps} \left( \frac{1}{6} -\frac{g^2}{\lambda}+ \frac{g^4}{2\lambda^2} \right)\\
+\left(\frac{5}{18} + \frac{1}{3} \gamma_E +\frac{1}{3}\ln \frac{R \mu}{2}\right)
-\frac{g^2}{\lambda}\left(\frac{7}{3} +  2\gamma_E +2\ln \frac{R \mu}{2}\right)
+\frac{g^4}{\lambda^2}\left(\frac{1}{2} +  \gamma_E +\ln \frac{R \mu}{2}\right)
\label{loopsRxi}
\end{multline}
 Note that the divergent terms agree with those in Fermi gauge, Eq.~\eqref{divsfermi},
 and when the scalar contribution is added (with divergence $\frac{3}{2\eps}$), the poles exactly cancel those in Eq.~\eqref{Sbd}. 
 
To perform the subtraction, we need to compute the contribution to the functional determinants in 4D from terms to second order in the couplings. 
For $s=0$, since $R_0^{AG'}=1$, there are no second-order contributions (or contributions at any order). 
For $s>0$, we note that in $R_\xi$ gauge the transverse modes have the same quadratic fluctuations as the ghosts and the two contributions exactly cancel. For the other photon polarizations and Goldstones, the fluctuation matrix with $\xi=1$ is
  \be
\cM_s^{SLG,R_\xi} =
 \begin{pmatrix}
-\Delta_s + \frac{3}{r^2}  + g^2 \phib^2& -\frac{ 2\sqrt{s(s+2)}}{r^2} & 2g \phib'  \\
- \frac{2\sqrt{s(s+2)}}{r^2} &  -\Delta_s - \frac{1}{r^2} + g^2 \phib^2 &0  \\ 
2 g \phib' & 0 & -\Delta_s  +( g^2+\lambda) \phib^2
 \end{pmatrix}
\ee
Changing basis, following~\cite{Isidori:2001bm} we find a convenient almost diagonal form
\be
U^{-1} \cM_s^{SLG,R_\xi}  
U
=
 \begin{pmatrix}
-\Delta_{s-1}   + g^2 \phib^2& 0 & g \frac{s}{s+1} \phib' \\
0  &  -\Delta_{s+1}+ g^2 \phib^2 &- g \frac{s+2}{s+1}\phib'  \\ 
2 g \phib' & - 2 g \phib' & -\Delta_s  +( g^2+\lambda) \phib^2
 \end{pmatrix}
 \label{mostlydiagonal}
\ee
where 
\be
U = \begin{pmatrix} 
1 & -1 & 0 \\
\sqrt{\frac{s+2}{s}} & \sqrt{\frac{s}{s+2}} & 0 \\
0  & 0 & 1 
\end{pmatrix}
\ee
In this form, we see that if we turn off the off-diagonal couplings,
each diagonal term is a 1D operator and the exact result  can then be read off, using Eq.~\eqref{RSx}:
\be
R_{s,\,\text{diag}}^{SLG,R_\xi}= R_{s+1}\Big(-\frac{ g^2}{3\lambda}\Big) R_{s-1} \Big(- \frac{g^2}{3\lambda}\Big) 
R_s\Big( -\frac{g^2+\lambda}{3\lambda} \Big)
\ee
The required subtractions to second order in the diagonal interactions then come from the expansions of these function to second order in their arguments:
\begin{multline}
\left[
\ln R_{s,\,\text{diag}}^{SLG,R_\xi} \right]_{\text{sub}}=
\frac{2 \left(s^3+4 s^2+4 s+2\right)}{s^2 (s+1)^2}
 - 4 \psi'(s)\\
+\left[\frac{2 \left(s^4+9 s^3+20 s^2+18 s+8\right)}{s^2 (s+1)^2 (s+2)}
- 8 \psi'(s)\right] \frac{g^2}{\lambda}\\
+\left[
\frac{2 (2 s+5) \left(3 s^4+12 s^3+18 s^2+12 s+4\right)}{s^2 (s+1)^2 (s+2)^2}
-12 \psi'(s)
 \right] \frac{g^4}{\lambda^2}
 \label{Rdiag}
\end{multline}
with $\psi(z)=\frac{\Gamma'(z)}{\Gamma(z)}$ the digamma function.
The remaining required subtractions involve the off diagonal couplings in Eq.~\eqref{mostlydiagonal}. Since these couplings are linear in $g$ there are no contributions to first order in $g$ (corresponding to no diagrams with one $g$ insertion). The contributions to second order in $g$ can then be computed turning all the diagonal interactions (mass term) to zero. We do this with the Gelfand-Yaglom method perturbative in $g$, following a similar
procedure to the one used in Section~\ref{sec:gts}. The result is
\be
\left[\ln R_{s,\,\text{off-diag}}^{SLG,R_\xi}\right]_{\text{sub}} = 
\left[\frac{8 \left(2 s^4+7 s^3+10 s^2+6 s+2\right)}{s^2 (s+1)^3}
- 16 \psi'(s) \right]\frac{g^2}{\lambda} 
\label{Roff}
\ee
In total, we get that 
\be
\left[
\ln R_{s}^{SLG,R_\xi} \right]_{\text{sub}} = \left[
\ln R_{s,\,\text{diag}}^{SLG,R_\xi} \right]_{\text{sub}}+\left[\ln R_{s,\,\text{off-diag}}^{SLG,R_\xi}\right]_{\text{sub}}
\label{eq:Rsubtot}
\ee
by adding Eqs.~\eqref{Rdiag} and~\eqref{Roff}. As a check we multiply Eq.~\eqref{eq:Rsubtot} by $(1+s)^2$ and expanding at large $s$, all of the divergent terms of the all-orders result $(1+s)^2\ln R_s^{AG}$, as shown in Eq.~\eqref{Rsubfull}, are reproduced exactly. 

Then, using Eq.~\eqref{Dsubs}  we compute
\begin{align}
S^{AG}_{\text{diff}}\left(-\frac{g^2}{3\lambda}\right)&=
\sum_{s=1}^\infty \left\{ (s+1)^2
\left[
\ln R_{s}^{SLG,R_\xi} \right]_{\text{sub}} \!\!\!\!\!\!\!- 3\Ssub^s\left(-\frac{g^2}{3\lambda}\right) - \Ssub^s\left(-\frac{1}{3}\right)
\right\}\\
&=\left(4\pi^2-\frac{121}{3} \right) \frac{g^2}{\lambda} -\frac{5 g^4}{2\lambda^2} 
\label{remainder}
\end{align}

Adding Eq.~\eqref{loopsRxi}, Eq.~\eqref{remainder} and the finite parts Eq.~\eqref{Dfinis}, we finally have
\be
\begin{boxed}{
\sqrt{ \frac{\det \wh \cO_{AG}}{\det' \cO_{AG}}} =
\exp\left[
-\frac{3}{2} \Sfin^+\left(-\frac{g^2}{3\lambda}\right)
-\frac{1}{2} \Sfin^{G+}-  S_\text{loops}^{AG}\left(-\frac{g^2}{3\lambda}\right)
+ \frac{1}{2} S^{AG}_\text{diff} \left(-\frac{g^2}{3\lambda}\right)\right]
}\end{boxed}
\label{withloops}
\ee
with $\Sfin^{+}(x)$ in Eq.~\eqref{Splus} 
and $\Sfin^{G+}=\Sfin^+\left(-\frac{1}{3}\right)$ 
in Eq. \eqref{SfinGp1}.

Getting to this formula involved a large number of steps, so let us summarize the journey. We first worked in Fermi gauges, where we derived the analytic form, Eq.~\eqref{Rrel},
for the functional determinant including gauge boson, ghost and Goldstone boson contributions:
$R_s^{AG} = (R_s^T)^3 R_s^G$ for $s>0$ and $R_0^{AG'} = 1$ for $s=0$. At fixed $s$ the determinant is finite, but there is a UV divergence when summing over $s$ at large $s$.
Subtracting the leading large $s$ behavior lets us do the sum, giving $\Sfin$.  We regulate the UV divergences in $d$ dimensions and compute the loops and counterterms in $\msbar$ using Feynman gauge, giving $S_\text{loops}$. The loops give the the correct large $s$ behavior, but have a different finite part from what we subtracted to get $\Sfin$, so we have
to add back in the difference between the loops and our subtraction. This is $S_\text{diff}$. Finally, there are IR divergences form the zero modes at $s=0$. We remove these from the functional determinant, finding $R_0^{AG'}=1$. Since the $s=0$ contribution is 1, there is no contribution at $s=0$ to the UV subtractions we did. Thus we simply have to remove the $s=0$ part from all of our finite sums. This changes $\Sfin$ to $\Sfin^+$ giving our final answer.

%
 
\subsection{Fermions}
Next, let us consider the addition of Dirac fermions. The Euclidean Lagrangian for a real scalar interacting with a Dirac fermion is
\be
\cL = \frac{1}{2}(\partial_\mu \phi)^2  + \frac{1}{4}\lambda \phi^4 + \bar{\psi} \slashed\partial \psi + \frac{y}{\sqrt{2}} \phi \bar{\psi} \psi
\ee
Around the bounce configuration, $\phi=\phib$, the fermion fluctuation operator is
\be
\cM_\psi = \slashed \partial  + \frac{y}{\sqrt{2}} \phib
\ee
To calculate the determinant of this operator, we expand in a basis of half-integer spin spherical harmonics.

 Including angular momentum, Dirac spinors transform in the direct sum of 
 $(k+\frac{1}{2},k)$ and $(k,k+\frac{1}{2})$ representations of the Lorentz algebra $\text{su}(2)\otimes \text{su}(2)\cong o(4)$.
 In a particular representation of the Euclidean Dirac algebra, the
 half-integer spherical harmonics take the form of hypergeometric functions (see Appendix A of~\cite{Avan:1985eg}).
  Expanding in this basis, $\cM_\psi$ reduces to a form
 which  depends only on the radial coordinate $r$ and only on two components of the Dirac spinor
 \be
 \cM_\psi^{k+} = \begin{pmatrix}
 \partial_r - \frac{2k}{r} & \frac{y}{\sqrt{2}} \phi_b \\
 \frac{y}{\sqrt{2}} \phi_b & \partial_r + \frac{2k+3}{r} 
 \end{pmatrix},
 \qquad
 \cM_\psi^{k-} = \begin{pmatrix}
 \partial_r + \frac{2k+3}{r} & \frac{y}{\sqrt{2}} \phi_b \\
 \frac{y}{\sqrt{2}} \phi_b & \partial_r - \frac{2k}{r} 
 \end{pmatrix}
 \ee
To match the literature, the first matrix here corresponds to Eq. (3.17) of~\cite{Avan:1985eg} with $K\to k$ and the second to (3.18) with $L=K-\frac{1}{2}\to k$.
The multiplicity of the $(a,b)$ representation is $(2a+1)(2b+1)$, so we have
\be
\det \cM_\psi = \prod_{k= 0,\frac{1}{2},1,\cdots}( \det \cM_\psi^{k-} \cM_\psi^{k+})^{(2k+1)(2k+2)}
\ee

Next, we  reduce the product of these operators to a quadratic form by conjugating with the unitary matrices
$U = \text{diag}(\frac{1}{r^{3/2}},\frac{1}{r^{3/2}})$
and
 $V = \text{diag}(\frac{1}{r^{3/2}},-\frac{1}{r^{3/2}})$:
\be
  U^{-1} \cM_\psi^{k-} U V^{-1} \cM_\psi^{k+}V 
 =\begin{pmatrix}
 \partial_r^2 - \frac{(4k+3)(4k+1)}{4r^2}  - \frac{y^2}{2}\phi_b^2 &-\frac{y}{\sqrt{2}} \phi_b' \\
-\frac{y}{\sqrt{2}} \phi_b' & \partial_r^2  - \frac{(4k+3)(4k+5)}{4r^2}- \frac{y^2}{2} \phi_b^2  \\
 \end{pmatrix}
\ee
This simplifies slightly by writing $k = \frac{j}{2} - \frac{3}{4}$. Then $j = 2k + \frac{3}{2} = \frac{3}{2}, \frac{5}{2}, \frac{7}{2}, \cdots$,
the multiplicity becomes $j^2-\frac{1}{4}$ and 
\be
\ln \det \cM_\psi = \sum_{j =  \frac{3}{2}, \frac{5}{2}, \frac{7}{2}, \cdots}\left({j^2 - \frac{1}{4}}\right) \ln \det \cM_{\bar{\psi}\psi}^j
\ee
where
\be
\cM_{\bar{\psi}\psi}^j= \begin{pmatrix}
\partial_r^2 - \frac{j(j-1)}{r^2} - \frac{y^2}{2} \phib^2 &    -  \frac{y}{\sqrt{2}} \phib' \\
- \frac{y}{\sqrt{2}} \phib' &\partial_r^2 - \frac{j(j+1)}{r^2} - \frac{y^2}{2} \phib^2 
 \end{pmatrix}
 \label{Mpsipsi}
\ee
in agreement with~\cite{Isidori:2001bm}. 
The matrix for fluctuations around the false vacuum, $\wh \cM_{\bar{\psi}\psi}^j$ is the same as this one with $\phib = 0$. 

Following similar techniques to those described in previous sections, we find, 
\be
R_j^{\bar{\psi}{\psi}}= \det \frac{ \cM_{\bar{\psi}\psi}^j }{\wh\cM_{\bar{\psi}\psi}^j } = 
\left[
\frac{\Gamma(|j|+\frac{1}{2})^2}{
\Gamma\Big( |j| + \frac{1}{2} + \sqrt{ \frac{y^2}{\lambda}}\Big) \Gamma\Big( |j| + \frac{1}{2} - \sqrt{ \frac{y^2}{\lambda}}\Big)}
\right]^2
\ee
For the subtractions, it is helpful to have the result for the determinant when the off-diagonal terms in $\cM_{\bar{\psi}\psi}^j$ are set to zero. When the matrix is diagonal, the fermion case is a special case of the general formula in Section~\ref{sec:gendet}. The result is
\be
R^{\bar{\psi}\psi}_{j,\text{diag}} = R_{s=j- \frac{3}{2} } \Big( -\frac{y_t^2}{6\lambda}\Big) R_{s=j-\frac{1}{2}} \Big(- \frac{y_t^2}{6\lambda}\Big)
\ee
with $R_s(x)$ given in Eq.~\eqref{RSx}.

The subtractions required to sum over $j$ are given by the expansions of $R_j^{\bar{\psi}{\psi}}$ to second order in the diagonal couplings and second order in the off-diagonal couplings:
\begin{align}
\Ssub^{\bar{\psi}\psi} &= \Big(j^2-\frac{1}{4}\Big)\left[ \ln R_j^{\bar{\psi}{\psi}}\right]_{y^2}  + \Big(j^2-\frac{1}{4}\Big)\left[ \ln R_{j,\text{diag}}^{\bar{\psi}{\psi}}\right]_{y^4}
\\
&= \Big(j^2 - \frac{1}{4}\Big) \left[ -\frac{2y^2}{\lambda}\psi'\Big(|j|+\frac{1}{2}\Big) 
+ \frac{8y^4}{\lambda^2}\frac{(j+1)(4j^2+1)}{(4j^2-1)^2}
- \frac{2y^4}{\lambda^2} \psi'\Big(|j|+ \frac{1}{2}\Big)\right]
\end{align}
Using this, we find
\be
\sum_{j=\pm \frac{3}{2}, \pm \frac{5}{2},\cdots}^{\infty}
\left[\Big(j^2-\frac{1}{4}\Big) \ln R_j^{\bar{\psi}{\psi}} - \Ssub^{\bar{\psi}\psi} \right] =  \Sfin^{\bar{\psi}\psi} \left(\sqrt{\frac{y^2}{\lambda}}\right)
\ee
where
\begin{multline}
\Sfin^{\bar{\psi}\psi}(z)=
16 \psi^{(-4)}(2) - \frac{8}{3} \psi^{(-2)}(2) +
\frac{4z^2}{3} \Big(1 -\gamma_E\Big)
- \frac{z^4}{3} \Big( 1  - 2 \gamma_E \Big)\\
-\frac{4z}{3}\Big(1-z^2\Big) 
\left[  \psi ^{(-1)}\Big(2+ z\Big)-  \psi ^{(-1)}\Big(2- z\Big)\right]
%
+\frac{4}{3}\Big(1 -3 z^2 \Big)
\left[\psi ^{(-2)}\Big(2+z\Big)  +  \psi ^{(-2)}\Big(2 - z\Big)\right]
   \\
   +8z \left[\psi ^{(-3)}\Big(2+  z\Big)- \psi^{(-3)}\Big(2- z\Big)\right]
 %
   -8\left[ \psi^{(-4)}\Big(2+ z\Big)+\psi ^{(-4)}\Big(2- z\Big)\right]
   \label{Sfinz}
\end{multline}
This function is real for imaginary $z$ and contains only even powers of $z$ when expanded around $z=0$.

The UV divergent part is added back in through a dimensionally regulated calculation quadratic in the interactions. 
We do this by evaluating the functional determinant as in \cite{Isidori:2001bm}:
\begin{align}
-S_{\bar{\psi}\psi}^\text{loops}\left(\frac{y^2}{\lambda}\right) &=\frac{1}{2}\left[\ln \frac{\det \cM_{\bar{\psi}\psi}^j }{\det	\wh\cM_{\bar{\psi}}}\right]_{W^2}=\frac{1}{2}\left[\ln \frac{\det(-\Box+W) }{\det(-\Box)}\right]_{W^2}\\
&=-\frac{1}{2}\Tr\left[\Box^{-1}W\right]-\frac{1}{4}\Tr\left[\Box^{-1}W\Box^{-1}W\right]
\end{align}
where $W=\frac{y^2}{2} \phib^2 + \frac{y}{\sqrt{2}} ({\slashed \partial} \phib)$ 
are the interactions from  Eq.~\eqref{Mpsipsi}, and the subscript $W^2$ indicates that we are truncating the expansion in $W$ to second order.
The traces can be rewritten in momentum space
\begin{align}
\Tr\left[\Box^{-1}W\right]&=\int\frac{d^dk}{(2\pi)^d}\frac{-1}{k^2}\int d^4x\Tr [W(x)]\\
\Tr\left[\Box^{-1}W\Box^{-1}W\right]&=\int\frac{d^dq}{(2\pi)^d}\frac{d^dk}{(2\pi)^d}\frac{1}{k^2(k+q)^2}\Tr\left[\widetilde{W}(q)\widetilde{W}(-q)\right]
\end{align} 
The single $W$-trace is zero in dimensional regularization, and we can evaluate the other one using $\widetilde{W}(q)=\frac{y^2}{2}\widetilde{\phib^2}(q)+i\frac{y}{\sqrt{2}} {\slashed q}\widetilde{\phib}(q)$ 
\begin{align}
\Tr\left[\Box^{-1}W\Box^{-1}W\right]&=\int\frac{d^dq}{(2\pi)^d}\frac{d^dk}{(2\pi)^d}\frac{1}{k^2(k+q)^2}\left[y^4\widetilde{\phib^2}(q)+2y^2 q^2\widetilde{\phib}(q)^2\right]\\
&=\int\frac{d^4q}{(2\pi)^4}\left[y^4\widetilde{\phib^2}(q)+2y^2 q^2\widetilde{\phib}(q)^2\right]B_0(q)
\end{align}
with $B_0(q)$ defined in Eq.~\eqref{B0}.
In $\msbar$ the result is
\be
-S_{\bar{\psi}\psi}^\text{loops}\left(\frac{y^2}{\lambda}\right)  
= \frac{1}{\eps} \left( \frac{y^2}{3\lambda} - \frac{y^4}{6\lambda^2}\right)
+\frac{y^2}{\lambda}\left( \frac{13}{18} + \frac{2}{3}\gamma_E + \frac{2}{3}\ln \frac{R\mu}{2}\right)
- \frac{y^4}{\lambda^2}\left(  \frac{5}{18} +  \frac{1}{3}\gamma_E + \frac{1}{3}\ln \frac{R \mu}{2}\right)
\label{Sloopspsi}
\ee

The final result for a single fermion is then
\be
\sqrt{ \frac{\det \cM_{\bar{\psi}\psi}}{\det  \wh \cM_{\bar{\psi}\psi}}}
= \exp\left[\frac{1}{2}\Sfin^{\bar{\psi}\psi} \left(\sqrt{\frac{y^2}{\lambda}}\right)
 -S_{\bar{\psi}\psi}^\text{loops}\left(\frac{y^2}{\lambda}\right)  
\right] 
\label{detpsifinal}
\ee

As a check, we note that the action on the bounce in this Yukawa theory in terms of the renormalized couplings is
\be
S[\phib]  
=-\frac{8\pi^2}{3\lambda_R} + \frac{1}{\eps}\left(\frac{3}{2} +\frac{y_R^2}{3\lambda_R}- \frac{y_R^4}{6\lambda_R^2} \right)
 +\cdots
 \label{Sbdpsi}
\ee
The UV divergences in this action exactly cancel those in Eq.~\eqref{detpsifinal} and Eq.~\eqref{imform}. 

\subsection{Summary of Results \label{sec:summ}}
To provide a convenient reference, we summarize here the main results of this section: the functional determinants for scalars, vectors and fermions. 

For a real scalar, there are zero modes corresponding to dilatations and translations, with Jacobian factors given in Eqs.~\eqref{Jd} and~\eqref{JT}:
\be
J_d = \frac{1}{ R} \sqrt{\frac{6S[\phib]}{5 \pi}},
\qquad
J_T 
= \frac{1}{ R} \sqrt{\frac{6 S[\phib]}{5 \pi}}
\ee
The fluctuation determinant with zero modes removed is in Eq.~\eqref{imform}:
\be
\im \sqrt{\frac{\det \wh \cO_\phi}{\det^\prime \cO_\phi} }  =
 \frac{25}{36} \sqrt{\frac{5}{6}} 
\exp \left[ \frac{3}{2\eps}- \frac{5}{4}  + 6 \zeta'(-1) +3  \ln \frac{R\mu}{2} \right]
\label{realscalarsumm}
\ee
In collective coordinates, we must integrate over $d^4 x\, d R$.

For a complex scalar field, there is a global $U(1)$ invariance spontaneously broken by the bounce. 
There is a zero mode corresponding to phase rotations. The Jacobian for changing to collective coordinates is given in Eq.~\eqref{JG}
\be
J_G = \sqrt{\frac{6S[\phib]}{\pi}}
\ee
and the fluctuation operator for the Goldstone bosons is  in Eq.~\eqref{detGfinal}
\be
\sqrt{\frac{\det \wh \cO_G}{\det^\prime \cO_G}}
=
\sqrt{\frac{3}{2}} 
\exp \left[ \frac{1}{6\eps}  -\frac{1}{12} + 2  \zeta'(-1) +\frac{1}{3}  \ln \frac{R\mu}{2} \right]
\ee
In collective coordinates this must be integrated over the volume $V=2\pi$ of $U(1)$. 

For a $U(1)$ gauge theory with a complex scalar, namely the Coleman-Weinberg model, the dilatation, translation and phase rotation modes are still present.  The functional determinant over gauge and Goldstone fluctutations with zero modes removed is in Eq.~\eqref{withloops}
\be
\sqrt{ \frac{\det \wh \cO_{AG}}{\det' \cO_{AG}}} =
\exp\left[
-\frac{3}{2} \Sfin^+\left(-\frac{g^2}{3\lambda}\right)
-\frac{1}{2} \Sfin^{G+}-  S_\text{loops}^{AG}\left(-\frac{g^2}{3\lambda}\right)
+ \frac{1}{2} S^{AG}_\text{diff} \left(-\frac{g^2}{3\lambda}\right)\right]
\label{AGsumm}
\ee
With $\Sfin^+(x)$ in Eq.~\eqref{Splus}, $\Sfin^{G+}$ in Eq.~\eqref{SfinGp1}, $S_\text{diff}^{AG}$ in Eq.~\eqref{remainder} and $S_\text{loops}^{AG}$ in Eq.~\eqref{loopsRxi}. The determinant over the real scalar fluctuations in this theory is the same as in Eq.~\eqref{realscalarsumm}. 

For a Dirac fermion, the fluctuation determinant is in Eq.~\eqref{detpsifinal}:
\be
\sqrt{ \frac{\det  \cM_{\bar{\psi}\psi}}{\det \wh \cM_{\bar{\psi}\psi}}}
= \exp\left[ \frac{1}{2}\Sfin^{\bar{\psi}\psi} \left(\sqrt{\frac{y^2}{\lambda}}\right)
 -S^{\bar{\psi}\psi }_{\text{loops}}\left(\frac{y^2}{\lambda}\right)
\right] 
\label{DiracFsumm}
\ee
with $\Sfin^{\bar{\psi}\psi}(z)$ in Eq.~\eqref{Sfinz} and $S^{\bar{\psi}\psi }_{\text{loops}}$ in Eq.~\eqref{Sloopspsi}. If the fermion is colored, then we get $N_c$ copes of Eq.~\eqref{DiracFsumm}.

\section{Vacuum Stability in the Standard Model \label{sec:SM}} 
Now we have all the ingredients necessary to compute the next-to-leading order decay rate in the Standard Model. The relevant part of
the Standard Model Lagrangian is
\begin{multline}
\cL_{\text{SM}} =  ( D_\mu H)^\dagger (D_\mu H) + \lambda (H^\dagger H)^2- \frac{1}{4} (W_{\mu\nu}^a)^2 - \frac{1}{4} B_{\mu\nu}^2 \\
+i \bar{Q} \slashed D Q + i\bar{t}_R \slashed D t_R+ i\bar{b}_R \slashed D b_R
- y_t \bar{Q} H t_R - y_t^\star \bar{t}_R H^\dagger Q 
- y_b \bar{Q} \widetilde H b_R - y_b^\star \bar{b}_R {\widetilde H}^\dagger Q + \cdots
\end{multline}
where $H$ is the Higgs doublet, $\widetilde H = i \sigma_2 H$,  $W_\mu$ are the $SU(2)$ gauge bosons, $B_\mu$ is the hypercharge gauge boson, $Q$ is the 3rd generation left-handed quark doublet, and $t_R$ and $b_R$ are the right-handed top  and bottom quarks. Contributions from other fermions are negligible and gluons have no effect at next-to-leading order. 
We have set the Higgs mass parameter $m^2$ to zero; $m^2 \ne 0$ corrections will be discussed in Section~\ref{sec:mass}. 

From this Lagrangian we see that there are only five parameters relevant to the NLO decay rate: $\lambda, y_t, y_b$ and the $SU(2)\times U(1)$ couplings $g$ and $g'$. All these parameters depend on scale. As explained in Section~\ref{sec:siprob}, for a consistent power counting the tunneling calculation has to be done near the scale $\mu^\star$ where $\beta_\lambda(\mu^\star) =0$. In the SM this scale is $\mu^\star \approx 10^{17} \GeV$. The five parameters are determined at much lower scales, $\mu \sim 100 ~\GeV$ (or $\mu\sim m_b$ for $y_b$). In determining the parameters matching conversions (also known as threshold corrections) are made from a physical scheme (like the pole-mass scheme where the $W$ and $Z$ masses are measured). The ingredients for this matching step are known at NNLO and depend on additional SM parameters, such as $\alpha_s$. After matching, one must run the couplings up to
$\mu \sim 10^{17}~\GeV$. The RG equations for this running are known at three or four loops  and involve additional parameters like $\alpha_s$ as well. 
To be clear, the goal of the matching and running is to get $\lambda(\mu^\star), y_t(\mu^\star), y_b(\mu^\star), g(\mu^\star)$ and $g'(\mu^\star)$ in $\msbar$. 
Thus, it is perfectly consistent to match at NNLO, run at three or four loops and compute the decay rate at NLO -- each step is a separate well-defined calculation.

\subsection{NLO Tunneling Rate Formula}
To compute the NLO tunneling rate in the SM we need to combine the formulas from Section~\ref{sec:funcdet}. 

The bounce spontaneously breaks translation and scale invariance as well as $SU(2)\times U(1)_Y \to U(1)_\text{EM}$. 
The zero modes for translations and dilatations must be integrated over with collective coordinates with appropriate Jacobian factors.
The three broken internal generators produce three zero modes which must be integrated over the volume of the broken gauge group. 
As we work to NLO only, we only need the action quadratic in the fluctuations around the bounce.
For gauge bosons, this means the non-Abelian interactions are irrelevant and each gauge boson can be treated independently. 
Thus, each gauge group collective coordinate produces a factor of $J_G$ in Eq.~\eqref{JG} as for a $U(1)$. 
In addition, since the $U(1)$ representing electromagnetism is unbroken, fluctutations of the photon are the same around the false vacuum and the bounce and therefore do not contribute to the rate. 
We can therefore compute the gauge-boson fluctuations by integrating over $W^\pm$ and $Z$ boson fluctuations and their associated
Goldstone bosons.

Resolving the integral over instanton size $R$ through the technique described in Section~\ref{sec:siprob}
and using Eq.~\eqref{Rint},  the NLO rate formula in the SM
is therefore
\begin{multline}
\frac{\Gamma}{ V} =
e^{-S[\phib]} 
\im 
\,V_{SU(2)}
 J_G^3
 (R J_T)^4 (R J_d)
 \sqrt{\frac{\det \wh \cO_h}{\det^\prime \cO_h}}
\sqrt{\frac{\det \wh \cO_{ZG}}{\det^\prime \cO_{ZG}}}
{\frac{\det \wh \cO_{WG}}{\det^\prime \cO_{WG}}}
\sqrt{ \frac{\det  \cO_{\bar{t}t}}{\det \wh \cO_{\bar{t}t}}}
\sqrt{ \frac{\det  \cO_{\bar{b}b}}{\det \wh \cO_{\bar{b}b}}}
\\
\times\mstar^4 \sqrt{-\frac{\pi \Sbs \lstar}{  \bps}}  e^{-\frac{4\lstar}{ \Sbs \bps}}
 \left[ \frac{\lstar}{\lol(\hat{\mu})}-1 - \frac{4 \lambda_\star}{\Sbs ^2\bps} \right]
 \label{bigSM}
\end{multline}
This formula is valid for $R^{-1}=\mu = \mu^\star$, with $\mstar$ the scale where $\beta_\lambda (\mstar)=0$. 
For other values of $\mu$, there are additional factors of $\beta_\lambda$ not shown, as in Eq.~\eqref{Gamma2mu}. The scale $\hat{\mu}$ is in Eq.~\eqref{muhatdef}.

The Jacobian factors for dilatations and translations are in Eqs.~\eqref{Jd} and~\eqref{JT}:
\be
( J_d)(J_T)^4  =   \frac{1}{R^5} \left(\frac{6S[\phib]}{5\pi}\right)^{5/2}
\ee
With $SU(2)$, the group theory volume factor is\footnote{
$SU(2)$ are matrices $\begin{pmatrix} a & b \\ - b^\star & a^\star \end{pmatrix}$ with $|a|^2 + |b|^2=1$, thus the volume is that of the 4-sphere, $2\pi^2$. 
}
\be
\frac{ V_{SU(2)\times U(1)}}{V_{U(1)}} = V_{SU(2)} =\int d\Omega_{SU(2)} =2\pi^2 
\ee
Thus 
\be
\int d\Omega_{SU(2)} J_G^3 = 2\pi^2 \left( \frac{6 S[\phib]}{\pi}\right)^{3/2}
\ee

For the real Higgs scalar, the determinant with zero modes removed  is in Eq.~\eqref{realscalarsumm}. Setting $R=\frac{1}{\mu}$ gives
\be
\im \sqrt{\frac{\det \wh \cO_h}{\det^\prime \cO_h} }  = \frac{25}{288} \sqrt{\frac{5}{6}} 
\exp \left[ \frac{3}{2\eps}- \frac{5}{4}  + 6 \zeta'(-1)\right]
\ee
For gauge bosons, 
we note that the $W$ and $Z$ bosons couple to $\phib$ with strengths $g_W=2\frac{m_W}{v}=g$ and $g_Z =2\frac{m_Z}{v}= \sqrt{g^2 + (g')^2}$ respectively. 
Including the conventional factor of $\frac{1}{2}$ normalizing Abelian versus non-Abelian generators, the gauge bosons and Goldstone fluctuations give the result summarized in Eq.~\eqref{AGsumm}
\begin{align}
\sqrt{ \frac{\det \wh \cO_{ZG}}{\det' \cO_{ZG}}}
 &=
\exp\left[
-\frac{3}{2} \Sfin^+\left(-\frac{g_Z^2}{12\lambda}\right) -\frac{1}{2} \Sfin^{G+}
+ \frac{1}{2} S_\text{diff}^{AG}\left(-\frac{g_Z^2}{12\lambda}\right) -  S_\text{loops}^{AG} \left(-\frac{g_Z^2}{12\lambda}\right)
\right]\\
\frac{\det \wh \cO_{WG}}{\det' \cO_{WG}} &=  \exp\left[
-3 \Sfin^+\left(-\frac{g_W^2}{12\lambda}\right)- \Sfin^{G+}
+ S_\text{diff}^{AG}\left(-\frac{g_W^2}{12\lambda}\right) - 2 S_\text{loops}^{AG} \left(-\frac{g_W^2}{12\lambda}\right)
\right]
\end{align}
The top quark contributes as in Eq.~\eqref{DiracFsumm} with a factor of $N_C=3$ for color
\be
\sqrt{ \frac{\det  \cM_{\bar{t}t}}{\det \wh \cM_{\bar{t}t}}}
= \exp\left[ \frac{N_C}{2}\Sfin^{\bar{\psi}\psi} \left(\sqrt{\frac{y_t^2}{\lambda}}\right)
 -N_CS^{\bar{\psi}\psi }_{\text{loops}}\left(\frac{y_t^2}{\lambda}\right)
\right] 
\ee
The bottom quark contribution is identical with $y_t \to y_b$ and we omit $y_b$ for simplicity in the next set of formulas. 

The UV divergences from the product of these functional determinants in $4-2\eps$ dimensions is
\begin{align}
 \sqrt{\frac{\det \wh \cO_{hZWt}}{\det^\prime \cO_{hZWt}}}
&= \exp \left[ \frac{1}{\eps}\left(
2-\frac{2g_W^2+g_Z^2}{4\lambda}
+ \frac{2g_W^4+g_Z^4}{32\lambda^2} 
+ \frac{N_C y_t^2}{3\lambda}  - \frac{N_C y_t^4}{6\lambda^2}
\right) + \cO(\eps^0)\right]\\
&=\exp \left[ \frac{1}{\eps}\left(
2-\frac{3g^2+g'{}^2}{4\lambda}+ \frac{3g^4+2g^2g'{}^2+g'{}^4}{32\lambda^2} 
+ \frac{N_C y_t^2}{3\lambda}  - \frac{N_C y_t^4}{6\lambda^2}
\right) + \cO(\eps^0)\right]\
\nonumber
\end{align}
These are exactly canceled by the renormalized tree-level action on the bounce
\be
-S[\phib] = \frac{8\pi^2}{3\lambda}-\left[
\frac{1}{\eps}\left(
2-\frac{3g^2+g'{}^2}{4\lambda}+ \frac{3g^4+2g^2g'{}^2+g'{}^4}{32\lambda^2} 
+ \frac{N_C y_t^2}{3\lambda}  - \frac{N_C y_t^4}{6\lambda^2}
\right)\right]
\label{SSM}
\ee

In the SM, the one loop $\beta$ function for $\lambda$ is
\be
\beta_\lambda^0 =  \frac {d\lambda}{ d \ln \mu} = \frac{1}{16\pi^2} \left(
24 \lambda^2 + \frac{9}{8} g^4 + \frac{3}{8} g'{}^4 + \frac{3}{4} g^2 g'{}^2 - 9 g^2 \lambda -3 g'{}^2 \lambda + 4 N_Cy_t^2 \lambda - 2 N_C y_t^4
\right)
\label{betalamSM}
\ee
This one loop $\beta$ function is of course linearly related to the $\frac{1}{\eps}$ poles in Eq.~\eqref{SSM}. 
The derivative of $\beta_\lambda$, required for Eq.~\eqref{bigSM} is
\begin{multline}
\beta_\lambda^0{}' =\mu \frac{d\beta_\lambda^0}{d \mu} =
\frac{1}{(16\pi^2)^2} \Big[
1152 \,\lambda^3 - 648 \,g^2 \lambda^2 -216 \,g'{}^2 \lambda^2 + 192\, g^4 \lambda
+90 \,g^2 g'{}^2 \lambda -14\,g'{}^4 \lambda
\\
-\frac{195}{8} g^6 - \frac{119}{8} g^4 {g'}^2  + \frac{37 }{8}g^2 g'{}^4 +\frac{73}{8} g'{}^6+ N_C^2 \Big(16\, \lambda  y_t^4 -8\, y_t^6\Big)\\
+N_C \Big(\frac{9 }{2}g^4 y_t^2  + 3\, g^2 g'{}^2 y_t^2 + 36\, g^2 y_t^4 - 90 g^2 \lambda  y_t^2 + \frac{3 }{2}g'{}^4 y_t^2
+ \frac{52 }{3}g'{}^2 y_t^4 \hspace{30mm}\\
 - \frac{106}{3} g'{}^2 \lambda  y_t^2 - 64\, g_s^2 y_t^2 \lambda
 + 64\, g_s^2 y_t^4 - 36\, y_t^6 - 60 \lambda  y_t^4 + 288\, \lambda^2 y_t^2\Big)
\Big]
\label{betaprimeSM}
\end{multline}
Note that although $\beta_\lambda'$ is formally two-loop order, it depends only on one loop $\beta$-function coefficients. Thus in the consistent NLO calculation all that is needed is one loop results.

\subsection{Absolute stability \label{sec:abstab}}
Absolute stability means that $\Gamma=0$ and our electroweak vacuum will never decay.
A naive criterion for $\Gamma \ne 0$ is that $\lambda^\star < 0$. That is
\be
\beta_\lambda(\mu^\star) = \lambda(\mu^\star) = 0 \qquad (\text{naive absolute stability})
\label{lostab}
\ee
This criterion has been been used in many treatments to establish the stability boundaries of phase space. For example, 
in~\cite{Bezrukov:2012sa,Bednyakov:2015sca}, this boundary is fixed by $\lambda(\mu^\text{cri}) = \beta_\lambda(\mu^\text{cri}) = 0$ (their $\mu^{\text{cri}}$ is our $\mu^\star$). We call this criterion ``naive" because is not systematically improvable: it only depends on the running $\lambda$. 
For example, if $\lambda^\star$ is positive but very very small, the rate can still be nonzero due to loop corrections but the naive criterion would miss this
possibility.\footnote{The way we compute  $\Gamma$ in this paper  is to expand around the bounce solution which requires $\lambda^\star <0$. 
If $\lambda^\star > 0$ and the electroweak vacuum is still unstable, one would have to modify the procedure to compute the rate (see~\cite{Weinberg:1992ds}).}
 
 To compute the absolute-stability phase space boundary, a gauge-invariant systematically-improvable procedure was developed in~\cite{Andreassen:2014eha,Andreassen:2014gha}.
 The starting point is that absolute stability is equivalent to the electroweak vacuum being the absolute minimum of the effective potential.
 Although the exact value of the potential at a minimum is known to be gauge invariant~\cite{Nielsen:1975fs,Fukuda:1975di}, some care has to be exercised in extracting this
 minimum value in perturbation theory.
 Because the effective potential having a minimum requires tree-level and one-loop effects to be comparable, the standard loop power counting cannot be used to establish stability (it violates gauge-invariance). A self-consistent gauge-invariant procedure for establishing absolute stability was developed in~\cite{Andreassen:2014eha}. 
Briefly, one starts with the leading-order effective potential
\begin{multline}
V^{\text{LO}}(h) = \frac{1}{4} \lambda h^4 + h^4 \frac{1}{2048 \pi^2}
\Big[-5 g'{}^4+6 (g'{}^2+g^2)^2 \ln \frac{h ^2 (g'{}^2+g^2)}{4 \mu ^2} \\
-10 g'{}^2 g^2-15 g^4+12 g^4 \ln \frac{g^2 h ^2}{4 \mu ^2}+144 y_t^4-96 y_t^4 \ln \frac{y_t^2 h ^2}{2 \mu ^2}
\Big]
\label{VeffLO}
\end{multline}
The first term of this potential is tree-level and rest comprises all of the one-loop corrections consistent with the power counting established in~\cite{Andreassen:2014eha,Andreassen:2014gha}. The point of the power counting is that since the one-loop contribution must overwhelm the tree-level contribution to turn the potential over, $\lambda$ must be the size of the one loop corrections.  Remarkably, one {\it must} impose this power counting consistently for gauge invariance to hold order-by-order in perturbation theory.
The minimum of $V^{\text{LO}}$ is where the couplings satisfy
\be
\lambda = \frac{1}{256\pi^2} \left[  g^4 +  3 g'{}^4 + 2 g^2 g'{}^2 +3 (g^2 + g'{}^2)^2 \ln \frac{4}{g^2 + g'{}^2} + 6 g^4 \ln \frac{4}{g^2} - 16 N_C y_t^4 \left(\ln \frac{2}{y_t^2}+1\right) \right]
\label{lmux}
\ee
We denote by $\mu_X$ the $\msbar$ renormalization-group scale where this equation holds. The NLO  effective potential with this consistent power counting is then computed by combining one-loop, two-loop and an infinite set of higher-loop daisy diagrams. Since the stability bound with this procedure is gauge invariant (as checked explicitly in scalar QED in~\cite{Andreassen:2014eha}), we can choose any gauge. Landau gauge ($\xi=0$) is particularly convenient as all the daisy diagrams vanish. The NLO effective potential in Landau gauge is extracted from~\cite{Degrassi:2012ry,Buttazzo:2013uya}. We present it in Appendix~\ref{app:vNLO} for completeness.

One cannot be certain that our universe is absolutely stable,
as quantum gravity or new physics coming in at an arbitrary high scale can open up new tunneling directions that can destabilize the universe~\cite{Branchina:2014rva,Branchina:2013jra,Lalak:2014qua,Branchina:2014efa,Branchina:2014usa,DiLuzio:2015iua,Andreassen:2016cvx}. So there is no sensible way of estimating a lower bound on the lifetime of our vacuum including new physics. 
The best one can do is to put an upper bound on the lifetime, and
the only question we can reasonably ask about new physics is at what scale, $\LNP$, it could come in to {\it stabalize} our vacuum? That is, how strong would it have to be to raise the upper bound on the lifetime to make it absolutely stable? To determine this scale, we add to the effective potential a gauge-invariant operator
\be
\Delta V_\text{eff} = \frac{1}{\LNP^2} h^6
\label{LNP}
\ee
This operator  contributes to $V^{\text{LO}}$ and modifies the equation for $\mu_X$, Eq.~\eqref{lmux}. Then we ask for given SM couplings, what
value of $\LNP$ will lift the minimum of $V_\text{eff}$ to zero.  The curves for this condition in the SM are shown in Fig.~\ref{fig:lambdaNP}.

\subsection{Numerical Results \label{sec:numbers}}
For numerical calculations, we take as inputs $G_F$, $m_W^{\text{pole}}$, $m_Z^{\text{pole}}$, $m_b^{\text{pole}}$, $m_t^{\text{pole}}$,
$m_h^{\text{pole}}$ and $\alpha_s(m_Z)$. These inputs are converted to $\msbar$  at a scale $\mu_0 =m_t^{\text{pole}}$ using 
threshold corrections known to two loops in all SM couplings~\cite{Degrassi:2012ry,Buttazzo:2013uya,Bezrukov:2012sa}, including mixed strong/electroweak contributions,
and partially to three and four loops in $\alpha_s$. The couplings are then run to high energy using the three-loop renormalization group 
equations with four-loop running included for $\alpha_s$~\cite{Chetyrkin:1997sg,Chetyrkin:2004mf,Chetyrkin:2012rz}. All of these threshold and running calculations are conveniently performed using the {\sc MR} package of Kniehl, Pikelner and Veretin~\cite{Kniehl:2016enc}.  

The numerical values are taken from the 2017 Particle Data Group~\cite{Olive:2016xmw}. We take as inputs
\be
G_F = 1.115 \times 10^{-5} \GeV^{-2};
\quad
m_W^{\text{pole}} = 80.385~\GeV,
\quad
m_Z^{\text{pole}} = 91.1876~\GeV,
\quad
m_b^{\text{pole}} = 4.93~\GeV
\ee
The uncertainty on these have a negligible effect on the rate so we set their uncertainties to zero. We also take current world averages~\cite{Olive:2016xmw}~\footnote{The most precise top quark mass measurements are currently done by matching experimental measurements to Monte Carlo (MC) simulators, and hence it is $m_t^{\text{MC}}$ that is being measured, \textit{not} $m_t^{\text{pole}}$. The uncertainty in translating from $m_t^{\text{MC}}$ to a well defined short-distance mass scheme has been studied, and early estimates were of order $1~\text{GeV}$ \cite{Hoang:2008xm}, although it may be much smaller, perhaps below $100~\text{MeV}$ \cite{Butenschoen:2016lpz,Hoang:2017suc}. For this analysis we will only use the standard PDG values for our central value and uncertainty, and do not include the $m_t^{\text{MC}}$ vs. $m_t^{\text{pole}}$ uncertainty. See also~\cite{Bezrukov:2014ina, Beneke:2016cbu, Andreassen:2017ugs}.}
\be
m_t^{\text{pole}} = 173.1 \pm 0.6~\GeV,
\quad
\mhpole = 125.09 \pm 0.24~\GeV
\quad
\alpha_s(m_Z) = 0.1181 \pm 0.0011 
\label{merrors}
\ee
These uncertainties will be propagated through to the final results. 

With these values, we find that $\lambda$ has a minimum at
\be
\mstar= 3.11 \times 10^{17} ~\GeV
\label{mstarSM}
\ee
At this scale, there is an instability ($\lambda < 0$) and
\be
 \lambda(\mstar) =- 0.0138, \quad y_t(\mstar) = 0.402,\quad g(\mstar) = 0.515,\quad g'(\mstar) = 0.460,
 \quad g_Z(\mstar) = 0.691
\ee
Note that the gauge couplings are quite large at this scale.

And, as needed for Eq.~\eqref{bigSM},
\be
\bps = 5.50 \times 10^{-5},\quad
 \hat{\mu} = 0.76\, \mstar ,\quad 
 \lol(\hat{\mu}) =  0.99993\, \lstar
 \ee
The action on the bounce is
\be
S[\phib] =- \frac{8\pi^2}{3\lstar} = 1900
\ee

The terms in Eq.~\eqref{bigSM} evaluate to
\be
\underbrace{e^{-S[\phib]}}_{10^{-826}} 
\underbrace{V_{SU(2)}}_{10^{1}}
\underbrace{ J_G^3}_{10^{5}}
\underbrace{ (R J_T)^4 (R J_d)}_{10^{7}}
\underbrace{ \sqrt{\frac{\det \wh \cO_h}{\det^\prime \cO_h}}}_{10^{-2}}
\underbrace{\sqrt{\frac{\det \wh \cO_{ZG}}{\det^\prime \cO_{ZG}}}}_{10^{-38}}
\underbrace{{\frac{\det \wh \cO_{WG}}{\det^\prime \cO_{WG}}}}_{10^{-14}}
\underbrace{\sqrt{ \frac{\det  \cO_{\bar{t}t}}{\det \wh \cO_{\bar{t}t}}}}_{10^{25}}
\underbrace{\sqrt{ \frac{\det  \cO_{\bar{b}b}}{\det \wh \cO_{\bar{b}b}}}}_{0.995}
\ee
and
\be
\underbrace{\mstar^4}_{10^{70} ~\GeV^4}
\underbrace{ \sqrt{-\frac{\pi  \lstar}{ \Sbs \bps}}  e^{-\frac{4\lstar}{ \Sbs \bps}}}_{1.09}
\underbrace{ \Sbs\left[ \frac{\lstar}{\lol(\hat{\mu})}-1 - \frac{4 \lambda_\star}{\Sbs ^2\bps} \right]}_{0.653}
 \ee
Multiplying everything together, we find the decay rate per unit volume is
\begin{align}
\frac{\Gamma}{V} &=  10^{-773 } \, \GeV^{4}
 \times \begin{pmatrix} 10^{-364} \\ 10^{198}\end{pmatrix}_{m_t}
 \!\!\!\!\!
 \times \begin{pmatrix} 10^{-48} \\ 10^{43}\end{pmatrix}_{m_h}
 \!\!\!\!\!
 \times \begin{pmatrix} 10^{-240} \\ 10^{156}\end{pmatrix}_{\alpha_s}
  \!\!\!\!\!
 \times \begin{pmatrix} 10^{-77} \\ 10^{125}\end{pmatrix}_{\text{thr.}}
  \!\!\!\!\!
 \times \begin{pmatrix} 10^{-2} \\ 10^{2}\end{pmatrix}_{\text{NNLO}}
 \nonumber
\\
 &= 10^{-773^{-638}_{+239}} \, \GeV^{4} \label{uncert}
\end{align}
The first three uncertainties are from variation of $m_t$, $m_h$ and $\alpha_s$ respectively according to Eq.~\eqref{merrors}.
The fourth uncertainty is theory uncertainty from varying the threshold matching scale $\mu_\text{thr} = \xi m_t^\text{pole}$ 
with $\frac{1}{2} < \xi < 2$ 
used in converting observables to $\msbar$ and as the starting point for RGE evolution. The final uncertainty marked NNLO represents the unknown two loop contributions to the functional determinant around the bounce. We estimate this error by scale variation around $\mu^\star$ by a factor of $\frac{1}{2}$ or $2$. 
Noting that the  NLO $t\bar{t}$ functional determinant contributes in the exponent at around 3\%  of the tree-level bounce action; therefore  our NNLO estimate of 7\% compared to NLO seems reasonable. 

The variations in the first line of Eq.~\erqef{uncert} are not independent and the dependence of $\Gamma$ on the masses and scales is highly non-linear. Nevertheless, since we can compute the effect on $\Gamma$ for any combination of their variations, we can determine their total correlated effect on the rate. To do this, we maximize or minimize the rate over $\chi^2 = 1$ hypersurface. We find that at 68\% confidence
$10^{-1411} < \frac{\Gamma}{V \GeV^4} < 10^{-533}$.
The range of decay rates allowed at 95\% confidence is $e^{-5660} < \frac{\Gamma}{V \GeV^4} < 10^{-386}$.

Thus, the lifetime of the Standard Model universe is
\be
\tau_\text{SM} = \left(\frac{\Gamma}{V} \right)^{-1/4} =10^{161^{+160}_{-59}}
\ee
That is, to 68\% confidence, $10^{102} < \frac{\tau_\text{SM}}{\text{years}} < 10^{321}$. 
To 95\% confidence $10^{65} < \frac{\tau_\text{SM}}{\text{years}} <10^{1383}$. 

To be more clear about what the lifetime means, we can ask a related question: what is the probability that we would have seen a bubble of decaying universe by now?  Using the space-time volume of our past lightcone~\cite{Buttazzo:2013uya},
 ${(VT)}_{\text{light-cone}} =\frac{0.15}{H_0^4} = 3.4 \times 10^{166}~\GeV^{-4}$ and
 the Hubble constant $H_0 = 67.4 \frac{\text{km}}{\text{s Mpc}} = 1.44 \times 10^{-42}~\GeV$, the probability that we should have seen a bubble by now is
   \be
 P = \frac{\Gamma}{V} {(VT)}_{\text{light-cone}} =  10^{-606^{-638}_{+239}}
 \label{Plc}
 \ee
Since the bubbles expand at the speed of light, chances are if we saw such a bubble we would have been destroyed by it; thus it is reassuring to find the probability of this happening to be exponentially small.

\begin{figure}[h!]
\centering
\includegraphics[scale = 0.47, trim = {0 -10mm -5mm 0mm}]{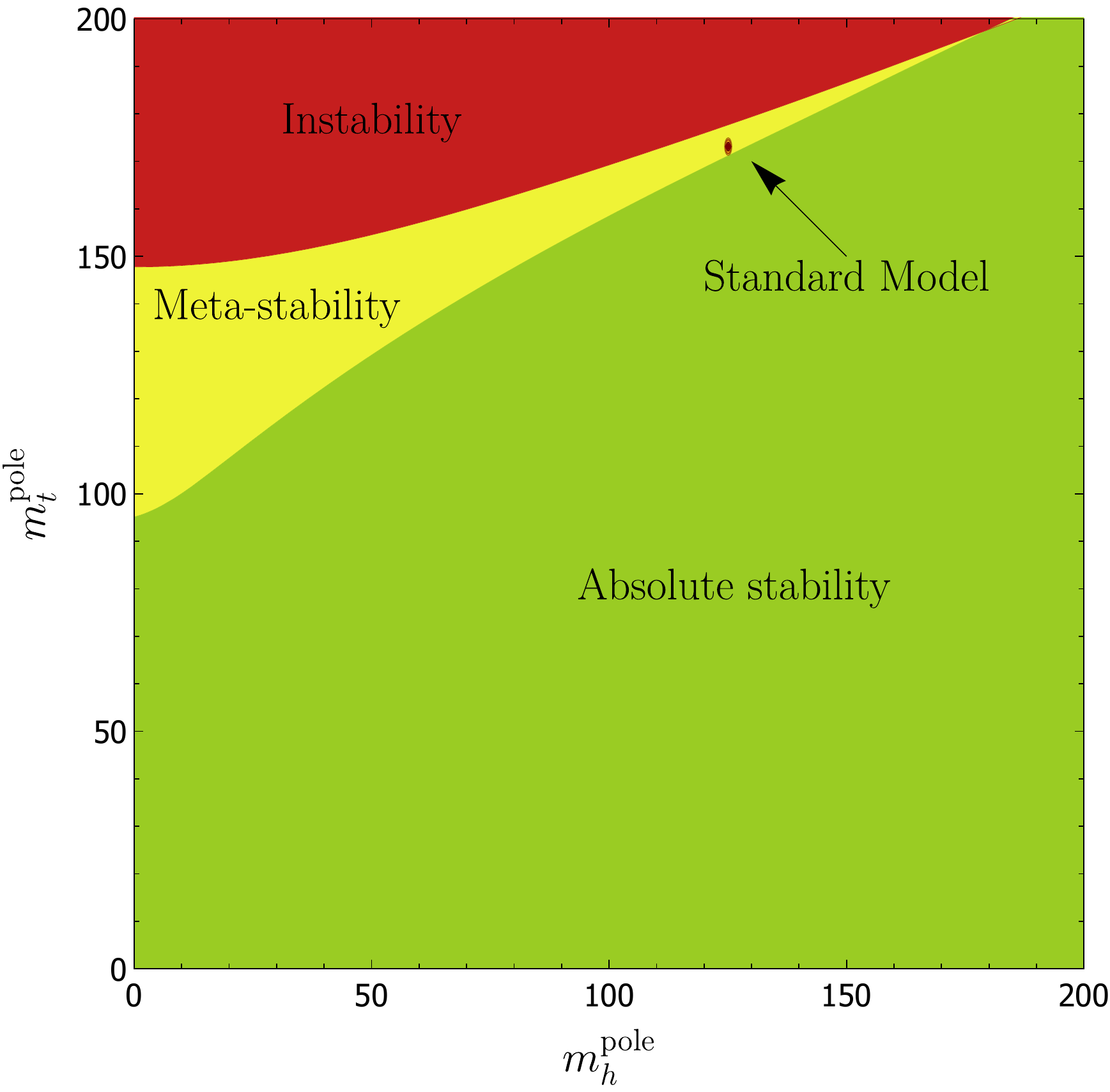}~~
\includegraphics[scale = 0.46, trim = {0 -10mm 0 0mm}]{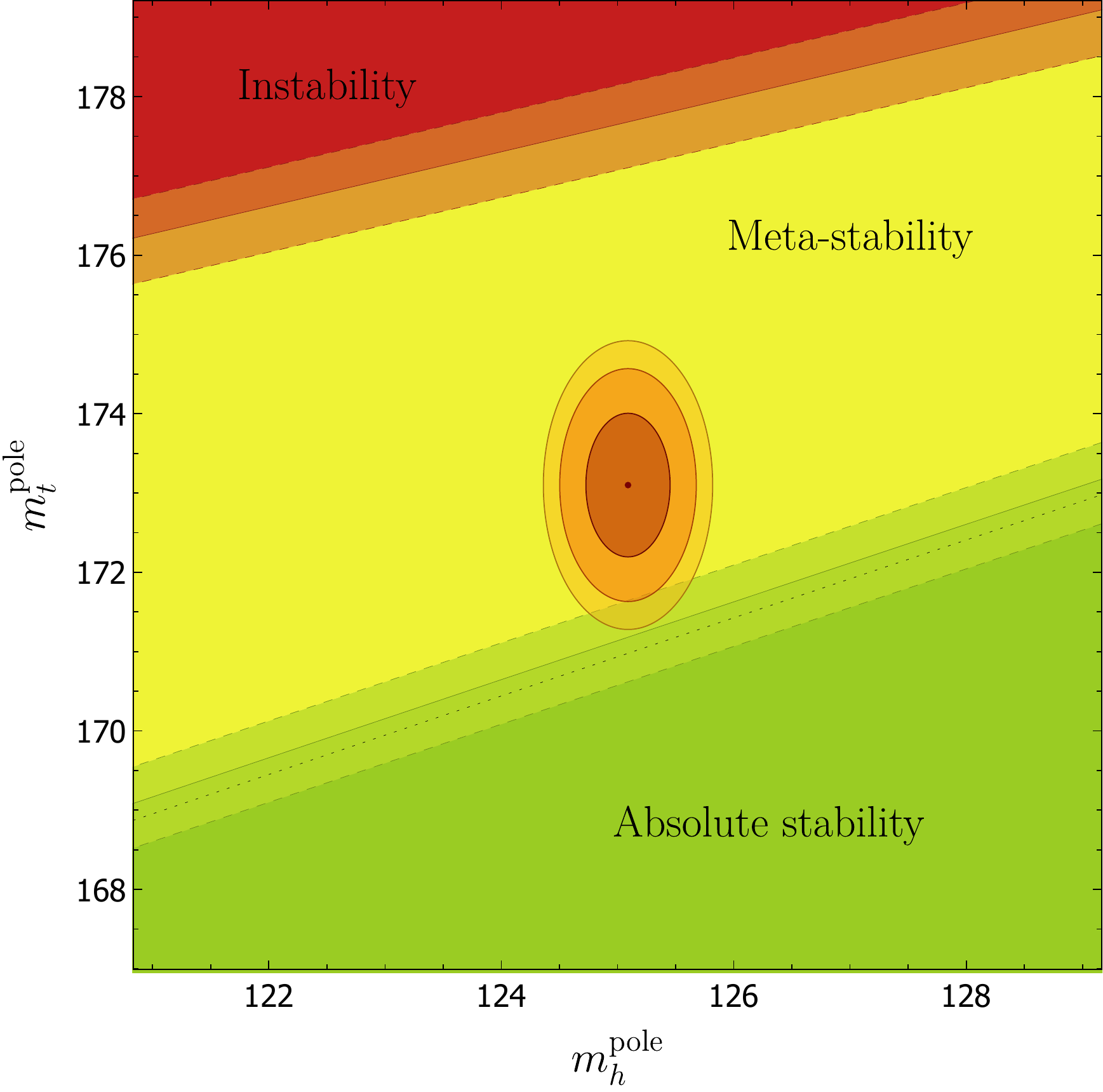}
\includegraphics[scale = 0.47, trim = {0 0 -5mm 0mm}]{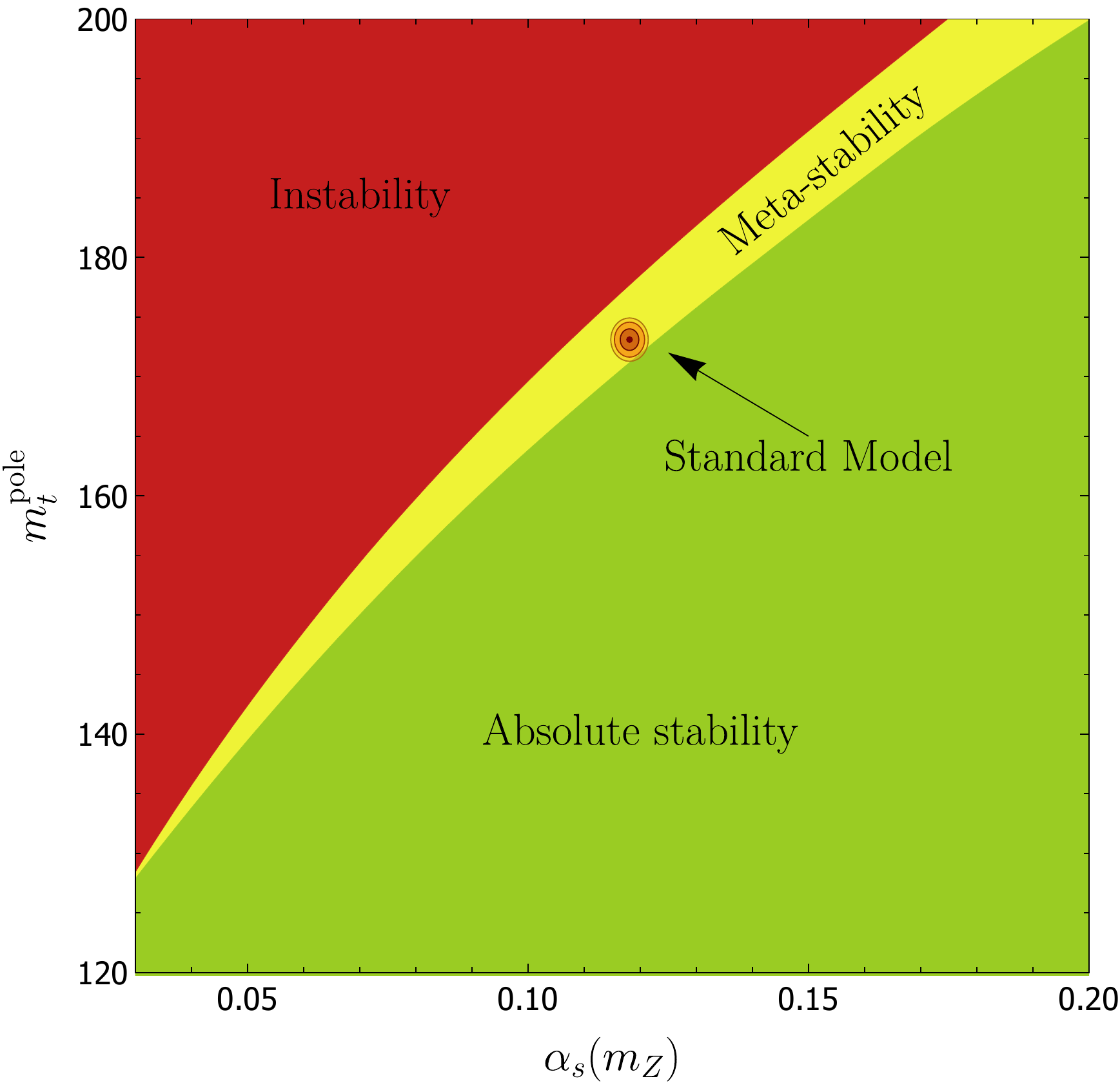}~~
\includegraphics[scale = 0.46]{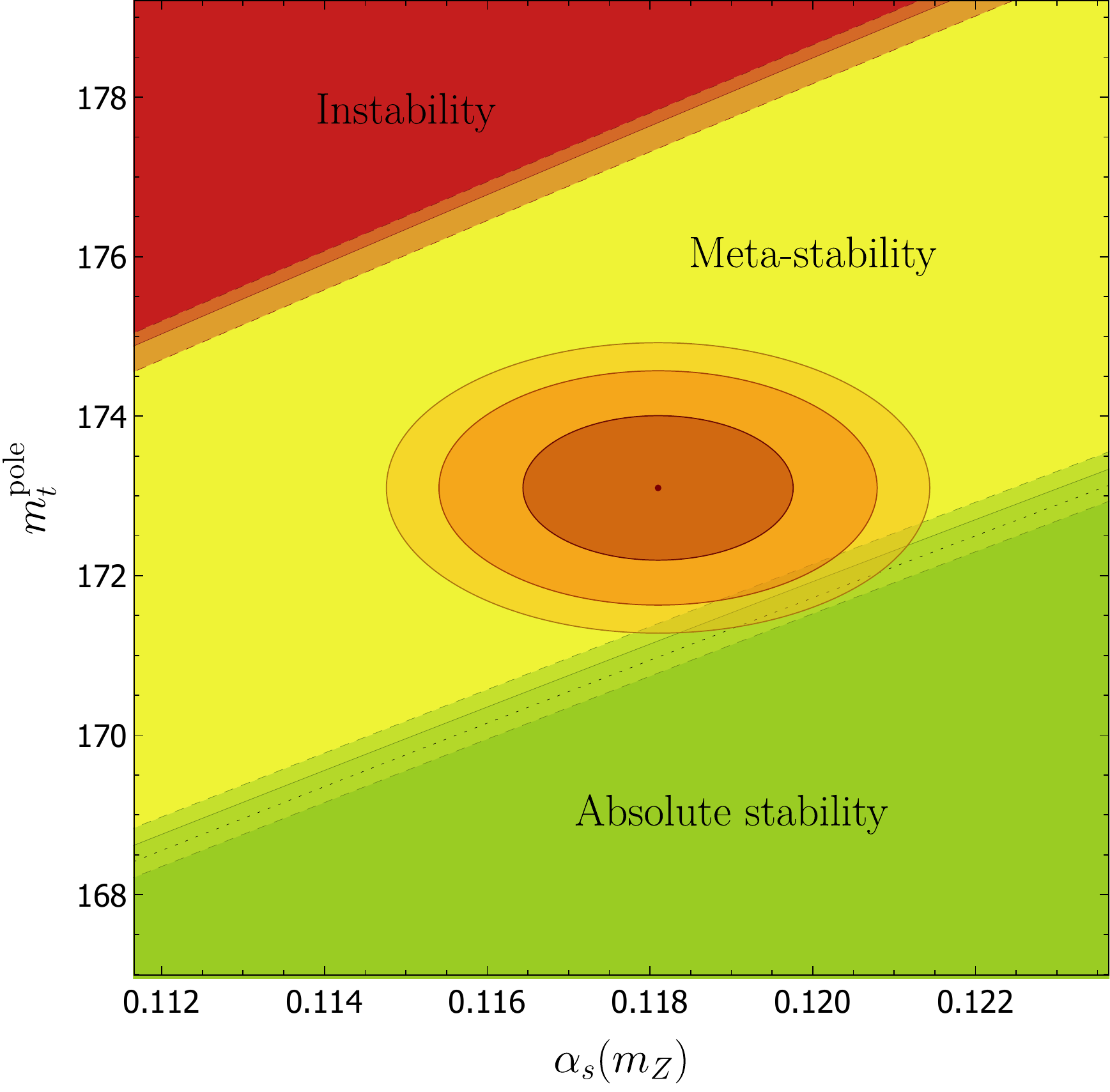}
\caption{(Top) phase diagram for stability in the $m_t^{\text{pole}}/m_h^{\text{pole}}$ plane and closeup of the SM region.
Ellipses show the 68\%, 95\% and 99\% contours based on the experimental uncertainties on $m_t^{\text{pole}}$ and $m_h^{\text{pole}}$. The shaded bands on the phase boundaries, framed by the dashed lines and centered on the solid lines, are combinations of the
$\alpha_s$ experimental uncertainty and the theory uncertainty.
(Bottom) phase diagram in the $m_t^{\text{pole}}/\alpha_s(m_Z)$ plane, with uncertainty on the boundaries given by combinations of
uncertainty on $m_h^{\text{pole}}$ and theory. 
The dotted line on the right plots is the naive absolute stability prediction using Eq.~\eqref{lostab}.
\label{fig:phase}}
\end{figure}

The phase diagrams in the $m_t/m_h$ and $m_t/\alpha_s$ planes are shown in Fig.~\ref{fig:phase}. In these diagrams, the boundary between metastability and instability is fixed by $P=1$, where $P$ is the probability that a bubble of true vacuum should have formed without our past lightcone, as in Eq.~\eqref{Plc}. The boundary between metastability and instability is determined by the gauge-invariant consistent procedure detailed Section~\ref{sec:abstab} (and in~\cite{Andreassen:2014eha,Andreassen:2014gha}). Although the absolute stability boundary is close to the condition $\lambda^\star=0$ in Eq.~\eqref{lostab}, it is systematically higher and  a better fit to the curve for $\lambda^\star = -0.0013$.

Varying one parameter holding the others fixed, we find that the range of $\mtpole$, $\mhpole$ or $\alpha_s$ for the SM to be in the metastability window are
\be
171.18 < \frac{\mtpole}{\GeV} < 177.68,
\quad
 129.01 > \frac{\mhpole}{\GeV} > 111.66,
\quad
 0.1230 > \alpha_s(m_Z) > 0.1077
\qquad
\ee
Numbers on the left in these ranges are for absolute stability and on the right for metastability. 

To be absolutely stable, the bounds on the parameters are
\begin{align}
\frac{\mtpole}{\GeV}&<171.18 +0.12\left(\frac{\mhpole/\GeV-125.09}{0.24}\right)+0.43\left(\frac{\alpha_s(m_Z)-0.1181}{0.0011}\right)+ (\text{th.})\begin{matrix}+0.17\\-0.35 \end{matrix}\nonumber
\\
\frac{\mhpole}{\GeV}&>129.01 +1.2\left(\frac{\mtpole/\GeV-173.1}{0.6}\right)+0.89\left(\frac{\alpha_s(m_Z)-0.1181}{0.0011}\right)+(\text{th.})\begin{matrix}+0.34\\-0.72 \end{matrix}\nonumber
\\
\alpha_s(m_Z)&>0.1230 +0.0016\left(\frac{\mtpole/\GeV-173.1}{0.6}\right)+0.0003\left(\frac{\mhpole/\GeV-125.09}{0.24}\right)+(\text{th.})\begin{matrix}+0.0005\\-0.0010 \end{matrix}
\end{align}
Absolute stability is currently excluded at 2.48$\sigma$, which translates to a one-sided confidence of 99.3\%. To exclude absolute stability to the one-sided confidence for 3$\sigma$, the top quark mass uncertainty must be  reduced below $250~\text{MeV}$. Similarly for $\alpha_s$ for a 3$\sigma$ uncertainty must be less than $\Delta \alpha_s<0.00025$.

\begin{figure}[t!]
\centering
\includegraphics{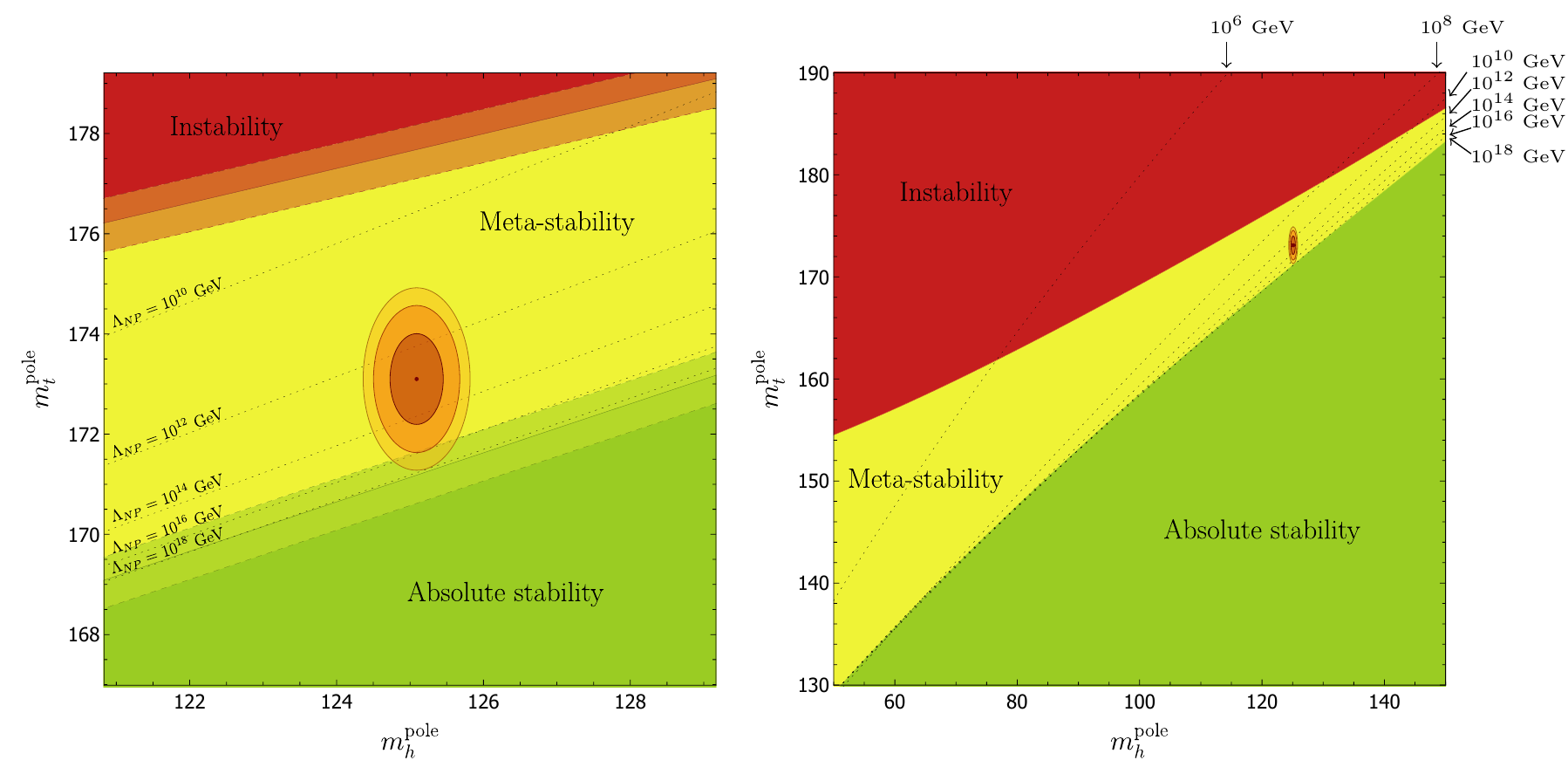}
\caption{
Phase diagram for stability in the $m_t^{\text{pole}}/m_h^{\text{pole}}$ plane with  dotted lines indicating the scale at which the addition of higher-dimension operators could stabilize the SM.
Note that
the curves 
 accumulate on the stability/metastability boundary. $\LNP$ curves in the $\alpha_s/\mtpole$ plane (not shown) are similar. 
\label{fig:lambdaNP}}
\end{figure}

The dashed lines Fig.~\ref{fig:lambdaNP} indicate the scale at which new physics operators at the scale $\LNP$ can stabilize the SM, added as in Eq.~\eqref{LNP}.
Recall that because tunneling is a non-perturbative phenomenon, higher-dimension operators do not decouple: new physics at an arbitrary high scale can destabilize the SM my opening up new tunneling directions~\cite{Branchina:2014rva,Branchina:2013jra,Branchina:2014efa,Branchina:2014usa,DiLuzio:2015iua,Andreassen:2016cvx}. To {\it stabalize} the SM, they have to be strong enough to lift the potential from negative to positive. 
In Fig.~\ref{fig:lambdaNP} we see that the density of $\LNP$ curves increases near the absolute stability line. 
This happens because the absolute stability region is necessarily insensitive to the addition of a positive operator.

\section{Mass Corrections \label{sec:mass}}
One remaining technical detail is how to handle the fact that the Higgs
potential in the Standard Model is not exactly scale invariant, since there is
a finite mass term for the Higgs field. We saw in Section~\ref{sec:scale} that with a scale-invariant classical potential, quantum corrections naturally
pick out the scale $\mu^\star$ where $\lambda(\mu)$ is minimal so that the action
is dominated by bounces of a size $R^\star = \frac{1}{\mu^\star}$. 
One hopes that because the Higgs mass parameter $m \sim 10^2$ GeV is much much smaller than
 $\mu^{\star} \sim 10^{17}$ GeV,  the corrections to the
decay rate from the mass term will be completely negligible.
Although normally classical effects, like the Higgs mass term, dominate over quantum effects,
in this case
 the quantum scale violation can be dominant since it scales as an inverse power of $\hbar$ (see Eq.~\eqref{Gamma2def}). 
Despite this convincing logic,  producing a quantitative estimate of the effect on the decay rate of a finite mass term
is surprisingly challenging.

\subsection{A Bound on the $m^2$ Correction}

Consider the potential $V (\phi) = \frac{1}{2} m^2 \phi^2  +\frac{1}{4} \lambda \phi^4$ with $\lambda <0$ and $m>0$.
Trying to solve the Euclidean equations of motion for this potential,
$\phi'' + \frac{3}{r} \phi' - m^2 \phi - \lambda \phi^3=0$, one quickly discovers that the only solution is $\phi =
0$. There are many ways to see this~\cite{Frishman:1978xs,Affleck:1980mp,Nielsen:1999vq,DiLuzio:2015iua} such as with Derrick's 
theorem~\cite{Derrick:1964ww}.
An intuitive way is to use Coleman's trick of thinking of the solution the Euclidean equations of motion as
a ball rolling with friction down a hill shaped like $-V(\phi)$ starting at $\phi(0)$ and ending at $\phi=0$ at ``time" $r=\infty$.
For $m=0$, the potential is scale invariant, so no matter where the ball starts it will get to $\phi=0$ only asymptotically at infinite time;
different starting points correspond to different $R$ for the bounce solutions in Eq.~\eqref{phibdef}.
Now, when we add $\frac{1}{2} m^2 \phi^2$ to the potential, it creates a
depression in $- V (\phi)$ near $\phi = 0$. Since for any $R$ the bounces just barely got to $\phi = 0$
at infinite time, adding even an infinitesimal
depression prevents solutions to the equations of motion from ever reaching $\phi = 0$. Thus there are no
bounces when $m^2>  0$.

Assuming $m$ is small compared to $\mu^{\star}$, one might think we can write
$\phi = \phib + m^2  \Delta \phi+ \cdots$ and evaluate the corrections to the action perturbatively. Trying
this, one immediately finds
\begin{equation}
  \Delta S = \int d^4 x \frac{1}{2} m^2 \phib (x)^2 = \infty
\end{equation}
This behavior is due to the non-normalizabilty of $\phi_b$.
Thus $\Gamma \sim e^{- S} = 0$ confirming that even an infinitesimal $m^2$ seems to prevent
vacuum decay.

To understand this unintuitive result, let us consider the alternative, more
physical, treatment of tunneling described in~\cite{Andreassen:2016cff,Andreassen:2016cvx}. There, a formula for the
tunneling rate was derived inspired by the understanding of tunneling in
non-relativistic quantum mechanics. In quantum field theory, the exponential
factor determining the decay rate along a path parameterized by $\phi(\vec{x}, \tau)$ is the integral
\be
  - \ln \Gamma_{\phi} = 4\int_{-\infty}^{0} d \tau U [\phi (\tau)] = \int d s \sqrt{2 U [\phi(s)]} 
  \label{partialgamma}
\ee
where the energy functional is~\cite{Coleman:1987rm,Lee:1985uv,Brown:2011um}
\be
  U [\phi (\tau)] = \int d^3 x \Big[\frac{1}{2}(\nabla \phi)^2 + V (\phi)\Big]
\ee
In Eq.~\eqref{partialgamma} $\tau$ is the Euclidean time and $s$ is the proper time,
determined by $(\frac{ds}{d\tau})^2 = 2U[\phi]$. Using $s$
gives a formula exactly like the WKB exponent formula $\int d x \sqrt{2 V
(x)}$ in quantum mechanics, but now with a contribution from gradient energy.

With this formulation let us now revisit the perturbative solution. If we try
to calculate $\Gamma_{\phi}$ along the $m=0$ bounce path
 $\phib (r=\sqrt{\vec{x}^2+ \tau^2})$, we find
\begin{equation}
  U [\phib(\tau) ] =- \frac{2\pi^2}{\lambda} \frac{R^2 \tau^2}{(R^2 +
  \tau^2)^{5 / 2}} - m^2 \frac{4\pi^2}{\lambda} \frac{R^2}{\sqrt{R^2 +
  \tau^2}}
\end{equation}
The integral over the first term gives $S [\phib]=-\frac{8\pi^2}{3\lambda}$, as for $m=0$. 
The second term, however, shifts $U (\tau)$ along this path up by a
finite positive amount at each $\tau$, and the resulting integral over $\tau$ is infinite.

One fault of using the path through the $\phib$ bounces in a non-scale-invariant potential
is that  conservation of energy is violated. Since bubbles are produced at rest, 
we can see this through $U [\phi (\tau=0)] \ne U [\phi (\tau=\infty)]$.
If the Euclidean equations of motion are satisfied, energy is conserved and
this cannot happen.

Let us consider instead a path through field space of the form
\begin{equation}
  \phi_G = \phi_0 \exp\left[- \frac{\tau^2 + \vec{x}^2}{R^2}\right]
\end{equation}
These Gaussian bubbles were previously introduced and discussed in~\cite{Andreassen:2016cvx}. Their energy is
\begin{equation}
  U [\phi_G(\tau)] = \frac{3 \sqrt{2}  \pi^{3 / 2} }{8 } R e^{- 2 \frac{\tau^2}{R^2}}\phi_0^2
    +
  \frac{\pi^{3 / 2}}{32} \lambda R^3 e^{- 4 \frac{\tau^2}{R^2}} \phi_0^4 +
  \frac{\sqrt{2} \pi^{3 / 2}}{8} m^2 R^3 e^{- 2 \frac{\tau^2}{R^2}} \phi_0^2 
\end{equation}
Setting $U [\phi_G (\tau = 0)] = U [\phi_G (\tau = \infty)]$ gives
\begin{equation}
  \phi_0 = \frac{1}{R} 2^{5 / 4} \sqrt{\frac{3 + m^2 R^2}{-\lambda}}
\end{equation}
With this value for $\phi_0$, the partial width for decaying to Gaussian bubbles is
\begin{equation}
  \ln \Gamma_{\phi_G} = - \int_{- \infty}^{\infty} d \tau U [\phi_G (\tau)] =
  \frac{\pi^2 ( \sqrt{2} - 1 )}{2 \lambda} (3 + m^2 R^2)^2 
\end{equation}
This is finite. Moreover, it has a nonzero maximum at $R = 0$. We conclude that
the exponent is bounded from above. In other words,
\begin{equation}
\frac{8 \pi^2}{3 \lambda} \leqslant  \ln \Gamma \leqslant 
\frac{8 \pi^2}{3  \lambda} \frac{27(\sqrt{2}-1)}{16}  \approx (0.7) \frac{8\pi^2}{3  \lambda}
\end{equation}
The lower bound comes from $m = 0$ and the upper bound comes from the Gaussian bubbles as
$R \rightarrow 0$. We conclude that the rate is finite.

\subsection{Constrained Instantons}

Now that we know the rate is finite, what is it? The Gaussian bubbles are in
fact far from producing the optimal path through field space, as we will see.
What we would like to do is directly minimize $\int d \tau U [\phi (\tau)]$
over field configurations with $\phi (0) = 0$. Luckily, we do not have to
find the absolute minimum; if we find any path which gives a finite rate close
enough to the $m = 0$ case that we can neglect it for the Standard Model, we
can conclude that ignoring $m$ when $m \ll \mu^{\star}$ is justified.

One way to find a finite-action path through field space is through the
constrained instanton approach~\cite{Frishman:1978xs,Affleck:1980mp}. In brief, the idea is that instead of
minimizing the action absolutely, we find the minimum along some surface. For
example, we can look along the surface where $\int d^4 x \phi^n = k$ for some
$k$ and some $n$. This constraint can be imposed through a Lagrange multiplier
by writing the action as
\begin{equation}
  S_{\sigma} = \int d^4 x \left[ \frac{1}{2} (\partial_{\mu} \phi)^2 +
  \frac{1}{2} m^2 \phi^2 + \frac{1}{4}\lambda \phi^4 + \sigma \phi^n \right] - k
\end{equation}
Taking $n = 2$ or $n = 4$ does not produce anything helpful since the new term
is just like one of the old ones. Taking $n \geqslant 5$ is also unhelpful,
since for a normalizable solution we need $\phi \rightarrow 0$ at large
distance, but then $\phi^n$ term is subdominant to the $\lambda \phi^4$ term
which produced the non-normalizable mode in the first place. Thus $n = 3$ is
our only hope. (See~\cite{Nielsen:1999vq} for a thorough discussion of constraints on the  constraints).

For $n = 3$, the procedure for producing a normalizable well-behaved
constrained solution that reduces to $\phib$ at $m = 0$ is discussed in~\cite{Nielsen:1999vq} and~\cite{Glerfoss:2005si}.
The solution and Langrange multiplier can be expanded perturbatively in $m^2$:
\be
  \phi = \phib + \phi_2 + \cdots,\qquad \sigma = \sigma_2 + \cdots
\ee
It is helpful to write $\phi_2 = \phi_{2, a} + \phi_{2, b}$ with
\be
  \phi_{2, a} = \frac{m^2 R}{\sqrt{-2 \lambda} r^2} \left[ \frac{9 R^2
  r^2 - 3 r^4 + (R^2 - 10 R^2 r^2+r^4) \ln \left( 1 + \frac{r^2}{R^2}
  \right)}{R^2 + r^2} + 6 R^2 r^2 \frac{R^2 - r^2}{(R^2 + r^2)^2}
  \mathrm{Li}_2 \left( - \frac{r^2}{R^2} \right) \right]
\ee
satisfying $\phi_{2,a}'' + \frac{3}{r} \phi_{2,a}' -m^2 \phi_b^2 -3 \lambda \phi_b^2 \phi_{2,a} = 0$ and
\be
 \phi_{2,b} = -\frac{1}{\lambda} \sigma_2= \text{const.} \label{phi2b}
\ee
So that
$ \Box \phi - m^2 \phi - \lambda \phi^3 - 3 \sigma  \phi^2 = \mathcal{O}
  (m^4)$.
  
To determine the constant in Eq.~\eqref{phi2b} and the $\cO(m^2)$ value of the Lagrange multiplier, we note
that  perturbative solution is not normalizable.  This non-normalizability is easy to understand: a solution perturbative in $m$ can never
describe the asymptotic behavior for $r \gg \frac{1}{m}$.  
no matter how small $m$ is. 
Indeed, a normalizable solution should have $\phi \rightarrow 0$ as $r
\rightarrow \infty$ and therefore match on to $\phi_K (r) = K_0 \frac{1}{m
r} \mathcal{K}_1 (m r)$ which solves $(\Box + m^2) \phi = 0$. At large
$r$, $\phi_K (r) \sim K_0 \sqrt{\frac{\pi}{2}} \frac{1}{(m r)^{3 /
2}} e^{- m r}$ which is exponentially suppressed and gets contribution from
all orders in $m$. 
We choose $K_0 = \sqrt{\frac{8}{-\lambda}}\frac{m^2}{R}$ so that to order $m^0$, $\phi_K(r)$ matches
on to $\phib(r)$ at large $r$.
This fixes the value for $\phi_{2,b}$ in Eq.~\eqref{phi2b}  to be
\be
  \phi_{2, b} = \sqrt{\frac{2}{- \lambda}} R m^2
   \left( \ln \frac{m R}{2} +  \gamma_E + 1\right)
\ee
This is the unique solution allowing $\phi_2$ at large $r$ to match the $m^2$ terms of $\phi_K(r)$.
 The Lagrange multiplier in Eq.~\eqref{phi2b} is then  $\sigma =  -\lambda \phi_{2, b} + \cO(m^4)$
 so that $\phi$ satisfies the (constrained) equations of motion to order $m^2$.
Higher order terms can systematically computed
guaranteeing exponential suppression at large $r$~\cite{Nielsen:1999vq,Glerfoss:2005si}.

Given these results, we now need to check that the decay rate along the constrained-instanton path
is finite. If we work strictly to order $m^2$ we find the action gets
corrected by
\begin{equation}
  \Delta S = \frac{m^2}{2} \int d^4 x \phib^2 + \int d^4 x \Big[(\partial_{\mu}
  \phib) (\partial_{\mu} \phi_2) +  \lambda \phib^3 \phi_2\Big] = \infty
\end{equation}
This is not surprising as $\phib + \phi_2$ is not normalizble. The key to
getting a finite action is to be careful with the boundary term. Without
integrating by parts we can write
\begin{align}
  S [\phi] &= \int d^4 x \left[ \frac{1}{2} (\partial_{\mu} \phi)^2 +
  \frac{1}{2} m^2 \phi^2 +\frac{1}{4} \lambda \phi^4 \right]\\
  &= \int d^4 x \left[ \frac{1}{2} \partial_{\mu} (\phi \partial_{\mu} \phi) -
  \frac{1}{2} \phi (\Box \phi - m^2 \phi -  \lambda \phi^3 - 3 \sigma \phi^2)
  - \frac{1}{4}\lambda \phi^4 - \frac{3}{2} \sigma \phi^3 \right]
\end{align}
The equations of motion on the constraint surface imply
 $\Box \phi - m^2 \phi -  \lambda \phi^3 - 3 \sigma \phi^2 = 0$ so we drop this term. We also drop the total
derivative term using that the exact $\phi$ vanishes exponentially at infinity
(although not at any fixed order in $m^2$). The remainder can then be
evaluated perturbatively:
\begin{align}
  S [\phi] &= 
  \int d^4 x [-\frac{1}{4}\lambda \phib^4]
   + \int d^4 x \left[-\lambda \phib^3 \phi_2 - \frac{3}{2} \sigma \phib^3 \right]\\
  &=- \frac{8 \pi^2}{3 \lambda} \left[ 1 - \frac{3}{2} m^2 R^2 \left( \ln  \frac{m^2 R^2}{4} + 2 \gamma_E + 1 \right) \right] \label{S144}
\end{align}
For $m =\mhpole$ in Eq.~\eqref{merrors} and $R = R^{\star} \sim \frac{1}{\mu^{\star}}$ in Eq.~\eqref{mstarSM},
\begin{equation}
  S [\phi] = -\frac{8 \pi^2}{3 \lambda} [1 + 1.02 \times 10^{- 28}]
\end{equation}
This justifies neglecting the mass for the Standard Model lifetime, as we have done through the rest of this paper.

\subsection{Comments on Constrained Instantons}
Before concluding, we add some comments on the constrained instanton approach.

First, we made no claim that the constrained instanton produces the exact
decay rate, as there may be lower-action configurations satisfying different
constraints. For situations where $m \slashed \ll ~\frac{1}{R^{\star}}$ it may be important to
look for other solutions and Eq. (\ref{S144}) cannot be used in such contexts.

Second, we never try to integrate over the value for the constraint $k$.
Normally one sets $k$ at the outset and solves for the Lagrange multiplier
$\sigma$ as a function of $k$. Here we have found $\sigma$ by requiring the
solution have finite action; $k$ is then fixed by $\sigma$. Although at
order $m^2$, $k = \infty$, $k$ should be finite when the full solution is used, since 
in the full solution $\phi$ dies off exponentially at large distance. In any case, we do not
need $k$ to compute the tunneling rate, as we have seen.

Third, although the constrained instanton approach is useful to understand $m >0$, 
it does {\it not} help resolve the divergent
integral over instanton size $R$ for $m=0$. Since the true minimum of the action is known when $m = 0$,
the divergence must be resolved from higher order perturbative effects, as we
explained in Section~\ref{sec:siprob}. There has been some confusion about this point in the literature~\cite{Isidori:2001bm}.

Fourth, we note that there is an apparent contradiction that the Euler-Lagrange equations have no solution with $m^2>0$, but we have proven there is a finite, non-zero minimum to the action. The resolution is that Euler-Lagrange equations are derived dropping a boundary term, but the
behavior of the solutions at infinity are critical to finding a correct minimum. As we have seen, to any finite order in $m^2$, the boundary terms cannot be dropped, so one would never come upon a perturbative solution like ours using the Euler-Lagrange equations alone. The importance
of the boundary behavior is emphasized and discussed at length in~\cite{Affleck:1980mp,Nielsen:1999vq,Glerfoss:2005si} as a motivation 
for the constrained instanton approach. 

Finally, we have done the whole analysis here, following~\cite{Nielsen:1999vq,Glerfoss:2005si}  for the case $m^2
> 0$. The case with $m^2 < 0$ is also interesting. For $m^2 < 0$, the energy
function $U [\phi_b (\tau)]$ gets shifted down and, for $m R > 1$, the tunneling
rate is in fact infinite: there is no barrier to tunneling (as in quantum
mechanics with a potential like $V = - x^2$). This result is also wrong. The
argument is flawed, since $U [\phi_b (0)] \neq 0$ just like for the $m^2 > 0$
case, so the proposed tunneling path violates energy conservation is not allowed. Tunneling should speed
up, for $m^2 < 0$, but only by an  amount suppressed by factors of $m^2R^2$. By analytic continuation
we can still use Eq.~\eqref{S144}; for $m^2<0$ $S [\phi]$ now has a
small imaginary part, but this produces a tiny effect on the final result, since
$\Gamma \sim \im (i e^{- S [\phi]})$.
\section{Conclusions \label{sec:conc}}
In this paper we have produced the first complete calculation of the lifetime of the Standard Model. Previous treatments
were incomplete in a number of ways. First, there was a long-standing problem of how to perform instanton 
calculations when scale-invariance is spontaneously broken. The problem is that in a classically-scale invariant theory,
the integral over instanton size $R$ is divergent at next-to-leading order (NLO). We showed that in fact there are contributions
which seem higher-order in $\hbar$ but which in fact dominate over the NLO contribution after the integral over $R$ is performed. Including all the relevant terms, to all loop-order, we are able to integrate over instanton size exactly giving a finite result.

The second problem we resolved is also related to instanton size. Since fluctuations associated with changing the size $R$ are unsuppressed, one has to allow for large deviations in field space. Changing to collective coordinates allows the integral over all $R$ to be done, however, it generates an infinite Jacobian.  We showed that this infinite Jacobian is in fact compensated by an infinity in the functional determinant previously missed. To handle the infinity and the zero, we employ a judicious operator rescaling inspired by a conformal mapping to the  4-sphere. We find the spectrum of the rescaled operators exactly and give an analytic formula for the Jacobian (now finite) as well as the functional determinant with zero modes removed (also finite now).

The third problem we resolved has to do with fluctuations of vector bosons around the instanton background. 
When a global internal symmetry is spontaneously broken there are additional zero modes. In previous treatments the Jacobian for going to collective coordinates for these symmetries was found to be infinite. We show that this infinity was
an artifact of working in $R_\xi$ gauge where the symmetry is actually explicitly broken by the gauge-fixing. 
Instead we work in Fermi gauges, and using the same technique as for the dilatation zero mode, show that the Jacobian for internal symmetries is finite.

The next new result in our paper is a complete analytic computation of the functional determinant around the instanton background for real and complex scalar fields, vector bosons and fermions. Moreover, we showed that the final result is gauge-invariant (of the parameter $\xi$ in Fermi gauges and between Fermi and $R_\xi$ gauges). 
For the scalars, the insight which allowed for these exact results was to use the exact spectrum known from the operator rescaling and mapping to the 4-sphere~\cite{McKane:1978md,Drummond:1978pf,Shore:1978eq}. For the vector bosons, we exploited a remarkable simplification of the fluctuation equations discovered in~\cite{Endo:2017gal,Endo:2017tsz}. These authors found that the equations that couple the scalar and longitudinally polarized gauge bosons with the Goldstone bosons can be written in terms of a set of simplified equations using auxiliary fields. Although the treatment in~\cite{Endo:2017gal,Endo:2017tsz} assumed a mass term for the scalar, so that their results do not exactly apply to the case of the Standard Model, our treatment very closely parallels theirs. 

Combining all our results together we produced a complete prediction for the lifetime of our metastable vacuum in the Standard Model. We find the lifetime to be
\be
\tau_\text{SM} =10^{161^{+160}_{-59}}
\ee
The enormous uncertainty in this number is roughly equal parts uncertainty on the top quark mass, uncertainty on the value of the strong-coupling constant $\alpha_s$ and theory uncertainty from threshold corrections, that is, from matching between observable pole masses and $\msbar$ parameters at the electroweak scale. 
The uncertainty from error on the Higgs boson mass is small as is, thankfully, uncertainty associated with the unknown NNLO corrections to the decay rate.

Phase diagrams in the $m_t/m_h$ plane and the $m_t/\alpha_s$ plane are shown in Fig.~\ref{fig:phase}. 
This figure indicates that the SM seems to sit in a peculiarly narrow swath of metastability in the phase space of top quark mass,
Higgs boson mass, and strong-coupling constant. 
An important fact to keep in mind when interpreting this tuning is that phase diagram assumes no gravity and no physics beyond the SM. In fact, any arbitrarily high-scale physics can destabilize the SM by opening up new tunneling directions~\cite{Branchina:2014rva,Branchina:2013jra,Branchina:2014efa,Branchina:2014usa,DiLuzio:2015iua,Andreassen:2016cvx}.
Moreover, near the absolute stability boundary, operators at an arbitrarily high scale can also stabilize the SM, as can be seen from Fig.~\ref{fig:lambdaNP}.
For the SM, which appears not to be on the stability boundary, the relevant scale of new physics is around $10^{13}$ GeV.
  
Because of the importance of the top quark mass, the Higgs boson mass and  $\alpha_s$ in determining stability, it is interesting to look
at their allowed ranges. We find that, varying each parameter separatly, the bounds for the SM to lie in the metastability window are
 \be
171.18 < \frac{\mtpole}{\GeV} < 177.68,
\quad
 129.01 > \frac{\mhpole}{\GeV} > 111.66,
\quad
 0.1230 > \alpha_s(m_Z) > 0.1077
 \ee
If we hope to rule out absolute stability to 3$\sigma$ confidence, assuming nothing else changes,
we would need $\Delta \mtpole<250~\text{MeV}$ or $\Delta \alpha_s(m_Z) < 0.00025$. 

Finally,  we note that the predicted lifetime $10^{161}$ years, while enormously long, has an exponent of roughly the same order of magnitude as the current lifetime of the universe, $10^{9}$ years. Indeed, the long lifetime of the SM is due to the fact that the Higgs quartic coupling has a minimum value of $\lstar = -0.0138$. If the minimum of the coupling were smaller, say $\lstar = -0.1$, then the SM lifetime which scales like $\exp(\frac{8\pi^2}{3\lambda})$ would be only
$10^{-20}$ seconds! Furthermore, since the lifetime is finite and the universe infinite, there is likely a bubble of true vacuum already out there, far far away. It is sobering to envision this bubble, with its wall of negative energy, barreling towards us at the speed of light. It seems the long-term future of our universe is not going to be slow freezing due to cosmic acceleration but an abrupt collision with one of these bubble walls.

\section*{Acknowledgements}
We are grateful to A. Strumia for help understanding their paper~\cite{Isidori:2001bm} and to A. Pikelner for assistance with the threshold corrections and the  {\sc mr} package~\cite{Kniehl:2016enc}. We also thank Y. Shoji for carefully checking the first version of this manuscript and suggesting some corrections. This research was supported in part by the Department of Energy under Grant No. DE-SC0013607.

\appendix

\section{Removing Zero Modes without Rescaling \label{app:norescale}}
In this appendix, we explore what goes wrong when we try to calculate the determinant for fluctuations around the bounce
without rescaling the operators as in Section~\ref{sec:scale}. Recall from the discussion in that section that without the operator rescaling 
the Jacobian for going to collective coordinates for scale transformations is infinite (Eq.~\eqref{Jdinf}). Since the full functional determinant should be independent of the operator rescaling, this infinity must be compensated by something else. However previous investigations found a finite value for $\det'$. So something seems inconsistent. 

To connect to previous work, let us perform the angular-momentum decomposition as in  Section~\ref{sec:angular}. 
This lets us write the functional determinant as
\be
\sqrt{ \frac{\det \left[ \Box \right] } {\det' \left[ - \Box + V''[\phib]\right]}}
=
\left[{\RW}_0'
({\RW}_1')^4
 \prod_{s\geq 2} [\RW_s]^{(s+1)^2}
 \right]^{-1/2}
 \label{prodrstilde}
\ee
where
\be
\RW_s =
\frac{\det \left[ \Delta_s - 3 \lambda \phib^2 \right]}{\det \left[ \Delta_s \right] } 
\ee
with $\Delta_s$ in Eq.~\eqref{Deltasdef}. 
For $s=0$ there is one mode, the dilatation mode, which has zero eigenvalue, so $\RW_0=0$. For $s=1$ there are four zero modes corresponding to translations. For $s\ge 2$ all the eigenvalues are positive. Removing the zero modes from the numerator,
Ref.~\cite{Isidori:2001bm} found $\RW_0' \approx -1$ and $\RW_1' \approx 0.041$.

First, we look at $s\ge 2$. Here there are no zero modes, so there are no issues with rescaling the operators for these values of $s$.
That is,
\be
\RW_{s\ge 2} 
= \frac{\det \left[ 3 \lambda \phib^2 \right] \cdot \det\left[ \frac{1}{3\lambda \phib^2} \Delta_s -1 \right]}
{ \det \left[  3 \lambda \phib^2 \right] \cdot \det \left[ \frac{1}{3\lambda \phib^2}\Delta_s\right]}
=R_{s\ge 2}
\label{Vout}
\ee
with $R_{s\ge 2}$ in Eq.~\eqref{RSexact}. 

For $s=0$ and $s=1$ there are  zero modes. Since zero modes are still zero modes if the operator is rescaled, we know the explicit form of these modes. 
They  are in Eq.~\eqref{phins} with $n=1$ and $s=0$ or $s=1$
\be
\phi_{10} =  R\frac{R^2-r^2}{(R^2+r^2)^2}, \qquad
\phi_{11} = \frac{R^2r}{2(R^2+r^2)^2}
\ee
It is easy to check that $(\Delta_0  - 3 \lambda \phib^2) \phi_{10}=0$ and $(\Delta_1  -3 \lambda \phib^2) \phi_{11}=0$.
Because of the zero modes, we need to compute
\be
\RW_s' = \frac{\det' \left[ \Delta_s - 3 \lambda \phib^2 \right]}{\det \left[ \Delta_s \right] } = \lim_{\epsilon \to 0}  \frac{1}{\epsilon}\frac{\det' \left[ \Delta_s - 3 \lambda \phib^2  + \epsilon\right]}{\det \left[ \Delta_s \right] } 
\label{eop}
\ee
Note that the zero modes become modes with eigenvalue $\epsilon$ of the shifted operators, so the shifted determinant will be proportional to $\epsilon$ as desired.  

For $s=1$, the zero modes are translations and the Jacobian is finite (Eq.~\eqref{Jnormal}). Thus we expect $\RW_1'$ to be finite too. To compute $\RW_1'$ we can first try the Gelfand-Yaglom method as in Section~\ref{sec:GY}. The Gelfand-Yaglom method requires
us to find a solution to
\be
\left[\Delta_s  - 3\lambda \phib^2+ \epsilon\right] \wt \phi_{1\eps} = 0
\label{shifted}
\ee
that scales like the free solution, $ \wh \phi_1 = r$ near $r=0$ and $r=\infty$. 
Unfortunately, this does not work. At finite $\epsilon$,
$\wt{\phi}_{1\epsilon}$ is oscillatory at large $r$ 
while the free-theory solution $\wh\phi_1 =r$ is not. Thus the two cannot approach each other and the Gelfand-Yaglom method does not seem to give a sensible answer.

To understand the failure of the Gelfand-Yaglom method we note that adding the $\epsilon$ term as in Eq.~\eqref{eop} is equivalent to adding a mass term $\frac{1}{2} \epsilon \phi^2$ to the potential. One would think that a small mass would be a small change in the theory, but it actually has a dramatic effect: it removes all bounce solutions to the equations of motion. Thus the limit $\epsilon \to 0$ is not smooth. One can deal with small masses using the constrained instanton approach described in Section~\ref{sec:mass}, however, there is a simpler way to compute $\RW_1'$. 

Since $\RW_1'$ is supposed to be finite, we can rescale the operator as for $s\ge 2$:
\be
\RW_1' =
\lim_{\epsilon \to 0}
\frac{1}{\epsilon}
 \frac{\det \left[ 3 \lambda \phib^2 \right] \cdot \det\left[ \frac{1}{3\lambda \phib^2} \Delta_1 -1 + \frac{\epsilon}{3\lambda \phib^2}\right]}
{ \det \left[ 3 \lambda \phib^2 \right] \cdot \det \left[ \frac{1}{3\lambda \phib^2}\Delta_s\right]}
 \label{longcomp}
\ee
We can now evaluate these determinants in the basis of $s=1$ modes, $\phi_{n1}$ given in Eq.~\eqref{phins}. 
These functions satisfy
\be
\frac{1}{3\lambda \phib^2}\Delta_1  \phi_{n1} =
\lambda_n  \phi_{n1},
\qquad
\lambda_n =\frac{(n+1)(n+2)}{6}
\label{s1s0}
\ee
as in Eq.~\eqref{lvalsfv} and are normalized as
\be
\int dr r^3  \Big[
(-3\lambda) \phib^2 \phi_{n1}\, \phi_{m1} \Big] = N_{n1} \delta_{nm}, \qquad
N_{n1} = \frac{12(n-1)!}{(2n+3)(n+3)!}
\ee
Then,
\begin{align}
  \det\left[ \frac{1}{3\lambda \phib^2} \Delta_1 -1 + \frac{\epsilon}{3\lambda \phib^2}\right]
 &= \prod_{n\ge 1} \int d r r^3 (3\lambda \phib^2)\phi_{n1}
 \left[ \frac{1}{3\lambda \phib^2}\Delta_1-1 + \frac{\epsilon}{3\lambda \phib^2}
 \right] \phi_{n1}\\
  &=\prod_{n\ge1}\left[N_{n1}(\lambda_n - 1) + \frac{\epsilon R^2}{6n (n+3)} \right]
  \label{Nlam}
\end{align}
In this derivation we have used a property of Legendre polynomials, that
\be
\int dr r^3 \phi_{n1} \phi_{m1} = \frac{R^2}{6 n(n+3)} \delta_{nm} 
\ee
As $\epsilon\to 0$ the first term in Eq.~\eqref{Nlam} always dominants unless $\lambda_n=1$.
Thus the $n=1$ mode contributes $\frac{\varepsilon R^2}{24}$
to the product and we can set $\epsilon=0$ for the other modes. We then find
\be
\RW_1' =R_1'\int dr r^3 \phi_{11}^2=\frac{R^2}{24} R_1'
\ee
This factor of $\frac{1}{24} = 0.041$ matches the result from numerical calculations in~\cite{Isidori:2001bm}.

Now we try the same approach for the $s=0$ modes. In this case, the relevant mode is $\phi_{10}$ in 
Eq.~\eqref{s1s0}, corresponding to dilatations.
Attempting the same calculation as for $s=1$ we find
\be
\RW_0' = R_0' \int dr r^3 \phi_{10}^2 = \infty\cdot   R_0'
\label{R0inf}
\ee
Thus we conclude that the ratio of determinants in Eq.~\eqref{prodrstilde} is zero, 
even after the zero mode is removed. 
This result, while in disagreement with finite numerical extractions in~\cite{Isidori:2001bm,Branchina:2014rva}, is expected from our treatment using rescaled operators in Section~\ref{sec:scale}.

 In conclusion, we find the infinite Jacobian for dilatations found in
previous work to be  compensated by an infinitity of the determinant  {\it after} zero modes are removed.
We therefore find no
inconsistency in the functional determinant calculated with or without rescaling the operators. 
\section{Divergent Graphs in Fermi  Gauge \label{app:Fermi}}
In Fermi gauge, the Lagrangian is given in Eq.~\eqref{LFermi}. We treat all the mass terms as interactions.
 The Feynman rules are
\be
\begin{gathered}
\includegraphics{FeynmanRuleXi1}
\end{gathered}
=- \lambda \widetilde{\phib^2}(q),
~~
\begin{gathered}
\includegraphics{FeynmanRuleXi2}
\end{gathered}
=-g^2 \widetilde{\phib^2}(q) \delta_{\mu\nu},
~~
\begin{gathered}
\includegraphics{FeynmanRuleXi3}
\end{gathered}
=i g \widetilde{\phib}(q)( q^\mu - p^\mu)
\ee
Here, the dashed lines are background fields, sold lines are Goldstones and wavy lines are photons. 
The Fourier transform of the bounce-squared is given in Eq.~\eqref{FTbs}. The Fourier transform of the bounce is
\be
\widetilde{\phib}(q) = \sqrt{\frac{2}{-\lambda}} \frac{8\pi^2 R^2}{q} \cK_1(q R)
\ee

At second order in the interactions, there are 3 divergent loops. One with just Goldstones
\begin{align}
-S_{GG} = 
\begin{gathered}
\includegraphics{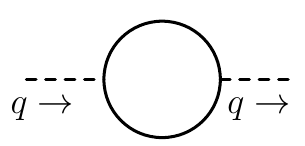}
\end{gathered}
\hspace{-1mm}
&=\frac{1}{4}\lambda^2 \int \frac{d^d q}{(2\pi)^d} \widetilde{\phib^2}(q)   \widetilde{\phib^2}(-q)\int \frac{d^d k}{(2\pi)^d} 
\frac{1}{k^2}\frac{1}{(k+q)^2}\\
&=  \frac{1}{6\eps}  + \frac{5}{18} + \frac{1}{6} \gamma_E + \frac{1}{6}\ln \pi R^2\mu^2
\end{align}
one with just photons
\begin{align}
-S_{AA} = 
\begin{gathered}
\includegraphics{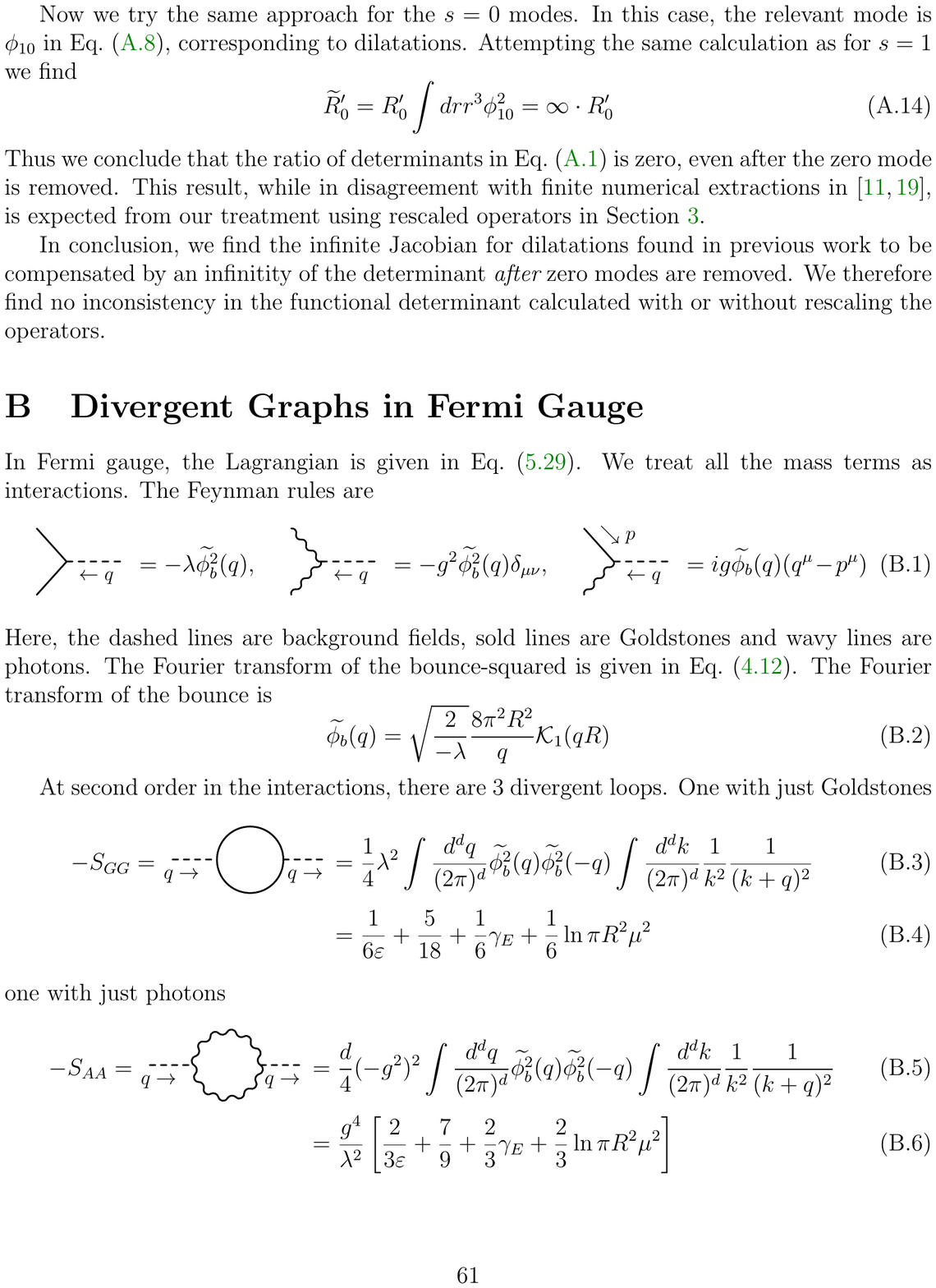}
\end{gathered}
\hspace{-1mm}
&=\frac{d}{4}(-g^2)^2 \int \frac{d^d q}{(2\pi)^d} \widetilde{\phib^2}(q)   \widetilde{\phib^2}(-q)\int \frac{d^d k}{(2\pi)^d} \frac{1}{k^2}\frac{1}{(k+q)^2}\\
&= \frac{g^4}{\lambda^2} \left[ \frac{2}{3\eps}  + \frac{7}{9} + \frac{2}{3} \gamma_E + \frac{2}{3}\ln \pi R^2\mu^2 \right]
\end{align}
and one with Goldstone-photon mixing:
\be
-S_{AG} = 
\begin{gathered}
\includegraphics{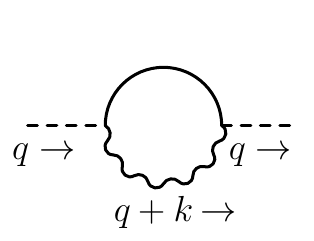}
\end{gathered}
\hspace{-1mm}
=
\frac{1}{2} (ig)^2\int \frac{d^d q}{(2\pi)^d} \widetilde{\phib}(q)   \widetilde{\phib}(-q)\int \frac{d^d k}{(2\pi)^d}
 \frac{(q^\mu-k^\mu)}{k^2}\frac{(-q^\mu + k^\mu)}{(k+q)^2}
 \vspace{3mm}
 \ee
 This graph is quadratically divergent. We can write it as the sum of scaleless integrals and a logarithmically divergent integral by noting that the numerator is $-(q-k)^2=(q+k)^2 -2k^2 - 2 q^2$. Then we evaluate the logarithmically divergent integral using Eq.~\eqref{B0} and
\begin{align}
\int \frac{d^4q}{(2\pi)^4} q^2  \widetilde{\phib}(q)   \widetilde{\phib}(-q)   &= -\frac{32\pi^2}{3\lambda} \\ 
\int \frac{d^4q}{(2\pi)^4} q^2  \widetilde{\phib}(q)   \widetilde{\phib}(-q)  \ln \frac{q^2}{\mu^2}  &= \frac{16\pi^2}{9\lambda} \left[1+12\gamma_E +12 \ln \frac{R\mu}{2}\right]
\label{usefulintegralsApp2}
\end{align}
to find
\be
-S_{A G}= -\frac{g^2}{\lambda} \left[ \frac{2}{3\eps}  + \frac{13}{9} + \frac{2}{3} \gamma_E +\frac{2}{3} \ln \pi R^2\mu^2 \right]
\ee

In Fermi gauge there are additional divergent integrals. At cubic order in couplings, there are two divergent graphs:
\be
-S_{AGG} = 
\begin{gathered}
\includegraphics{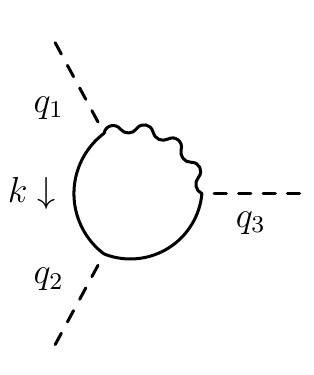}
\end{gathered}
\hspace{-1mm}
=\begin{gathered}
\\
\\
-\frac{1}{2}(i g)^2 \lambda \int\frac{d^d k}{(2\pi)^d} \frac{d^4 q_1 d^4 q_2 d^4 q_3}{(2\pi)^{12}}   (2\pi)^4\delta^4(q_1+q_2+q_3)\\
\times \widetilde{\phib}(q_1) \widetilde{\phib^2}(q_2)\widetilde{\phib}(q_3)  \frac{(q_1^\mu + k^\mu)(q_3^\mu - k^\mu - q_2^\mu)}{k^2 (k+q_2)^2(k-q_1)^2}
\end{gathered}
\ee
This is a difficult graph to evaluate completely. However, the UV divergent part can be extracted relatively simply by taking the leading behavior at large $k$. We get
\be
-S_{AGG} = -\frac{g^2}{\lambda} \frac{1}{3 \eps} + \text{finite} 
\ee
And the other 3-point graph gives
\begin{align}
-S_{AAG} = 
\begin{gathered}
\includegraphics{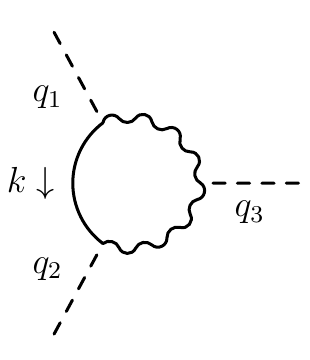}
\end{gathered}
\hspace{-1mm}
&=\begin{gathered}
\\
\\
\frac{1}{2}(i g)^2 (-g^2) \int\frac{d^d k}{(2\pi)^d} \frac{d^4 q_1 d^4 q_2 d^4 q_3}{(2\pi)^{12}}   (2\pi)^4\delta^4(q_1+q_2+q_3)\\
\times \widetilde{\phib}(q_1) \widetilde{\phib}(q_2)\widetilde{\phib^2}(q_3)  \frac{(q_1^\mu + k^\mu)(q_2^\mu - k^\mu)}{k^2 (k+q_2)^2(k-q_1)^2}
\end{gathered}\\
&=-\frac{g^4}{\lambda^2} \frac{1}{3 \eps} + \text{finite} 
\end{align}
Finally, there is a divergent box diagram
\begin{align}
-S_{AGAG} &= 
\begin{gathered}
\includegraphics{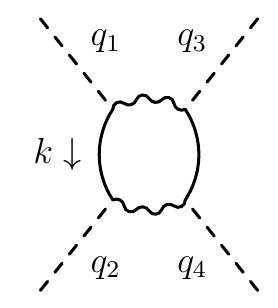}
\end{gathered}
\nonumber
\\
&=
\frac{1}{4}(i g)^4 \int\frac{d^d k}{(2\pi)^d} \frac{d^4 q_1 d^4 q_2 d^4 q_3d^4 q_3}{(2\pi)^{16}}   (2\pi)^4\delta^4(q_1 +q_2+q_3+q_4)
\nonumber
\\[2mm]
&\times \widetilde{\phib}(q_1) \widetilde{\phib}(q_2)\widetilde{\phib}(q_3) \widetilde{\phib}(q_4) 
 \frac{[(q_3^\mu-k^\mu - q_2^\mu - q_4^\mu)(q_1^\mu + k^\mu)][(q_2^\nu - k^\nu)(q_4^\mu+k^\nu +q_2^\nu +q_4^\nu)]}
 {k^2 (k-q_1)^2(k+q_2)^2(k+q_2+q_4)^2}
\nonumber\\
&=\frac{g^4}{\lambda^2} \frac{1}{6 \eps} + \text{finite} 
\end{align}
The sum of all the divergent graphs in Fermi gauges
\be
-S_{GG} - S_{AA} - S_{AGG} - S_{AAG} - S_{AGAG} = \frac{1}{\eps} \left( \frac{1}{6} -\frac{g^2}{\lambda}+ \frac{g^4}{2\lambda^2} \right)
+ \text{finite}
\label{divsfermi}
\ee
These are identical to the UV divergences in $R_\xi$ gauges, in Eq.~\eqref{loopsRxi}. 

\section{NLO effective potential \label{app:vNLO}}
In this Appendix we list the effective potential at NLO according to the consistent power counting developed in~\cite{Andreassen:2014eha,Andreassen:2014gha}. 
The LO effective potential is in Eq.~\eqref{VeffLO}. 
In the consistent power counting of~\cite{Andreassen:2014eha,Andreassen:2014gha}, there are contributions from the one loop effective potential to the NLO effective potential. However, in Landau gauge these vanish. The full two loop effective potential was first computed in~\cite{Ford:1992pn},
and simplified in~\cite{Degrassi:2012ry}. With the consistent power-counting, in Landau gauge, we only need the terms of order $\lambda^0$ in the
two loop potential; these can be found in~\cite{Buttazzo:2013uya}. 
In the consistent power counting, one cannot resum the Higgs field strength with the factor $\Gamma$ as in~\cite{Buttazzo:2013uya};
instead, one must include the single and double logarithmic terms only. The complete NLO effective potential in Landau gauge is therefore
\begin{eqnarray}
V^{\text{NLO}}(h)
 &=& \frac{h^4}{4}\frac{1}{(4\pi)^4}\bigg[ 
          8g_s^2y_t^4 \left(3r_t^2-8  r_t+9\right) +\frac{1}{2} y_t^6 \left(-6 r_t r_W-3 r_t^2+48 r_t-6 r_{tW}-69-\pi ^2\right)
                      \nonumber\\&&+
   \frac{3y_t^2 g^4}{16}    \left(8 r_W+4 r_Z-3  r_t^2-6 r_t r_Z-12 r_t+12 r_{tW}+15+2 \pi ^2\right)
   \nonumber\\&&+
   \frac{y_t^2g'{}^4}{48}
   \left(27 r_t^2-54 r_t  r_Z-68r_t -28 r_Z+189\right)+
   \frac{y_t^2 g^2 g'{}^2 }{8} (9 r_t^2-18 r_t r_Z+4r_t+44 r_Z-57)
      \nonumber\\&&+
      \frac{g^6}{192}  (
    36 r_t r_Z\!+\!54 r_t^2\!-\!414 r_W r_Z\!+\!69 r_W^2\!+\!1264 r_W\!+\!156 r_Z^2\!+\!632 r_Z-\!144r_{tW}-2067\!+\!90 \pi ^2)  
      \nonumber\\&&+
     \frac{ g^4 g'{}^2}{192} (12 r_t r_Z-6 r_t^2-6 r_W (53 r_Z+50)+213 r_W^2+4 r_Z
   (57 r_Z-91)+817+46 \pi ^2)
      \nonumber\\&&+
     \frac{ g^2 g'{}^4}{576} (132 r_t r_Z-66 r_t^2+306 r_W r_Z-153 r_W^2-36 r_W+924 r_Z^2-4080 r_Z+4359+218 \pi ^2)\nonumber
      \\&&+
      \frac{g'{}^6}{576} (6
   r_Z (34 r_t+3 r_W-470)-102 r_t^2-9 r_W^2+708 r_Z^2+2883+206 \pi ^2)
\nonumber       \\&&+
     \frac{   y_t^4 }{6}
\left(4 g'{}^2 (3 r_t^2-8r_t+9)-9 g^2 \left(r_t-r_W+1\right)\right)+
   \frac{3}{4} \left(g^6-3 g^4 y_t^2+4 y_t^6\right)\mathrm{Li}_2\frac{g^2}{2y_t^2} 
         \nonumber\\&&+
      \frac{y_t^2}{48}  \xi(\frac{g^2+g'{}^2}{2 y_t^2} )
         \left(9 g^4
-6 g^2 g'{}^2+17 g'{}^4+2y_t^2 \big(7 g'{}^2-73 g^2+\frac{64 g^4}{g'{}^2+g^2}\big) \right)
   \nonumber\\&&+
    \frac{g^2}{64} \xi(\frac{g^2+g'{}^2}{g^2}) \left(18 g^2 g'{}^2+g'{}^4-51 g^4-\frac{48
   g^6}{g'{}^2+g^2}\right)\nonumber \\&&
   + \frac{1}{48} V^{(1)}_{\text{logs}} \ln \frac{h}{\mu} + V^{(2)}_{\text{logs}} \ln^2 \frac{h}{\mu}
   \bigg]\ . 
\label{large-h-effeP}
\end{eqnarray}
with
\begin{align}
V^{(1)}_{\text{logs}}  =&
-12 g^4 \ln g \Big(-9 {g'}^2+11 g^2+36 y_t^2\Big)+ 2{g'}^6 \Big(-235-91 \ln{8}\Big) 
+192  \Big(4 {g'}^2+48 g_s^2-9 y_t^2\Big) y_t^4 \ln y_t \nonumber \\
&+3 (g'{}^2+g^2) \Big[91 g'{}^4-36 y_t^2 (g'{}^2+g^2)+36 g'{}^2 g^2-11 g^4\Big] \ln (g'{}^2+g^2)\nonumber \\
&-2g'{}^4 \Big[g^2 (343+127 \ln{8})+12 y_t^2 (8-3 \ln{8})\Big]\nonumber \\
&-2g'{}^2 \Big[g^4 (166+43 \ln{8})-72 g^2 y_t^2 (4+\ln{8})+64 y_t^4 (4+\ln{8})\Big]\nonumber \\
&+2g^6 (474+33 \ln{8})+216 g^4 y_t^2 \ln{8}+96 y_t^4 \Big[3 y_t^2 (8+\ln{8})-16 g_s^2 (4+\ln{8})\Big]
\end{align}
and
\begin{multline}
V^{(2)}_{\text{logs}}  
= 8 y_t^4 \left(g'{}^2+12 g_s^2\right)-\frac{9}{4} y_t^2 \left(g'{}^4 +2 g'{}^2 g^2+3 g^4\right) \\
+\frac{1}{16} \left(91 g'{}^6+127 g'{}^4 g^2+43 g'{}^2 g^4-33 g^6\right)
-18 y_t^6
\end{multline}
and where
\be
r_W \equiv \ln\frac{g^2}{4}, \qquad
r_Z \equiv \ln\frac{g^2+g'{}^2}{4},\qquad
r_t \equiv \ln\frac{y_t^2}{2}, \qquad
r_{tW} \equiv \left(r_t-r_W\right)\ln\left[\frac{y_t^2}{2}-\frac{g^2}{2}\right]
\ee
and
\be
\xi(z) \equiv\sqrt{z^2-4z}\left[2\ln^2\frac{z-\sqrt{z^2-4z}}{2z}-\ln^2 z-4\text{Li}_2\left(\frac{z-\sqrt{z^2-4 z}}{2 z}\right)+\frac{\pi^2}{3}\right]
\ee

\bibliography{NLOscale}

\providecommand{\href}[2]{#2}\begingroup\raggedright\begin{thebibliography}{100}

\bibitem{Langer:1967ax}
J.~S. Langer, ``{Theory of the condensation point},''
  \href{http://dx.doi.org/10.1016/0003-4916(67)90200-X}{{\em Annals Phys.}
  {\bfseries 41} (1967) 108--157}.
[Annals Phys.281,941(2000)].

\bibitem{Kobzarev:1974cp}
I.~Kobzarev, L.~B. Okun, and M.~B. Voloshin, ``{Bubbles in Metastable
  Vacuum},'' {\em Sov. J. Nucl. Phys.} {\bfseries 20} (1975) 644--646.
[Yad. Fiz.20,1229(1974)].

\bibitem{Coleman:1977py}
S.~R. Coleman, ``{The Fate of the False Vacuum. 1. Semiclassical Theory},''
  \href{http://dx.doi.org/10.1103/PhysRevD.15.2929,
  10.1103/PhysRevD.16.1248}{{\em Phys. Rev.} {\bfseries D15} (1977)
  2929--2936}.
[Erratum: Phys. Rev.D16,1248(1977)].

\bibitem{Callan:1977pt}
C.~G. Callan, Jr. and S.~R. Coleman, ``{The Fate of the False Vacuum. 2. First
  Quantum Corrections},''
\href{http://dx.doi.org/10.1103/PhysRevD.16.1762}{{\em Phys. Rev.} {\bfseries
  D16} (1977) 1762--1768}.

\bibitem{Hammer:1978xu}
C.~L. Hammer, J.~E. Shrauner, and B.~DeFacio, ``{Alternate Derivation of Vacuum
  Tunneling},''
\href{http://dx.doi.org/10.1103/PhysRevD.19.667}{{\em Phys. Rev.} {\bfseries
  D19} (1979) 667}.

\bibitem{Frampton:1976kf}
P.~H. Frampton, ``{Vacuum Instability and Higgs Scalar Mass},''
  \href{http://dx.doi.org/10.1103/PhysRevLett.37.1378}{{\em Phys. Rev. Lett.}
  {\bfseries 37} (1976) 1378}.
[Erratum: Phys. Rev. Lett.37,1716(1976)].

\bibitem{Frampton:1976pb}
P.~H. Frampton, ``{Consequences of Vacuum Instability in Quantum Field
  Theory},''
\href{http://dx.doi.org/10.1103/PhysRevD.15.2922}{{\em Phys. Rev.} {\bfseries
  D15} (1977) 2922}.

\bibitem{Sher:1988mj}
M.~Sher, ``{Electroweak Higgs Potentials and Vacuum Stability},''
\href{http://dx.doi.org/10.1016/0370-1573(89)90061-6}{{\em Phys.Rept.}
  {\bfseries 179} (1989) 273--418}.

\bibitem{Casas:1994qy}
J.~Casas, J.~Espinosa, and M.~Quiros, ``{Improved Higgs mass stability bound in
  the standard model and implications for supersymmetry},''
  \href{http://dx.doi.org/10.1016/0370-2693(94)01404-Z}{{\em Phys.Lett.}
  {\bfseries B342} (1995) 171--179},
\href{http://arxiv.org/abs/hep-ph/9409458}{{\ttfamily arXiv:hep-ph/9409458
  [hep-ph]}}.

\bibitem{Espinosa:1995se}
J.~Espinosa and M.~Quiros, ``{Improved metastability bounds on the standard
  model Higgs mass},''
  \href{http://dx.doi.org/10.1016/0370-2693(95)00572-3}{{\em Phys.Lett.}
  {\bfseries B353} (1995) 257--266},
\href{http://arxiv.org/abs/hep-ph/9504241}{{\ttfamily arXiv:hep-ph/9504241
  [hep-ph]}}.

\bibitem{Isidori:2001bm}
G.~Isidori, G.~Ridolfi, and A.~Strumia, ``{On the metastability of the standard
  model vacuum},'' \href{http://dx.doi.org/10.1016/S0550-3213(01)00302-9}{{\em
  Nucl.Phys.} {\bfseries B609} (2001) 387--409},
\href{http://arxiv.org/abs/hep-ph/0104016}{{\ttfamily arXiv:hep-ph/0104016
  [hep-ph]}}.

\bibitem{Espinosa:2007qp}
J.~Espinosa, G.~Giudice, and A.~Riotto, ``{Cosmological implications of the
  Higgs mass measurement},''
  \href{http://dx.doi.org/10.1088/1475-7516/2008/05/002}{{\em JCAP} {\bfseries
  0805} (2008) 002},
\href{http://arxiv.org/abs/0710.2484}{{\ttfamily arXiv:0710.2484 [hep-ph]}}.

\bibitem{Ellis:2009tp}
J.~Ellis, J.~Espinosa, G.~Giudice, A.~Hoecker, and A.~Riotto, ``{The Probable
  Fate of the Standard Model},''
  \href{http://dx.doi.org/10.1016/j.physletb.2009.07.054}{{\em Phys.Lett.}
  {\bfseries B679} (2009) 369--375},
\href{http://arxiv.org/abs/0906.0954}{{\ttfamily arXiv:0906.0954 [hep-ph]}}.

\bibitem{Degrassi:2012ry}
G.~Degrassi, S.~Di~Vita, J.~Elias-Miro, J.~R. Espinosa, G.~F. Giudice,
  G.~Isidori, and A.~Strumia, ``{Higgs mass and vacuum stability in the
  Standard Model at NNLO},''
  \href{http://dx.doi.org/10.1007/JHEP08(2012)098}{{\em JHEP} {\bfseries 08}
  (2012) 098},
\href{http://arxiv.org/abs/1205.6497}{{\ttfamily arXiv:1205.6497 [hep-ph]}}.

\bibitem{Buttazzo:2013uya}
D.~Buttazzo, G.~Degrassi, P.~P. Giardino, G.~F. Giudice, F.~Sala, A.~Salvio,
  and A.~Strumia, ``{Investigating the near-criticality of the Higgs boson},''
  \href{http://dx.doi.org/10.1007/JHEP12(2013)089}{{\em JHEP} {\bfseries 12}
  (2013) 089},
\href{http://arxiv.org/abs/1307.3536}{{\ttfamily arXiv:1307.3536 [hep-ph]}}.

\bibitem{Lalak:2014qua}
Z.~Lalak, M.~Lewicki, and P.~Olszewski, ``{Higher-order scalar interactions and
  SM vacuum stability},'' \href{http://dx.doi.org/10.1007/JHEP05(2014)119}{{\em
  JHEP} {\bfseries 05} (2014) 119},
\href{http://arxiv.org/abs/1402.3826}{{\ttfamily arXiv:1402.3826 [hep-ph]}}.

\bibitem{Andreassen:2014gha}
A.~Andreassen, W.~Frost, and M.~D. Schwartz, ``{Consistent Use of the Standard
  Model Effective Potential},''
  \href{http://dx.doi.org/10.1103/PhysRevLett.113.241801}{{\em Phys. Rev.
  Lett.} {\bfseries 113} no.~24, (2014) 241801},
\href{http://arxiv.org/abs/1408.0292}{{\ttfamily arXiv:1408.0292 [hep-ph]}}.

\bibitem{Bednyakov:2015sca}
A.~V. Bednyakov, B.~A. Kniehl, A.~F. Pikelner, and O.~L. Veretin, ``{Stability
  of the Electroweak Vacuum: Gauge Independence and Advanced Precision},''
  \href{http://dx.doi.org/10.1103/PhysRevLett.115.201802}{{\em Phys. Rev.
  Lett.} {\bfseries 115} no.~20, (2015) 201802},
\href{http://arxiv.org/abs/1507.08833}{{\ttfamily arXiv:1507.08833 [hep-ph]}}.

\bibitem{Branchina:2014rva}
V.~Branchina, E.~Messina, and M.~Sher, ``{Lifetime of the electroweak vacuum
  and sensitivity to Planck scale physics},''
  \href{http://dx.doi.org/10.1103/PhysRevD.91.013003}{{\em Phys. Rev.}
  {\bfseries D91} (2015) 013003},
\href{http://arxiv.org/abs/1408.5302}{{\ttfamily arXiv:1408.5302 [hep-ph]}}.

\bibitem{Iacobellis:2016eof}
G.~Iacobellis and I.~Masina, ``{Stationary configurations of the Standard Model
  Higgs potential: electroweak stability and rising inflection point},''
  \href{http://dx.doi.org/10.1103/PhysRevD.94.073005}{{\em Phys. Rev.}
  {\bfseries D94} no.~7, (2016) 073005},
\href{http://arxiv.org/abs/1604.06046}{{\ttfamily arXiv:1604.06046 [hep-ph]}}.

\bibitem{Nielsen:1975fs}
N.~Nielsen, ``{On the Gauge Dependence of Spontaneous Symmetry Breaking in
  Gauge Theories},''
\href{http://dx.doi.org/10.1016/0550-3213(75)90301-6}{{\em Nucl.Phys.}
  {\bfseries B101} (1975) 173}.

\bibitem{Fukuda:1975di}
R.~Fukuda and T.~Kugo, ``{Gauge Invariance in the Effective Action and
  Potential},''
\href{http://dx.doi.org/10.1103/PhysRevD.13.3469}{{\em Phys.Rev.} {\bfseries
  D13} (1976) 3469}.

\bibitem{Kang:1974yj}
J.~Kang, ``{Gauge Invariance of the Scalar-Vector Mass Ratio in the
  Coleman-Weinberg Model},''
\href{http://dx.doi.org/10.1103/PhysRevD.10.3455}{{\em Phys.Rev.} {\bfseries
  D10} (1974) 3455}.

\bibitem{Dolan:1974gu}
L.~Dolan and R.~Jackiw, ``{Gauge Invariant Signal for Gauge Symmetry
  Breaking},''
\href{http://dx.doi.org/10.1103/PhysRevD.9.2904}{{\em Phys.Rev.} {\bfseries D9}
  (1974) 2904}.

\bibitem{Frere:1974ia}
J.-M. Frere and P.~Nicoletopoulos, ``{Gauge Invariant Content of the Effective
  Potential},''
\href{http://dx.doi.org/10.1103/PhysRevD.11.2332}{{\em Phys.Rev.} {\bfseries
  D11} (1975) 2332}.

\bibitem{Kunstatter:1985yt}
G.~Kunstatter and H.~Leivo, ``{On the Gauge Dependence of the One Loop
  Effective Potential in Selfconsistent Dimensional Reduction},''
\href{http://dx.doi.org/10.1016/0370-2693(86)90808-7}{{\em Phys.Lett.}
  {\bfseries B166} (1986) 321}.

\bibitem{DoNascimento:1987mn}
J.~Do~Nascimento and D.~Bazeia, ``{Gauge Invariance of the Effective
  Potential},''
\href{http://dx.doi.org/10.1103/PhysRevD.35.2490}{{\em Phys.Rev.} {\bfseries
  D35} (1987) 2490--2494}.

\bibitem{Buchmuller:1994vy}
W.~Buchmuller, Z.~Fodor, and A.~Hebecker, ``{Gauge invariant treatment of the
  electroweak phase transition},''
  \href{http://dx.doi.org/10.1016/0370-2693(94)90953-9}{{\em Phys.Lett.}
  {\bfseries B331} (1994) 131--136},
\href{http://arxiv.org/abs/hep-ph/9403391}{{\ttfamily arXiv:hep-ph/9403391
  [hep-ph]}}.

\bibitem{Metaxas:1995ab}
D.~Metaxas and E.~J. Weinberg, ``{Gauge independence of the bubble nucleation
  rate in theories with radiative symmetry breaking},''
  \href{http://dx.doi.org/10.1103/PhysRevD.53.836}{{\em Phys.Rev.} {\bfseries
  D53} (1996) 836--843},
\href{http://arxiv.org/abs/hep-ph/9507381}{{\ttfamily arXiv:hep-ph/9507381
  [hep-ph]}}.

\bibitem{Boyanovsky:1996dc}
D.~Boyanovsky, D.~Brahm, R.~Holman, and D.~Lee, ``{The Gauge invariant
  effective potential: Equilibrium and nonequilibrium aspects},''
  \href{http://dx.doi.org/10.1103/PhysRevD.54.1763}{{\em Phys.Rev.} {\bfseries
  D54} (1996) 1763--1777},
\href{http://arxiv.org/abs/hep-ph/9603337}{{\ttfamily arXiv:hep-ph/9603337
  [hep-ph]}}.

\bibitem{Lin:1998up}
G.-L. Lin and T.-K. Chyi, ``{Gauge independent effective potential and the
  Higgs mass bound},'' \href{http://dx.doi.org/10.1103/PhysRevD.60.016002}{{\em
  Phys.Rev.} {\bfseries D60} (1999) 016002},
\href{http://arxiv.org/abs/hep-ph/9811213}{{\ttfamily arXiv:hep-ph/9811213
  [hep-ph]}}.

\bibitem{Binosi:2005yk}
D.~Binosi, J.~Papavassiliou, and A.~Pilaftsis, ``{Displacement operator
  formalism for renormalization and gauge dependence to all orders},''
  \href{http://dx.doi.org/10.1103/PhysRevD.71.085007}{{\em Phys.Rev.}
  {\bfseries D71} (2005) 085007},
\href{http://arxiv.org/abs/hep-ph/0501259}{{\ttfamily arXiv:hep-ph/0501259
  [hep-ph]}}.

\bibitem{Wainwright:2011qy}
C.~Wainwright, S.~Profumo, and M.~J. Ramsey-Musolf, ``{Gravity Waves from a
  Cosmological Phase Transition: Gauge Artifacts and Daisy Resummations},''
  \href{http://dx.doi.org/10.1103/PhysRevD.84.023521}{{\em Phys.Rev.}
  {\bfseries D84} (2011) 023521},
\href{http://arxiv.org/abs/1104.5487}{{\ttfamily arXiv:1104.5487 [hep-ph]}}.

\bibitem{Patel:2011th}
H.~H. Patel and M.~J. Ramsey-Musolf, ``{Baryon Washout, Electroweak Phase
  Transition, and Perturbation Theory},''
  \href{http://dx.doi.org/10.1007/JHEP07(2011)029}{{\em JHEP} {\bfseries 1107}
  (2011) 029},
\href{http://arxiv.org/abs/1101.4665}{{\ttfamily arXiv:1101.4665 [hep-ph]}}.

\bibitem{Garny:2012cg}
M.~Garny and T.~Konstandin, ``{On the gauge dependence of vacuum transitions at
  finite temperature},'' \href{http://dx.doi.org/10.1007/JHEP07(2012)189}{{\em
  JHEP} {\bfseries 1207} (2012) 189},
\href{http://arxiv.org/abs/1205.3392}{{\ttfamily arXiv:1205.3392 [hep-ph]}}.

\bibitem{Lalak:2016zlv}
Z.~Lalak, M.~Lewicki, and P.~Olszewski, ``{Gauge fixing and renormalization
  scale independence of tunneling rate in Abelian Higgs model and in the
  standard model},'' \href{http://dx.doi.org/10.1103/PhysRevD.94.085028}{{\em
  Phys. Rev.} {\bfseries D94} no.~8, (2016) 085028},
\href{http://arxiv.org/abs/1605.06713}{{\ttfamily arXiv:1605.06713 [hep-ph]}}.

\bibitem{Andreassen:2013hpa}
A.~Andreassen, ``{Gauge Dependence of the Quantum Field Theory Effective
  Potential},''
{\em master's thesis, NTNU-Trondheim} (2013) .

\bibitem{Andreassen:2014eha}
A.~Andreassen, W.~Frost, and M.~D. Schwartz, ``{Consistent Use of Effective
  Potentials},'' \href{http://dx.doi.org/10.1103/PhysRevD.91.016009}{{\em Phys.
  Rev.} {\bfseries D91} no.~1, (2015) 016009},
\href{http://arxiv.org/abs/1408.0287}{{\ttfamily arXiv:1408.0287 [hep-ph]}}.

\bibitem{Endo:2017gal}
M.~Endo, T.~Moroi, M.~M. Nojiri, and Y.~Shoji, ``{On the Gauge Invariance of
  the Decay Rate of False Vacuum},''
  \href{http://dx.doi.org/10.1016/j.physletb.2017.05.057}{{\em Phys. Lett.}
  {\bfseries B771} (2017) 281--287},
\href{http://arxiv.org/abs/1703.09304}{{\ttfamily arXiv:1703.09304 [hep-ph]}}.

\bibitem{Endo:2017tsz}
M.~Endo, T.~Moroi, M.~M. Nojiri, and Y.~Shoji, ``{False Vacuum Decay in Gauge
  Theory},''
\href{http://arxiv.org/abs/1704.03492}{{\ttfamily arXiv:1704.03492 [hep-ph]}}.

\bibitem{Kleinert}
H.~Kleinert, {\em Path Integrals in Quantum Mechanics, Statistics, Polymer
  Physics, and Financial Markets}.
\newblock World Scientific Publishing Co, 5th~ed., 2009.

\bibitem{Muller-Kirsten:2012wla}
H.~J.~W. Müller-Kirsten, {\em {Introduction to Quantum Mechanics}}.
\newblock World Scientific, 2012.
\newblock
\url{http://www.worldscientific.com/worldscibooks/10.1142/8428}.
\newblock

\bibitem{Zinn-JustinPI}
J.~Zinn-Justin, {\em Path Integrals in Quantum Mechanics}.
\newblock Oxford University Press, 2005.

\bibitem{ZinnJustin:2002ru}
J.~Zinn-Justin, ``{Quantum field theory and critical phenomena},''
{\em Int. Ser. Monogr. Phys.} {\bfseries 113} (2002) 1--1054.

\bibitem{Marino}
M.~Marino, {\em Instantons and Large N: An Introduction to Non-Perturbative
  Methods in Quantum Field Theory}.
\newblock Cambridge University Press, 2015.

\bibitem{Weinberg:2012pjx}
E.~J. Weinberg, {\em {Classical solutions in quantum field theory}}.
\newblock Cambridge Monographs on Mathematical Physics. Cambridge University
  Press, 2012.
\newblock
\url{http://www.cambridge.org/us/knowledge/isbn/item6813336/}.
\newblock

\bibitem{Behtash:2015zha}
A.~Behtash, G.~V. Dunne, T.~Schaefer, T.~Sulejmanpasic, and M.~Unsal,
  ``{Complexified path integrals, exact saddles and supersymmetry},''
\href{http://arxiv.org/abs/1510.00978}{{\ttfamily arXiv:1510.00978 [hep-th]}}.

\bibitem{Witten:2010cx}
E.~Witten, ``{Analytic Continuation Of Chern-Simons Theory},'' {\em AMS/IP
  Stud. Adv. Math.} {\bfseries 50} (2011) 347--446,
\href{http://arxiv.org/abs/1001.2933}{{\ttfamily arXiv:1001.2933 [hep-th]}}.

\bibitem{Andreassen:2016cff}
A.~Andreassen, D.~Farhi, W.~Frost, and M.~D. Schwartz, ``{Direct Approach to
  Quantum Tunneling},''
  \href{http://dx.doi.org/10.1103/PhysRevLett.117.231601}{{\em Phys. Rev.
  Lett.} {\bfseries 117} no.~23, (2016) 231601},
\href{http://arxiv.org/abs/1602.01102}{{\ttfamily arXiv:1602.01102 [hep-th]}}.

\bibitem{Andreassen:2016cvx}
A.~Andreassen, D.~Farhi, W.~Frost, and M.~D. Schwartz, ``{Precision decay rate
  calculations in quantum field theory},''
  \href{http://dx.doi.org/10.1103/PhysRevD.95.085011}{{\em Phys. Rev.}
  {\bfseries D95} no.~8, (2017) 085011},
\href{http://arxiv.org/abs/1604.06090}{{\ttfamily arXiv:1604.06090 [hep-th]}}.

\bibitem{Coleman:1973jx}
S.~R. Coleman and E.~J. Weinberg, ``{Radiative Corrections as the Origin of
  Spontaneous Symmetry Breaking},''
\href{http://dx.doi.org/10.1103/PhysRevD.7.1888}{{\em Phys.Rev.} {\bfseries D7}
  (1973) 1888--1910}.

\bibitem{Gervais:1974dc}
J.-L. Gervais and B.~Sakita, ``{Extended Particles in Quantum Field
  Theories},''
\href{http://dx.doi.org/10.1103/PhysRevD.11.2943}{{\em Phys. Rev.} {\bfseries
  D11} (1975) 2943}.

\bibitem{Gervais:1975pa}
J.-L. Gervais, A.~Jevicki, and B.~Sakita, ``{Perturbation Expansion Around
  Extended Particle States in Quantum Field Theory. 1.},''
\href{http://dx.doi.org/10.1103/PhysRevD.12.1038}{{\em Phys. Rev.} {\bfseries
  D12} (1975) 1038}.

\bibitem{Callan:1975yy}
C.~G. Callan, Jr. and D.~J. Gross, ``{Quantum Perturbation Theory of
  Solitons},''
\href{http://dx.doi.org/10.1016/0550-3213(75)90150-9}{{\em Nucl. Phys.}
  {\bfseries B93} (1975) 29}.

\bibitem{Jevicki:1976kd}
A.~Jevicki, ``{Treatment of Zero Frequency Modes in Perturbation Expansion
  About Classical Field Configurations},''
\href{http://dx.doi.org/10.1016/0550-3213(76)90403-X}{{\em Nucl. Phys.}
  {\bfseries B117} (1976) 365}.

\bibitem{Kusenko:1996bv}
A.~Kusenko, K.-M. Lee, and E.~J. Weinberg, ``{Vacuum decay and internal
  symmetries},'' \href{http://dx.doi.org/10.1103/PhysRevD.55.4903}{{\em Phys.
  Rev.} {\bfseries D55} (1997) 4903--4909},
\href{http://arxiv.org/abs/hep-th/9609100}{{\ttfamily arXiv:hep-th/9609100
  [hep-th]}}.

\bibitem{Affleck:1980mp}
I.~Affleck, ``{On Constrained Instantons},''
\href{http://dx.doi.org/10.1016/0550-3213(81)90307-2}{{\em Nucl. Phys.}
  {\bfseries B191} (1981) 429}.

\bibitem{Glerfoss:2005si}
P.~M. Glerfoss and N.~K. Nielsen, ``{Instanton constraints and
  renormalization},'' \href{http://dx.doi.org/10.1016/j.aop.2005.05.001}{{\em
  Annals Phys.} {\bfseries 321} (2006) 331--354},
\href{http://arxiv.org/abs/hep-th/0504178}{{\ttfamily arXiv:hep-th/0504178
  [hep-th]}}.

\bibitem{Nielsen:1999vq}
M.~Nielsen and N.~K. Nielsen, ``{Explicit construction of constrained
  instantons},'' \href{http://dx.doi.org/10.1103/PhysRevD.61.105020}{{\em Phys.
  Rev.} {\bfseries D61} (2000) 105020},
\href{http://arxiv.org/abs/hep-th/9912006}{{\ttfamily arXiv:hep-th/9912006
  [hep-th]}}.

\bibitem{Frishman:1978xs}
Y.~Frishman and S.~Yankielowicz, ``{Large Order Behavior of Perturbation Theory
  and Mass Terms},''
\href{http://dx.doi.org/10.1103/PhysRevD.19.540}{{\em Phys. Rev.} {\bfseries
  D19} (1979) 540}.

\bibitem{tHooft:1976snw}
G.~'t~Hooft, ``{Computation of the Quantum Effects Due to a Four-Dimensional
  Pseudoparticle},'' \href{http://dx.doi.org/10.1103/PhysRevD.18.2199.3,
  10.1103/PhysRevD.14.3432}{{\em Phys. Rev.} {\bfseries D14} (1976)
  3432--3450}.
[Erratum: Phys. Rev.D18,2199(1978)].

\bibitem{DiLuzio:2015iua}
L.~Di~Luzio, G.~Isidori, and G.~Ridolfi, ``{Stability of the electroweak ground
  state in the Standard Model and its extensions},''
  \href{http://dx.doi.org/10.1016/j.physletb.2015.12.009}{{\em Phys. Lett.}
  {\bfseries B753} (2016) 150--160},
\href{http://arxiv.org/abs/1509.05028}{{\ttfamily arXiv:1509.05028 [hep-ph]}}.

\bibitem{McKane:1978md}
A.~J. McKane and D.~J. Wallace, ``{Instanton calculations using dimensional
  regularization},''
\href{http://dx.doi.org/10.1088/0305-4470/11/11/013}{{\em J. Phys.} {\bfseries
  A11} (1978) 2285}.

\bibitem{Drummond:1978pf}
I.~T. Drummond and G.~M. Shore, ``{Dimensional Regularization and Instantons: A
  Scalar Field Theory Model},''
\href{http://dx.doi.org/10.1016/0003-4916(79)90097-6}{{\em Annals Phys.}
  {\bfseries 121} (1979) 204}.

\bibitem{Shore:1978eq}
G.~M. Shore, ``{Dimensional Regularization and Instantons},''
\href{http://dx.doi.org/10.1016/0003-4916(79)90206-9}{{\em Annals Phys.}
  {\bfseries 122} (1979) 321}.

\bibitem{Coleman:1978ae}
S.~R. Coleman, ``{Fate of the false vacuum},''
{\em Subnucl.Ser.} {\bfseries 15} (1979) 805.

\bibitem{Weinberg:1996kr}
S.~Weinberg, {\em {The quantum theory of fields. Vol. 2: Modern applications}}.
\newblock Cambridge University Press,
2013.
\newblock

\bibitem{Gamow:1928}
G.~Gamow, ``Quantum theory of the atomic nucleus,'' {\em Z. Phys.} {\bfseries
  51} (1928) 204.

\bibitem{Banks:1973ps}
T.~Banks, C.~M. Bender, and T.~T. Wu, ``{Coupled anharmonic oscillators. 1.
  Equal mass case},''
\href{http://dx.doi.org/10.1103/PhysRevD.8.3346}{{\em Phys.Rev.} {\bfseries D8}
  (1973) 3346--3378}.

\bibitem{Collins:1977dw}
J.~C. Collins and D.~E. Soper, ``{Large Order Expansion in Perturbation
  Theory},''
\href{http://dx.doi.org/10.1016/0003-4916(78)90084-2}{{\em Annals Phys.}
  {\bfseries 112} (1978) 209--234}.

\bibitem{Bender:1984jc}
C.~M. Bender, F.~Cooper, B.~Freedman, and R.~W. Haymaker, ``{Tunneling and the
  low momentum expansion of the effective action},''
\href{http://dx.doi.org/10.1016/0550-3213(85)90413-4}{{\em Nucl. Phys.}
  {\bfseries B256} (1985) 653}.

\bibitem{Bender:1990pd}
C.~M. Bender and T.~T. Wu, ``{Anharmonic oscillator. 2: A Study of perturbation
  theory in large order},''
\href{http://dx.doi.org/10.1103/PhysRevD.7.1620}{{\em Phys. Rev.} {\bfseries
  D7} (1973) 1620--1636}.

\bibitem{Jentschura:2010zza}
U.~D. Jentschura, A.~Surzhykov, and J.~Zinn-Justin, ``{Multi-instantons and
  exact results III: Unification of even and odd anharmonic oscillators},''
  \href{http://dx.doi.org/doi:10.1016/j.aop.2010.01.002}{{\em Annals Phys.}
  {\bfseries 325} (2010) 1135--1172}.

\bibitem{Jentschura:2011zza}
U.~D. Jentschura and J.~Zinn-Justin, ``{Multi-instantons and exact results. IV:
  Path integral formalism},''
\href{http://dx.doi.org/10.1016/j.aop.2011.04.002}{{\em Annals Phys.}
  {\bfseries 326} (2011) 2186--2242}.

\bibitem{Coleman:1987rm}
S.~R. Coleman, ``{Quantum Tunneling and Negative Eigenvalues},''
\href{http://dx.doi.org/10.1016/0550-3213(88)90308-2}{{\em Nucl. Phys.}
  {\bfseries B298} (1988) 178}.

\bibitem{Fubini:1976jm}
S.~Fubini, ``{A New Approach to Conformal Invariant Field Theories},''
\href{http://dx.doi.org/10.1007/BF02785664}{{\em Nuovo Cim.} {\bfseries A34}
  (1976) 521}.

\bibitem{Lipatov:1976ny}
L.~N. Lipatov, ``{Divergence of the Perturbation Theory Series and the
  Quasiclassical Theory},'' {\em Sov. Phys. JETP} {\bfseries 45} (1977)
  216--223.
[Zh. Eksp. Teor. Fiz.72,411(1977)].

\bibitem{Gelfand:1959nq}
I.~M. Gelfand and A.~M. Yaglom, ``{Integration in functional spaces and it
  applications in quantum physics},''
\href{http://dx.doi.org/10.1063/1.1703636}{{\em J. Math. Phys.} {\bfseries 1}
  (1960) 48}.

\bibitem{Kirsten:2004qv}
K.~Kirsten and A.~J. McKane, ``{Functional determinants for general
  Sturm-Liouville problems},''
  \href{http://dx.doi.org/10.1088/0305-4470/37/16/014}{{\em J. Phys.}
  {\bfseries A37} (2004) 4649--4670},
\href{http://arxiv.org/abs/math-ph/0403050}{{\ttfamily arXiv:math-ph/0403050
  [math-ph]}}.

\bibitem{Kirsten:2003py}
K.~Kirsten and A.~J. McKane, ``{Functional determinants by contour integration
  methods},'' \href{http://dx.doi.org/10.1016/S0003-4916(03)00149-0}{{\em
  Annals Phys.} {\bfseries 308} (2003) 502--527},
\href{http://arxiv.org/abs/math-ph/0305010}{{\ttfamily arXiv:math-ph/0305010
  [math-ph]}}.

\bibitem{Dunne:2005rt}
G.~V. Dunne and H.~Min, ``{Beyond the thin-wall approximation: Precise
  numerical computation of prefactors in false vacuum decay},''
  \href{http://dx.doi.org/10.1103/PhysRevD.72.125004}{{\em Phys. Rev.}
  {\bfseries D72} (2005) 125004},
\href{http://arxiv.org/abs/hep-th/0511156}{{\ttfamily arXiv:hep-th/0511156
  [hep-th]}}.

\bibitem{Avan:1985eg}
J.~Avan and H.~J. De~Vega, ``{Inverse scattering transform and instantons in
  four-dimensional Yukawa and phi**4 thoeories},''
\href{http://dx.doi.org/10.1016/0550-3213(86)90515-8}{{\em Nucl. Phys.}
  {\bfseries B269} (1986) 621--664}.

\bibitem{Bezrukov:2012sa}
F.~Bezrukov, M.~{\relax Yu}. Kalmykov, B.~A. Kniehl, and M.~Shaposhnikov,
  ``{Higgs Boson Mass and New Physics},''
  \href{http://dx.doi.org/10.1007/JHEP10(2012)140}{{\em JHEP} {\bfseries 10}
  (2012) 140}, \href{http://arxiv.org/abs/1205.2893}{{\ttfamily arXiv:1205.2893
  [hep-ph]}}.
[,275(2012)].

\bibitem{Weinberg:1992ds}
E.~J. Weinberg, ``{Vacuum decay in theories with symmetry breaking by radiative
  corrections},'' \href{http://dx.doi.org/10.1103/PhysRevD.47.4614}{{\em
  Phys.Rev.} {\bfseries D47} (1993) 4614--4627},
\href{http://arxiv.org/abs/hep-ph/9211314}{{\ttfamily arXiv:hep-ph/9211314
  [hep-ph]}}.

\bibitem{Branchina:2013jra}
V.~Branchina and E.~Messina, ``{Stability, Higgs Boson Mass and New Physics},''
  \href{http://dx.doi.org/10.1103/PhysRevLett.111.241801}{{\em Phys.Rev.Lett.}
  {\bfseries 111} (2013) 241801},
\href{http://arxiv.org/abs/1307.5193}{{\ttfamily arXiv:1307.5193 [hep-ph]}}.

\bibitem{Branchina:2014efa}
V.~Branchina, ``{Stability of the EW vacuum, Higgs boson, and new physics},''
\href{http://arxiv.org/abs/1405.7864}{{\ttfamily arXiv:1405.7864 [hep-ph]}}.

\bibitem{Branchina:2014usa}
V.~Branchina, E.~Messina, and A.~Platania, ``{Top mass determination, Higgs
  inflation, and vacuum stability},''
  \href{http://dx.doi.org/10.1007/JHEP09(2014)182}{{\em JHEP} {\bfseries 1409}
  (2014) 182},
\href{http://arxiv.org/abs/1407.4112}{{\ttfamily arXiv:1407.4112 [hep-ph]}}.

\bibitem{Chetyrkin:1997sg}
K.~G. Chetyrkin, B.~A. Kniehl, and M.~Steinhauser, ``{Strong coupling constant
  with flavor thresholds at four loops in the MS scheme},''
  \href{http://dx.doi.org/10.1103/PhysRevLett.79.2184}{{\em Phys. Rev. Lett.}
  {\bfseries 79} (1997) 2184--2187},
\href{http://arxiv.org/abs/hep-ph/9706430}{{\ttfamily arXiv:hep-ph/9706430
  [hep-ph]}}.

\bibitem{Chetyrkin:2004mf}
K.~G. Chetyrkin, ``{Four-loop renormalization of QCD: Full set of
  renormalization constants and anomalous dimensions},''
  \href{http://dx.doi.org/10.1016/j.nuclphysb.2005.01.011}{{\em Nucl. Phys.}
  {\bfseries B710} (2005) 499--510},
\href{http://arxiv.org/abs/hep-ph/0405193}{{\ttfamily arXiv:hep-ph/0405193
  [hep-ph]}}.

\bibitem{Chetyrkin:2012rz}
K.~G. Chetyrkin and M.~F. Zoller, ``{Three-loop beta-functions for top-Yukawa
  and the Higgs self-interaction in the Standard Model},''
  \href{http://dx.doi.org/10.1007/JHEP06(2012)033}{{\em JHEP} {\bfseries 06}
  (2012) 033},
\href{http://arxiv.org/abs/1205.2892}{{\ttfamily arXiv:1205.2892 [hep-ph]}}.

\bibitem{Kniehl:2016enc}
B.~A. Kniehl, A.~F. Pikelner, and O.~L. Veretin, ``{mr: a C++ library for the
  matching and running of the Standard Model parameters},''
  \href{http://dx.doi.org/10.1016/j.cpc.2016.04.017}{{\em Comput. Phys.
  Commun.} {\bfseries 206} (2016) 84--96},
\href{http://arxiv.org/abs/1601.08143}{{\ttfamily arXiv:1601.08143 [hep-ph]}}.

\bibitem{Olive:2016xmw}
{\bfseries Particle Data Group} Collaboration, C.~Patrignani {\em et~al.},
  ``{Review of Particle Physics},''
\href{http://dx.doi.org/10.1088/1674-1137/40/10/100001}{{\em Chin. Phys.}
  {\bfseries C40} no.~10, (2016) 100001}.

\bibitem{Hoang:2008xm}
A.~H. Hoang and I.~W. Stewart, ``{Top Mass Measurements from Jets and the
  Tevatron Top-Quark Mass},''
  \href{http://dx.doi.org/10.1016/j.nuclphysbps.2008.10.028}{{\em Nucl. Phys.
  Proc. Suppl.} {\bfseries 185} (2008) 220--226},
\href{http://arxiv.org/abs/0808.0222}{{\ttfamily arXiv:0808.0222 [hep-ph]}}.

\bibitem{Butenschoen:2016lpz}
M.~Butenschoen, B.~Dehnadi, A.~H. Hoang, V.~Mateu, M.~Preisser, and I.~W.
  Stewart, ``{Top Quark Mass Calibration for Monte Carlo Event Generators},''
  \href{http://dx.doi.org/10.1103/PhysRevLett.117.232001}{{\em Phys. Rev.
  Lett.} {\bfseries 117} no.~23, (2016) 232001},
\href{http://arxiv.org/abs/1608.01318}{{\ttfamily arXiv:1608.01318 [hep-ph]}}.

\bibitem{Hoang:2017suc}
A.~H. Hoang, A.~Jain, C.~Lepenik, V.~Mateu, M.~Preisser, I.~Scimemi, and I.~W.
  Stewart, ``{The MSR Mass and the ${\cal O}(\Lambda_{\rm QCD})$ Renormalon Sum
  Rule},''
\href{http://arxiv.org/abs/1704.01580}{{\ttfamily arXiv:1704.01580 [hep-ph]}}.

\bibitem{Bezrukov:2014ina}
F.~Bezrukov and M.~Shaposhnikov, ``{Why should we care about the top quark
  Yukawa coupling?},'' \href{http://dx.doi.org/10.1134/S1063776115030152}{{\em
  J. Exp. Theor. Phys.} {\bfseries 120} (2015) 335--343},
  \href{http://arxiv.org/abs/1411.1923}{{\ttfamily arXiv:1411.1923 [hep-ph]}}.
[Zh. Eksp. Teor. Fiz.147,389(2015)].

\bibitem{Beneke:2016cbu}
M.~Beneke, P.~Marquard, P.~Nason, and M.~Steinhauser, ``{On the ultimate
  uncertainty of the top quark pole mass},''
\href{http://arxiv.org/abs/1605.03609}{{\ttfamily arXiv:1605.03609 [hep-ph]}}.

\bibitem{Andreassen:2017ugs}
A.~Andreassen and M.~D. Schwartz, ``{Reducing the Top Quark Mass Uncertainty
  with Jet Grooming},''
\href{http://arxiv.org/abs/1705.07135}{{\ttfamily arXiv:1705.07135 [hep-ph]}}.

\bibitem{Derrick:1964ww}
G.~H. Derrick, ``{Comments on nonlinear wave equations as models for elementary
  particles},''
\href{http://dx.doi.org/10.1063/1.1704233}{{\em J. Math. Phys.} {\bfseries 5}
  (1964) 1252--1254}.

\bibitem{Lee:1985uv}
K.-M. Lee and E.~J. Weinberg, ``{Tunneling without barriers},''
\href{http://dx.doi.org/10.1016/0550-3213(86)90150-1}{{\em Nucl.Phys.}
  {\bfseries B267} (1986) 181}.

\bibitem{Brown:2011um}
A.~R. Brown and A.~Dahlen, ``{The Case of the Disappearing Instanton},''
  \href{http://dx.doi.org/10.1103/PhysRevD.84.105004}{{\em Phys.Rev.}
  {\bfseries D84} (2011) 105004},
\href{http://arxiv.org/abs/1106.0527}{{\ttfamily arXiv:1106.0527 [hep-th]}}.

\bibitem{Ford:1992pn}
C.~Ford, I.~Jack, and D.~R.~T. Jones, ``{The Standard model effective potential
  at two loops},'' \href{http://dx.doi.org/10.1016/0550-3213(92)90165-8,
  10.1016/S0550-3213(97)00532-4}{{\em Nucl. Phys.} {\bfseries B387} (1992)
  373--390}, \href{http://arxiv.org/abs/hep-ph/0111190}{{\ttfamily
  arXiv:hep-ph/0111190 [hep-ph]}}.
[Erratum: Nucl. Phys.B504,551(1997)].

\end{thebibliography}\endgroup


\providecommand{\href}[2]{#2}\begingroup\raggedright\endgroup
\bibliographystyle{utphys}

\end{document}